\pdfoutput=1
\documentclass[useAMS]{mn2e}
\usepackage{times}
\usepackage{amssymb}
\usepackage{amsmath}
\usepackage{rotating}
\usepackage{color}
\usepackage{epsfig}
\usepackage{graphicx}
\input{epsf}

\title[A Physical Model for NLS1 X-ray Time Lags]
{A Physical Model for the X-ray Time Lags of Narrow Line Seyfert Type 1 Active Galactic Nuclei}

\author[E. Gardner and C. Done]
{Emma Gardner and Chris Done\\
Department of Physics, University of Durham, South Road,
Durham DH1 3LE, UK\\}

\date{Submitted to MNRAS}
\pagerange{\pageref{firstpage}--\pageref{lastpage}} \pubyear{}

\begin{document}

\topmargin = -0.5cm

\maketitle

\label{firstpage}

\begin{abstract}

We study the origin of the soft X-ray excess seen in the 'simple' Narrow Line Seyfert 1 galaxy PG1244+026 using all available spectral-timing information. This object shows the now ubiquitous switch between soft leading the hard band on long timescales, to the opposite behaviour on short timescales. This is interpreted as a combination of intrinsic fluctuations propagating down through the accretion flow giving the soft lead, together with reflection of the hard X-rays giving the soft lag. We build a full model of the spectral and time variability including both propagation and reflection, and compare our model with the observed power spectra, coherence, covariance, lag-frequency and lag-energy spectra. We compare models based on a separate soft excess component with those based on reflection dominated soft emission.

Reflection dominated spectra have difficulty reproducing the soft lead at low frequency since reflection will always lag. They also suffer from high coherence and nearly identical hard and soft band power spectra in disagreement with the observations. This is a direct result of the power law and reflection components both contributing to the hard and soft energy bands, and the small radii over which the relativistically smeared reflection is produced allowing too much high frequency power to be transmitted into the soft band. 

Conversely, we find the separate soft excess models (where the inner disc radius is $>6R_g$) have difficulty reproducing the soft lag at high frequency, as reflected flux does not contribute enough signal to overwhelm the soft lead. However, reflection should also be accompanied by reprocessing and this should add to the soft excess at low energies. This model can quantitatively reproduce the switch from soft lead to soft lag seen in the data and reproduces well the observed power spectra and other timing features which reflection dominated models cannot. 

\end{abstract}


\section{Introduction} \label{sec:introduction}

The discovery of a lagged signal in the X-ray lightcurves of AGN has led to a breakthrough in our
ability to probe the structure of the emission region on the smallest scales.  
The simplest interpretation of this lag is that it is due to the light travel time delayed 
response of the disc to X-ray illumination (Fabian et al 2009). However, the lag behaviour is complex; 
at long timescales (low frequency) the hard X-rays lag behind the soft, while at high
frequencies the opposite is true (Papadakis et al 2001; Vaughan et al 2003; McHardy et al 2004; Fabian et al 2009; Zoghbi et al 2011; Emmanoulopoulos et al 2011; De Marco et al 2013). This is interpreted as the interplay of two different processes, 
with propagation of fluctuations on longer timescales giving rise to the hard lags, while 
reverberation from the disc takes over on shorter timescales. Reflection from the disc produces 
an iron line and Compton hump above 10~keV, but can also contribute to the soft X-ray band
if the disc is partially ionised, producing a soft lag at high frequencies (Fabian et al 2009; Zoghbi et al 2011; Cackett et al 2013). 

However, an alternative explanation for the soft lags at high frequency was suggested by Alston, Done \& Vaughan
(2014, hereafter ADV14, also Zoghbi et al 2011). They noted that another feature of X-ray illumination of the disc is that the non-reflected flux is reprocessed as thermal emission. 
This provides another component which reverberates behind the hard X-rays in the same way as reflection, and can contribute to the soft X-ray bandpass 
even if reflection itself does not produce much soft X-ray flux. This is important as there is a
long standing controversy over the origin of the soft X-ray emission seen in AGN. The 2-10~keV flux is dominated
by a power law component, but there is a ubiquitous excess over and above the extrapolation of the power law, at lower energies ($<1keV$). This soft X-ray excess has a smooth shape, with no clear atomic signatures, so is well fit by an
additional Comptonised component. However, the temperature of this is
remarkably constant at $\sim 0.2$~keV across a range in mass and mass accretion rates, requiring
some unknown fine tuning mechanism (Czerny et al 2003; 
Gierlinski \& Done 2004; Porquet et al 2004). An alternative model is that the 
soft X-ray excess is produced by reflection from a partially ionised disc. 
However this also requires fine tuning of the ionisation state of the disc in order to consistently generate sufficient change in opacity from partially ionised material at $\sim 0.7keV$ (Done \& Nayakshin 2007), 
and extreme relativistic effects are required in order to smear out the resulting emission lines, as these are not seen in the data (Fabian et al 2004; Crummey et al 2006).

While the temperature of the soft excess is remarkably constant, the fraction of power carried by this component is not, being systematically higher at lower $L/L_{Edd}$. It can contain as much as half the total accretion power in Broad Line Seyfert 1 galaxies
(BLS1: typically $10^8M_\odot$, with $L/L_{Edd}=0.1$), while it carries less than 10 per cent in the Narrow Line Seyfert 1s (NLS1s: typically $10^{6-7}M_\odot$, with $L/L_{Edd}\sim 1$). This is initially surprising as one of the defining characteristics of NLS1s is their strong soft X-ray emission (Boller, Brandt \& Fink 1996). 
However, their relatively 
small black hole masses and high accretion rates mean that the disc itself should contribute to the soft 
X-ray bandpass if it extends down close to the innermost stable circular orbit (ISCO). Nonetheless an additional soft excess component is still required to fit the spectra since the disc rolls over very rapidly past its maximum temperature, whereas the data show a much more gradual decline. 
Thus there is also soft excess emission in NLS1s, but the majority of their strong soft X-ray flux is from the disc
(Done et al 2012; 2013). 

Both models for the soft X-ray excess - the reflection and the additional Compton component - can give equally good fits to the X-ray spectra 
in the 0.3-10~keV CCD bandpass, though including higher energy NuStar data in the BLS1 Akn 120 strongly favours  the additional Comptonisation model (Matt et al 2014). Variability gives 
another way to distinguish between these two 
models, with long timescale correlated variability between the UV and soft excess in the BLS1 Mkn 509 also favouring the Compton model (Mehdipour et al 2011). 
However, short timescale variability is likely to be more constraining, in particular the existence of the lags. 
The switch in sign noted above, from soft leading at low frequencies to hard leading at higher frequencies, 
requires that there is some part of the soft variability which propagates down from soft to harder X-rays, while
the soft lag at high frequencies requires that some part of the soft emission is reprocessed from the hard X-rays.  A single X-ray power law and its partially ionised, highly smeared 
reflected emission cannot do this switch in behaviour, 
as a reflection origin for the entire soft X-ray excess means that it can only lag. 

Thus the spectrum must be more complex than a single power law and its reflection. 
In the most extreme NLS1s such as 1H0707-495, the spectrum is instead fit with two power law components and their reflection, together with a blackbody component at the very lowest energies (Zoghbi et al 2011, Kara et al 2013a). This then gives a potential origin
for the propagation lags where fluctuations propagate down from the disc to the soft power law to the hard power law and are then reflected. Similarly, in the ADV14 model, there is a separate soft excess component, so 
the fluctuations propagate down from the disc to the soft excess to the power law and are then reflected. 

In this paper we model the X-ray time lags at both high and low frequencies, looking especially at the
additional information which comes from the low frequency propagation lags as well as the high frequency
reverberation lags. We build a fully self consistent spectral and timing model, starting from fluctuations generated intrinsically in the accretion flow which then propagate through the system and including reflection/reprocessing with realistic transfer functions. We focus on explaining the data from the NLS1 PG1244+026 where a range of 
spectral and timing properties favour a separate soft excess model; in particular the low energy drop 
in the spectrum of fast variability correlated with the 4-10 keV spectrum shows that the soft excess provides the seed photons for the power law (Jin et al 2013, hereafter J13). We explore the lags in this model, and then compare them with the lags expected from the reflection models for the soft X-ray excess, including the one specifically for this object (Kara et al 2013b, hereafter K13). We show that the propagation lags strongly favour models where the soft excess is predominantly from an additional continuum component rather than being dominated by reflection, and that they rule out models where the soft X-ray excess is produced externally in the jet (K13). We show reflection dominated models struggle to match other key observed timing features such as power spectra, coherence and covariance spectra, whilst the separate soft excess model can quantitatively reproduce all these timing features. 

\begin{figure*}
\centering
\begin{tabular}{l|r}
\leavevmode  
\includegraphics[width=8cm]{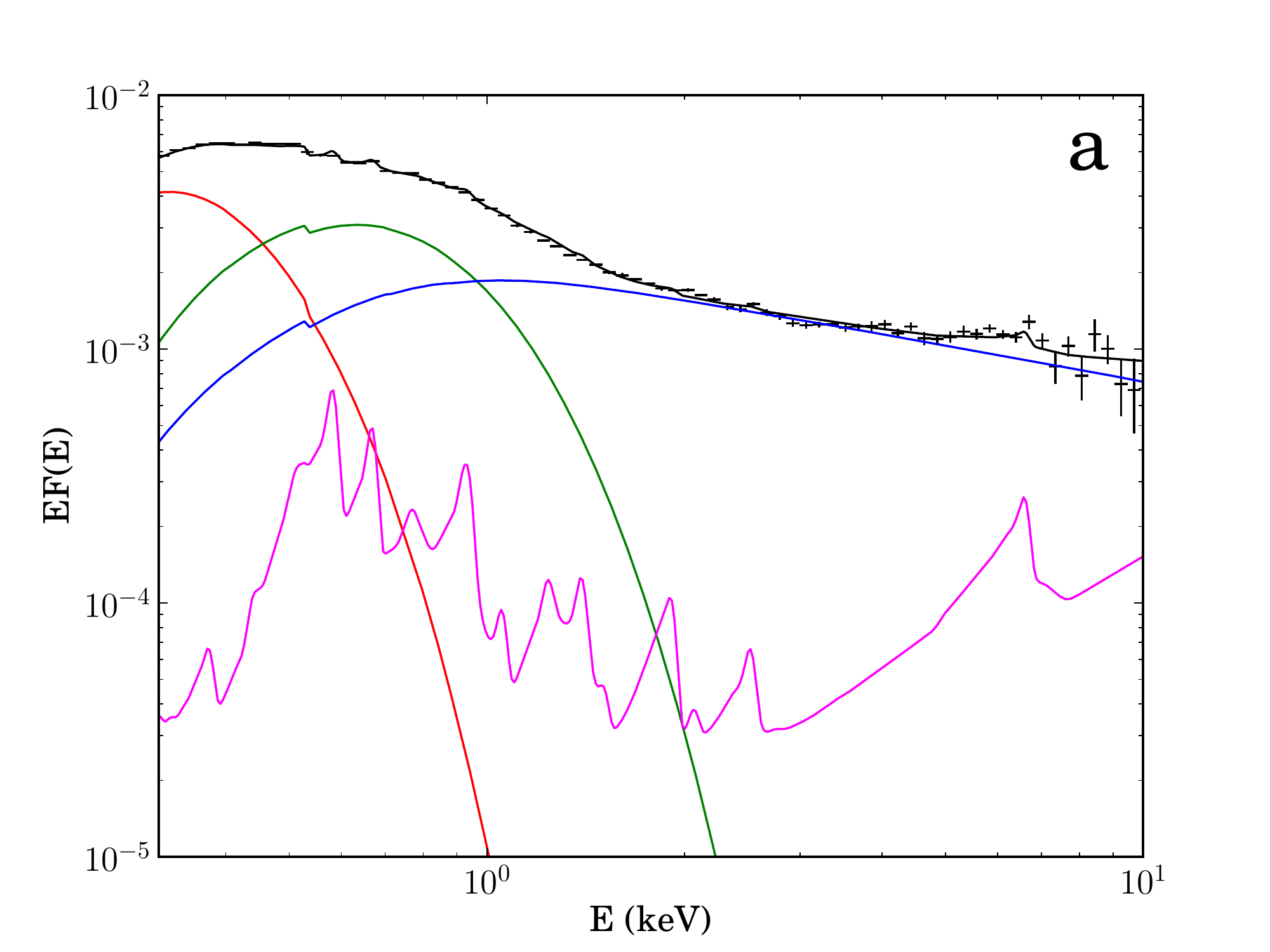} &
\includegraphics[width=8cm]{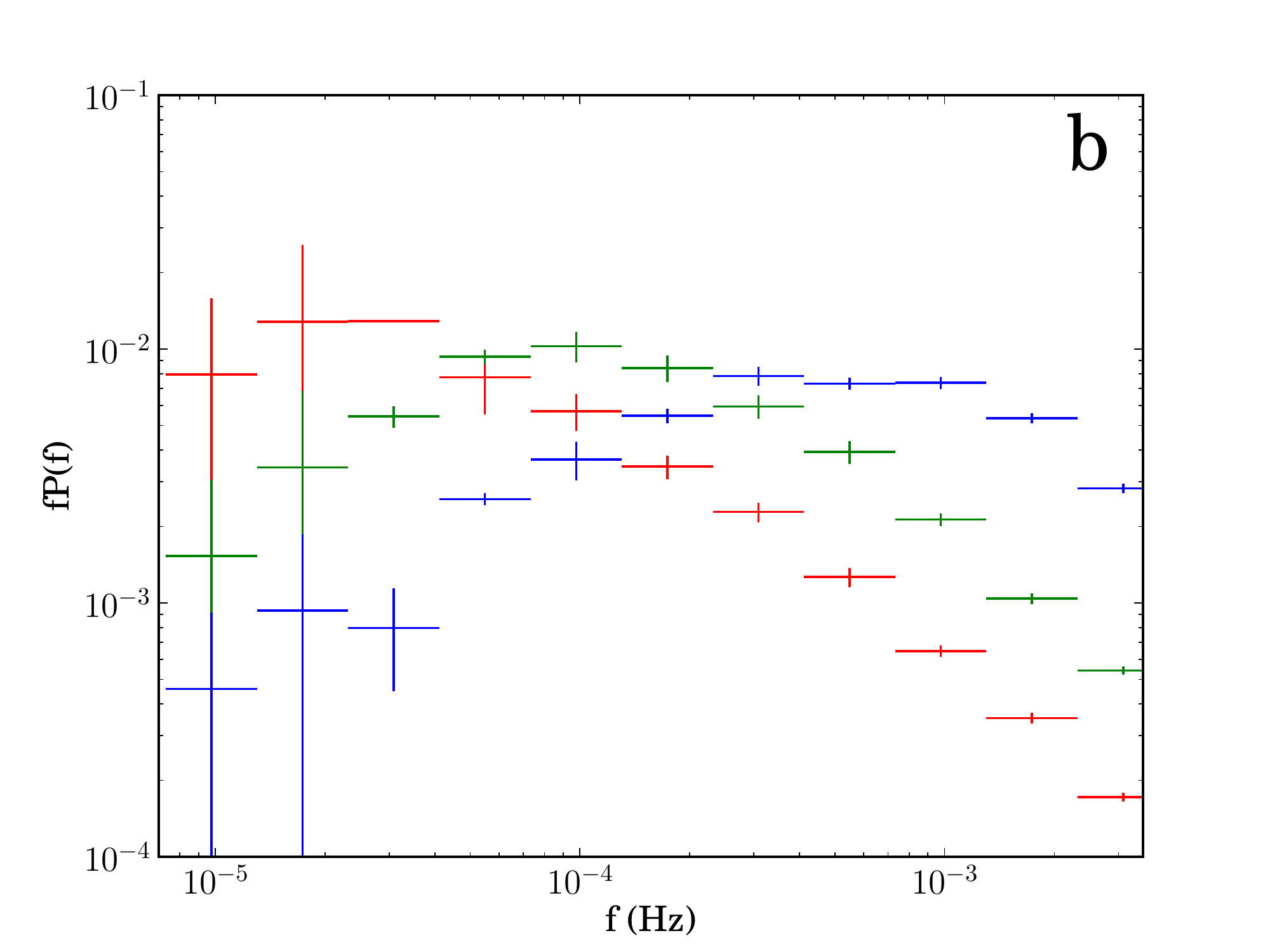} \\
\end{tabular}
\caption{a). Spectral decomposition for PG1244+026: disc (red), soft excess (green), coronal power law (blue), reflection (magenta), total (black). Data points show time averaged spectrum (OBS ID: 0675320101). b). Power spectrum of intrinsic fluctuations in each component: disc (red), soft excess (green), coronal power law (blue).}
\label{fig1}
\end{figure*}

\begin{table}
\begin{tabular}{lll}
\hline
Component & Parameter & Value \\
\hline
Galactic absorption & $N_h$ ($10^{22}cm^{-2}$) & $0.019$ \\
Intrinsic absorption & $N_h$ ($10^{22}cm^{-2}$) & $6.1\times10^{-11}$ \\
 & redshift & 0.048 \\
Blackbody & $kT (keV)$ & 0.062 \\
 & norm & $2.0\times10^{-4}$ \\
nthComp & $\Gamma$ & 2.4 \\
 & $kT_e (keV)$ & 100 \\
 & norm & $1.9\times10^{-3}$ \\
kdblur & index & 3.0 \\
 & $r_{in} (R_g)$ & 11 \\
 & $i$ ($^{\circ}$) & 30 \\
rfxconv & relative refl norm & -0.65 \\
 & $Z_{Fe}$ & 1.0 \\
 & $\log(x_i)$ & 1.3 \\
compTT & $kT_e (keV)$ & 0.15 \\
 & $\tau$ & 190 \\
 & norm & 0.073 \\
\hline
\end{tabular}
\caption{Parameters for the spectral model shown in fig 1a, with a separate optically thick Comptonisation component producing the soft excess emission: {\sc{wabs*zwabs(bbody+nthcomp+kdblur*rfxconv*nthcomp+comptt)}}.}
\label{table1}
\end{table}

\section{Spectral Decomposition}

We use the long (120ks) XMM-Newton observation of PG1244+026 (OBS ID: 0675320101, as studied by J13; ADV14 and K13). We use a similar spectral decomposition to J13 which assumes that the spectrum is composed of three components: a (colour temperature corrected) 
blackbody (BB) disc, 
a low temperature optically thick Compton component to describe the soft excess, and a second optically thin Compton component to describe the hard X-ray power law. We model these together using the 
publicly available model {\sc{optxagnf}}, which assumes that these three components are all powered by the
accretion flow, so the luminosity to power the soft excess and power law sets the truncation radius $R_{cor}$
of the standard BB disc emission (Done et al 2012). This model assumes that the soft excess 
arises at radii $<R_{cor}$, and 
a plausible origin is that this represents the inner regions of the standard disc, but that the emission
here does not completely thermalise, perhaps because of a larger scale height (and hence
lower density) expected if there are strong winds from the disc (Done et al 2012). 

We also include moderately ionised reflection of the power law off the disc in order to account for reflection features at $6-7keV$ due to iron. J13 showed two extreme fits, one where the disc was mostly neutral, the other where it was
highly ionised. ADV14 used these to explore the shape of the lag spectra, but here we use the
more likely (and better fitting) moderate ionisation reflection model. Fig 1a shows the model components compared to the data, with full parameters detailed in table 1. We fix the inclination of the reflector to $30^{\circ}$ and match the seed photon temperature for the high energy Comptonisation to the temperature of the soft excess.

\section{Time Dependent Model}

Ideally, the full (energy and frequency dependent) cross-spectrum 
should be directly fit to 
constrain the intrinsic components, both their energy spectra and
variability properties. Since this inverse
problem has not yet been solved, we do forward fitting instead,
using the spectral components from model fitting, and estimating
their variability from a combination of the observed power spectra and lag-frequency spectra. We then calculate the various spectral-timing properties, and qualitatively compare these to the observations. 

There have been many timing studies of PG1244+026. In order to maximise diagnostic power, we choose to use the power spectra and covariance measured by J13 and the coherence, lag-frequency and lag-energy spectra measured by ADV14. We note that ADV14 used slightly different soft and hard energy bands ($0.3-0.7keV$ and $1.2-4.0keV$) compared to J13 ($0.3-1.0keV$ and $2-10keV$). Therefore when calculating power spectra we use the hard and soft bands of J13 and use the energy bands of ADV14 when calculating lags and coherence, in order to match to the data. For completeness, we show equations for these calculations in appendix A.

We use the spectral components derived from the previous section
- BB disc, soft excess, power law and reflection - to determine the relative contributions of each component to a given energy band. This then determines the observed power spectrum of that energy band. Since the observed power spectra of the hard and soft bands are different, this implies different components dominate each band and these components have fluctuations at different characteristic frequencies. 

\subsection{Intrinsic Fluctuations: no propagation}

\begin{figure*} 
\centering
\begin{tabular}{l|r}
\leavevmode  
\includegraphics[width=8cm]{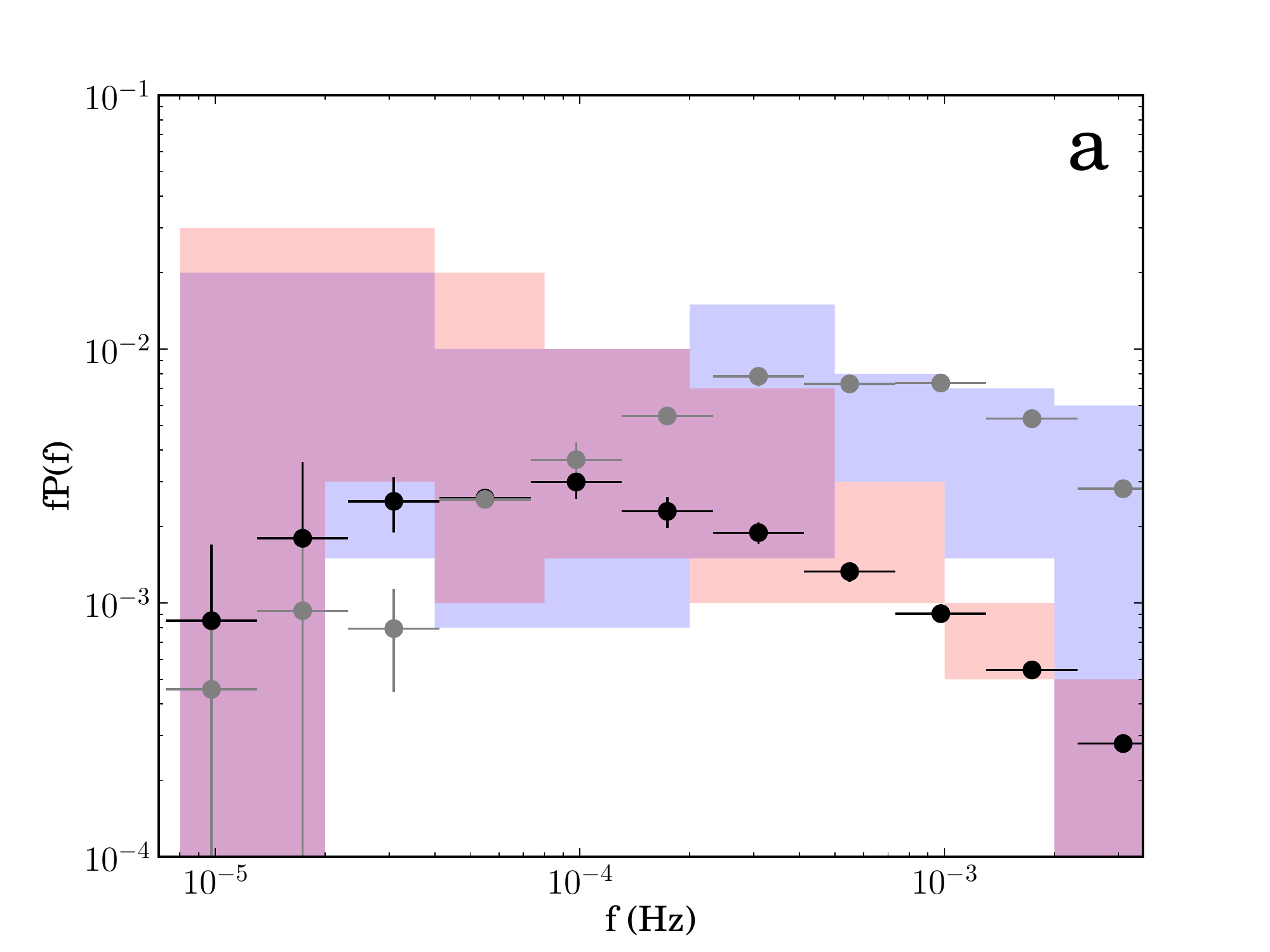} &
\includegraphics[width=8cm]{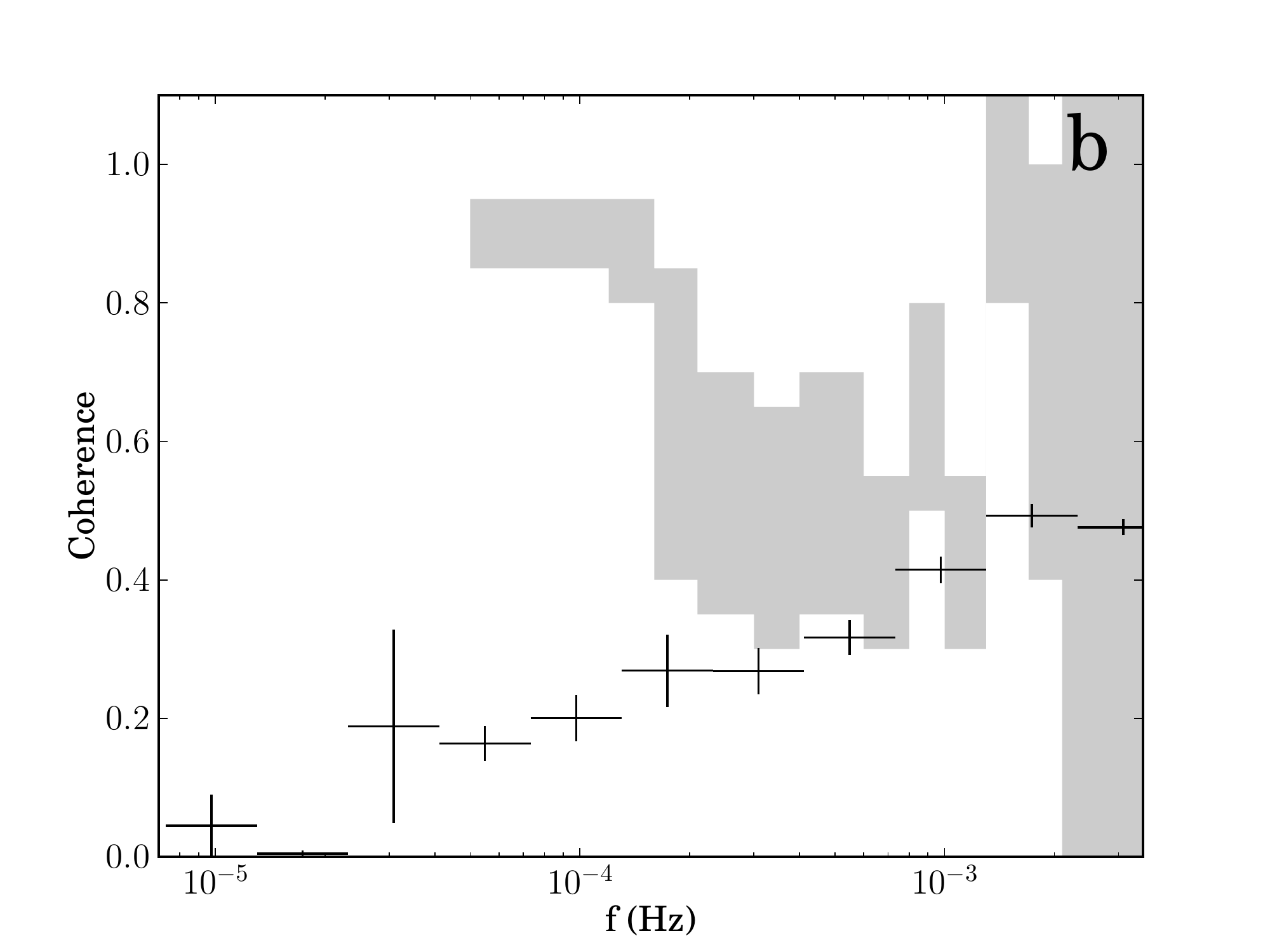} \\
\end{tabular}
\caption{Model with only intrinsic fluctuations, no propagation between components. a). Soft band ($0.3-1keV$, black) and hard band ($2-10keV$, grey) model power spectra. Red (blue) regions show range of error on soft (hard) band power spectra measured by J13 for PG1244+026. b). Coherence between hard ($1.2-4keV$) and soft ($0.3-0.7keV$) band lightcurves. Grey regions show range of error on coherence measured by ADV14 for PG1244+026.}
\label{fig2}
\end{figure*}

We assume fluctuations are generated intrinsically within the BB disc, soft excess and corona. We assume the power spectrum of these fluctuations takes the form of a Lorentzian centred around some characteristic frequency - $f_d$, $f_s$ or $f_p$ respectively. This characteristic frequency will be different for each component and can be associated with a viscous timescale in analogy with propagating fluctuation models in X-ray binaries (Kotov et al 2001; Arevalo \& Uttley 2006; Ingram \& Done 2011; 2012). Since our physical model describes an outer disc, inner soft excess and corona, the relevant radii for each component  
decreases from disc to soft excess to corona (and scale height probably increases also), so the characteristic frequency of fluctuations should increase. 

This increase in characteristic frequency can be seen in the power spectra of PG1244+026 (J13, ADV14). The soft band power spectra show more power at low frequencies ($f<10^{-4}Hz$), whilst the hard band power spectra show much more high frequency power ($f>10^{-4}Hz$). The corresponding lag-frequency spectrum (ADV14) shows the soft band leading the hard band at low frequencies and lagging it at high frequencies. This implies the low frequency fluctuations must be generated in the soft band components (or at least pass through them) before reaching the hard power law component, whilst the hard power law component must be the source of the high frequency fluctuations, which then reverberate in the soft band to produce soft lags at high frequencies. The switch between soft leads and soft lags occurs at $\sim10^{-4}Hz$ in PG1244+026, implying the characteristic frequency of intrinsic fluctuations in the soft band components (disc and soft excess) are at frequencies below $10^{-4}Hz$, whilst the hard power law generates frequencies above $10^{-4}Hz$.

Motivated by the observed power spectra and lag-frequency spectrum, we choose $f_d=3\times10^{-5}Hz$ as the characteristic frequency of the fluctuations in the disc, $f_s=1\times10^{-4}Hz$ for the soft excess, and use two Lorentzians ($f_{p,1}=3\times10^{-4}Hz$ and $f_{p,2}=1\times10^{-3}Hz$) to describe the breadth of the high frequency variability in the coronal power law. We use the method of Timmer \& Koenig (1995) to generate fluctuations in each component,  $\dot{M}_{d,s,p}(t)$. Each $\dot{M}_{d,s,p}(t)$ is normalised to a mean of unity and fractional variability $\sigma/I=F_{d,s,p}$. This represents fluctuations of the component around its time averaged value determined from the fit to the time averaged spectrum. 

For each time, $t$, the time averaged spectral components are multiplied by their fluctuations $\dot{M}_{d,s,p}(t)$ at that time, and summed to give the total spectrum. In order to allow for red noise leakage, we calculate the total spectrum for $2^{20}$ timesteps with $dt=100s$ and then split it into ten $102.4ks$ segments. For each segment we calculate the power spectrum in the hard ($2-10keV$) and soft ($0.3-1keV$) bands and average the ten power spectra to get a mean power spectrum. 

Fig 1b shows the power spectrum of the fluctuations in each component (red: disc, green: soft excess, blue: power law). Fig 2a shows the resulting hard and soft band power spectra in grey and black respectively, analogous to those in J13. The red and blue regions show the error on the soft and hard band power spectra measured by J13 for PG1244+026. The observed hard and soft power spectra are the same within a factor of 2 below $10^{-4}Hz$ and a factor of 10 different at $10^{-3}Hz$, with much less high frequency power in the soft band. In our model, the power in the soft band is much less than that in the individual disc and soft excess components which contribute to it, because the power in both components is incoherent, so summing them reduces the total power (Arevalo \& Uttley 2006). In contrast, the hard band is dominated by the power law with little dilution from other components, so the hard band power spectrum is very similar to the power spectrum of the power law component. 

Fig 2b shows the coherence between the hard and soft bands, where 1 is perfect coherence and 0 is incoherence (Vaughan \& Nowak 1997; Nowak et al 1999). The components themselves are incoherent, so the coherence between the soft excess and power law is 0. Although the hard band is dominated by the power law and the soft band by the soft excess, the soft band also contains a non negligible contribution from the power law. Thus some fraction of the soft band light curve is correlated with the hard band, due to the presence of the power law in both bands. Consequently the coherence at frequencies produced by the power law is equal to the fractional contribution of the power law to the soft band. 

The grey regions in fig 2b show the range of the error on the coherence measured by ADV14 for PG1244+026. The measured coherence clearly shows the opposite trend, with the coherence being highest for low frequencies and dropping between $10^{-4}-10^{-3}Hz$. In order to replicate this, we must allow fluctuations to propagate inwards from the slowly varying soft components to the faster varying coronal power law which dominates the hard band. 

\begin{figure*}
\centering
\begin{tabular}{l|r}
\leavevmode  
\includegraphics[width=8cm]{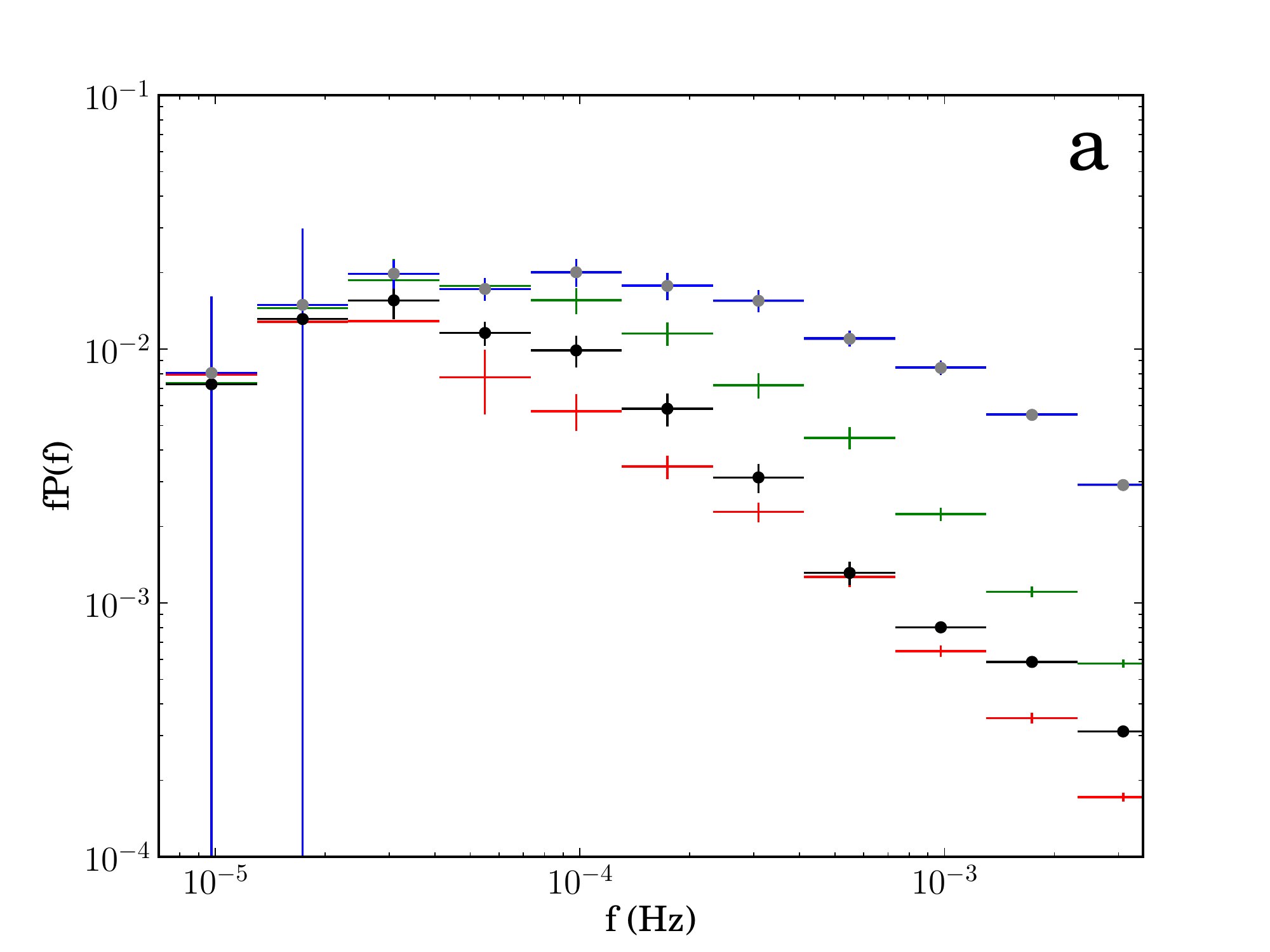} &
\includegraphics[width=8cm]{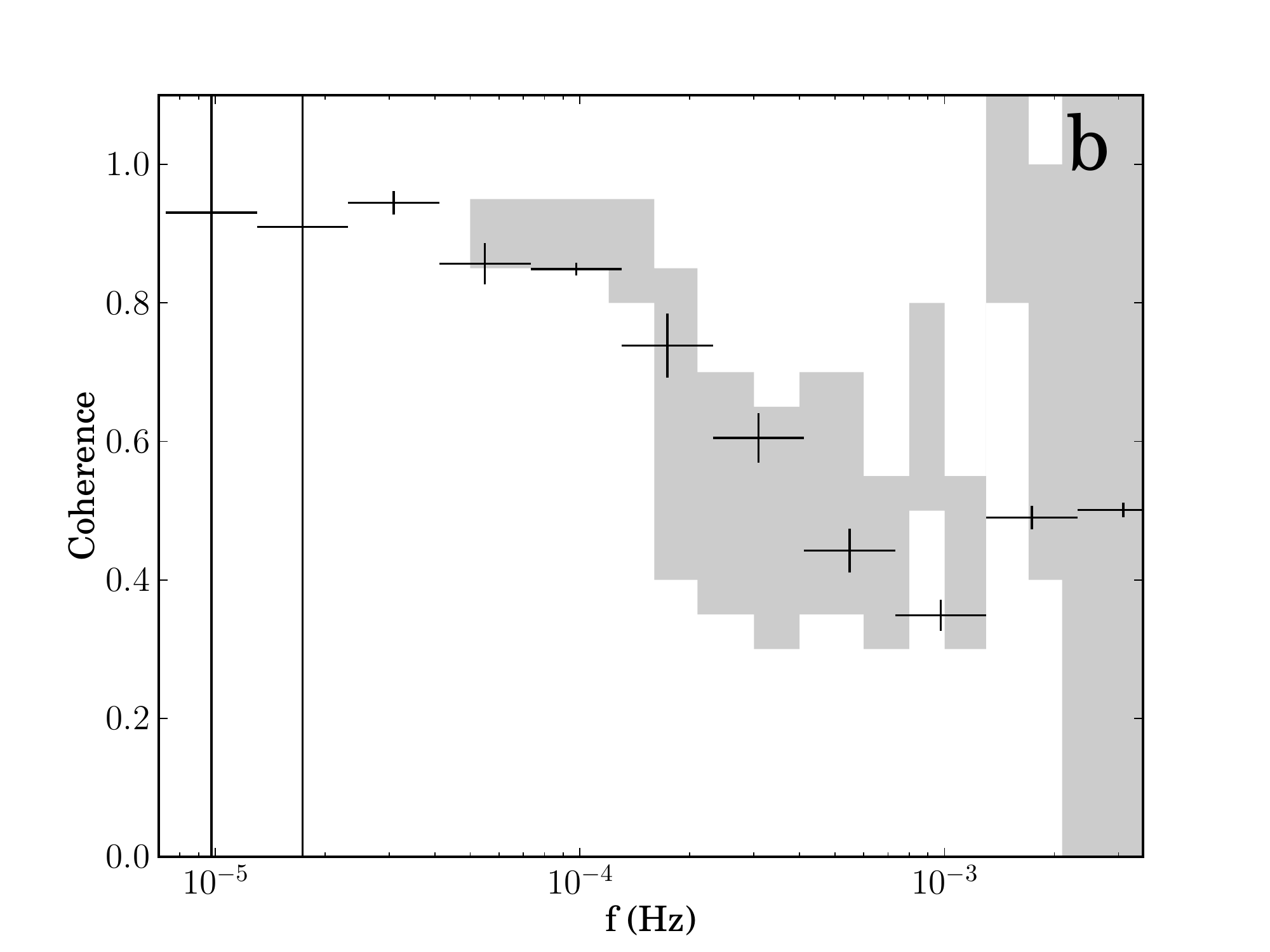} \\
\includegraphics[width=8cm]{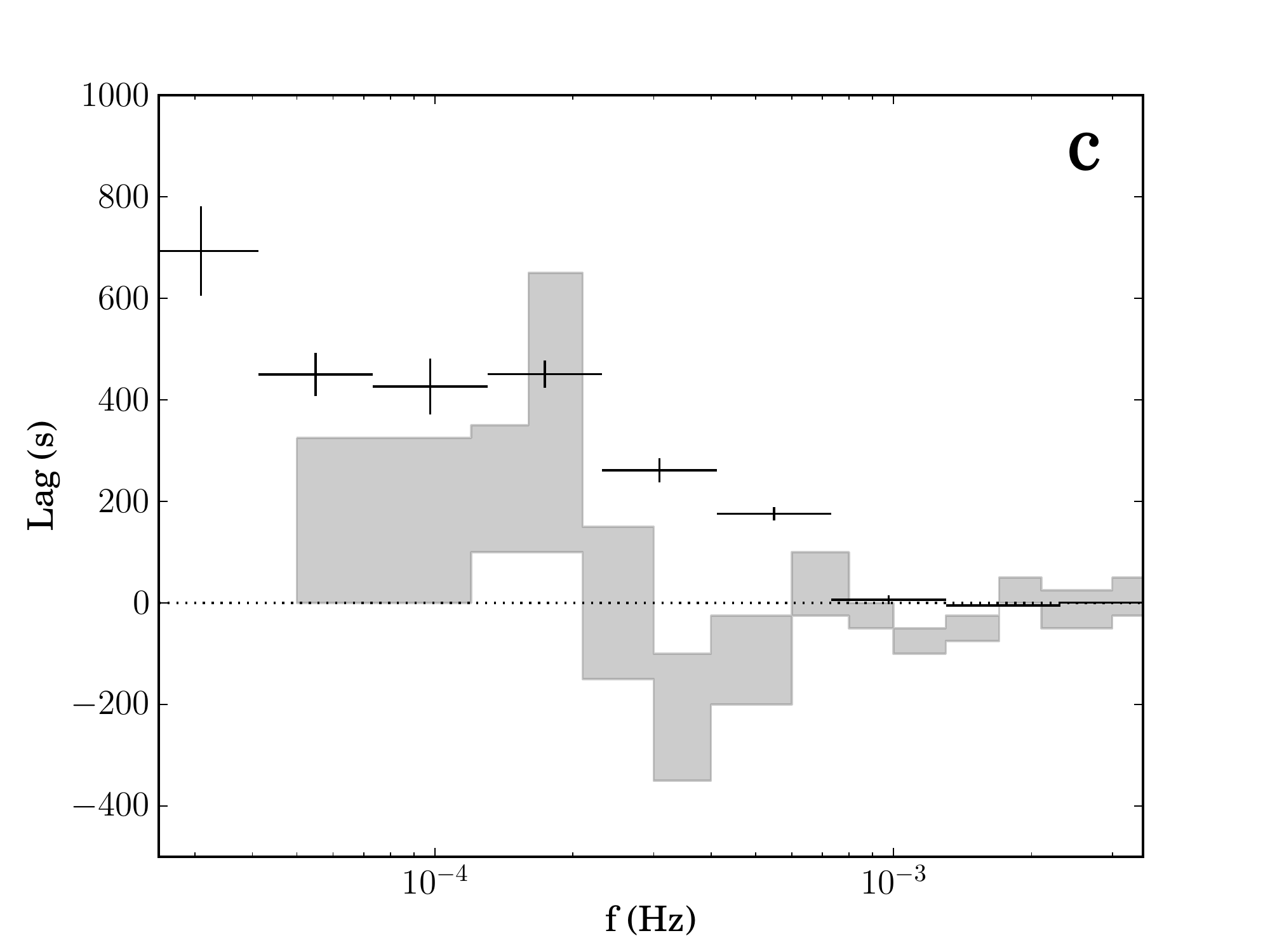} &
\includegraphics[width=8cm]{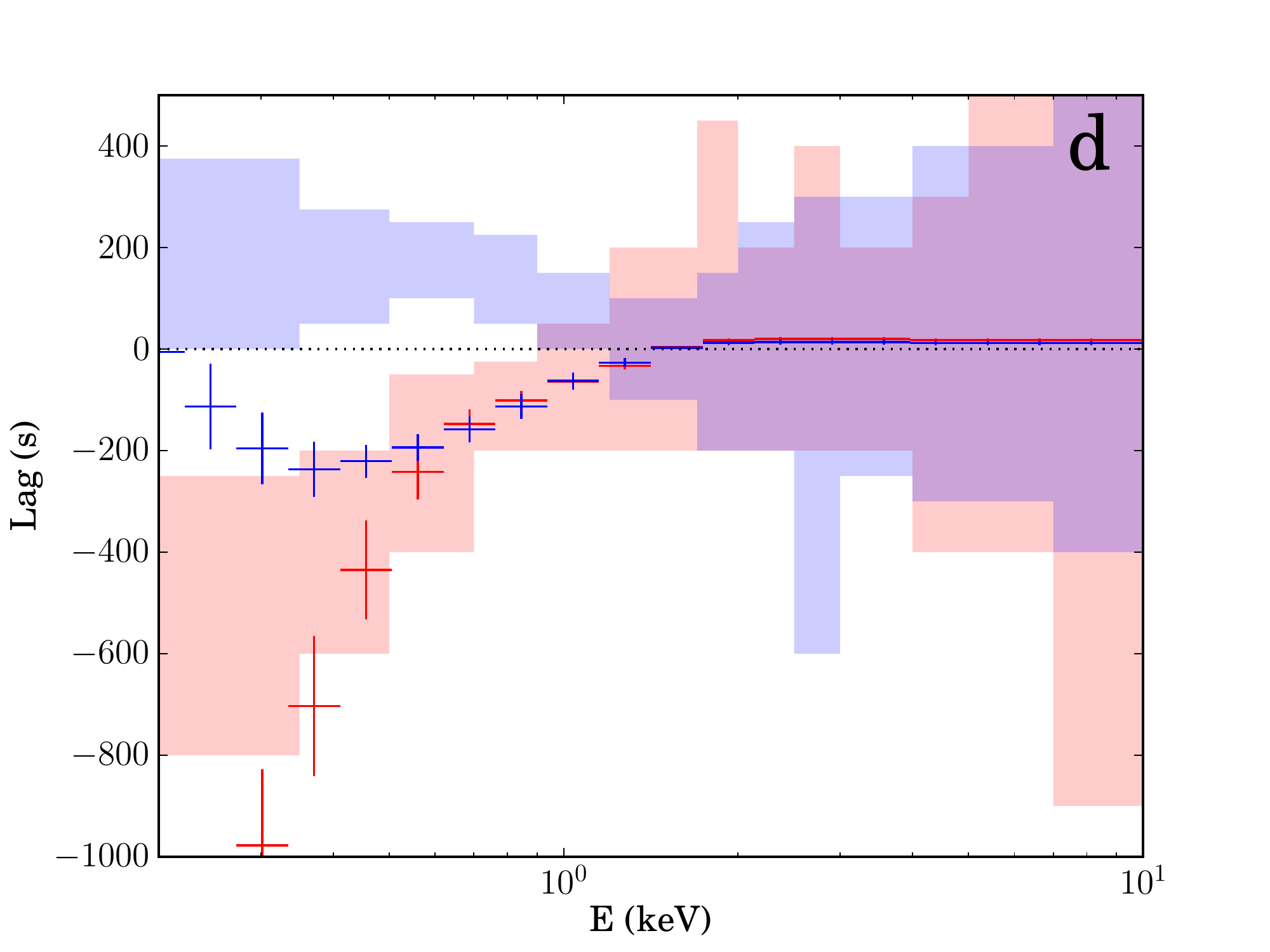} \\
\end{tabular}
\caption{Model allowing fluctuations to propagate from disc to soft excess ($t_{lag,s}=1000s$) to coronal power law ($t_{lag,p}=600s$). a). Power spectra: disc (red), soft excess (green), coronal power law (blue), soft band ($0.3-1.0keV$, black), hard band ($2-10keV$, grey). b). Coherence between hard ($1.2-4keV$) and soft ($0.3-0.7keV$) bands. c). Lag-frequency spectrum between hard ($1.2-4keV$) and soft ($0.3-0.7keV$) bands. d). Lag-energy spectrum calculated using $1.2-4keV$ reference band: low frequency lag ($2.3\times10^{-5} - 7.3\times10^{-5}Hz$), red points; high frequency lag ($2.3\times10^{-4} - 7.3\times10^{-4}Hz$), blue points. Shaded regions show range of error on values measured by ADV14 for PG1244+026.}
\label{fig3}
\end{figure*}

\subsection{Propagating Fluctuations}

We now allow the fluctuations to propagate inwards from the disc into the soft excess and then to the corona. We choose this scenario, rather than allowing fluctuations to pass directly from the disc into the corona, since Jin et al (2013) showed that the soft excess rather than the disc provides the seed photons for Comptonisation in the corona. This suggests that the soft excess and corona are more closely associated. Perhaps the disc and corona are spatially separated by the soft excess. Or more likely in PG1244+026 (where the inner disc radius derived from the spectral fit is $\sim12R_g$), the soft excess lies below the corona, with the corona formed from material that has evaporated from the optically thick soft excess region beneath it. A moderate scale height corona extended above the soft excess would then receive a greater seed photon flux from the soft excess than from the disc. This is in agreement with the characteristic frequencies of coronal fluctuations (which cannot vary on timescales faster than the source light crossing time), with the highest frequency fluctuations ($3\times10^{-3}Hz$) being generated at $\sim6R_g$ and the lower frequency fluctuations ($10^{-3}Hz$) being generated at larger radii $\sim10R_g$. 

In this scenario the fluctuations in the soft excess therefore consist of the intrinsic fluctuations ($\dot{M}_{s,int}(t)$) generated in the soft excess, modulated by fluctuations that have propagated inwards from the disc: 

\begin{equation}
\dot{M}_s(t) = \dot{M}_{s,int}(t)\dot{M}_{d,int}(t-t_{lag,s})
\end{equation}

\noindent where $t_{lag,s}$ is the time delay for propagation of fluctuations from the disc to the soft excess. We set $t_{lag,s}=1000s$, from the results of ADV14. As well as lagging, we also smooth the propagating fluctuations on this timescale using a sliding boxcar of width $t_{lag,s}$. We note that this is much shorter than the expected lag times for propagation, which should be related to the viscous timescale if NLS1s behaved as scaled up BHBs (Ingram \& Done 2011), ie. $t_{lag,s}\sim\frac{1}{f_{visc,d}}-\frac{1}{f_{visc,s}}\sim10^4s$. The light travel time between $20-12Rg$ is $\sim500s$ for PG1244+026 ($M_{BH}\sim10^7M_{\odot}$), so this could mean $t_{lag,s}$ represents a light travel time from the disc to the soft excess, ie. that the disc provides the seed photons for the optically thick Compton scattering in the soft excess. 

Similarly the fluctuations in the corona consist of the intrinsic coronal fluctuations ($\dot{M}_{p,int}(t)$) modulated by the smoothed and lagged fluctuations propagating inwards from the soft excess: 

\begin{equation}
\dot{M}_p(t) = \dot{M}_{p,int}(t)\dot{M}_s(t-t_{lag,p})
\end{equation}

\noindent where $t_{lag,p}$ is the time delay for propagation of fluctuations from the soft excess to the corona. We set $t_{lag,p}=600s$, again guided by ADV14, and again note that this is much closer to a light travel time than to a viscous propagation time. 

Fig 3a shows the power spectra of the individual components and of the hard and soft bands respectively. Due to accumulation of fluctuations, the hard band power spectrum now has more low frequency power than the soft band. The total power in the soft band has also increased, because propagation of fluctuations means more of the power in the soft excess is correlated with the disc power. Due to space, we no longer show the data but note that this model is indeed giving hard and soft power spectra which are within a factor of 2 at $10^{-4}Hz$ yet different by a factor of 10 at $10^{-3}Hz$ as required by the data. Fig 3b shows the coherence between the hard and soft bands, which is now highest at low frequencies due to the inward propagation of the slower outer fluctuations. This gives a much better match to the data. The coherence drops above $10^{-4}$ since this is the maximum typical frequency of fluctuations generated in the soft excess that can propagate down to the coronal power law, which dominates the hard band. 

In fig 3c and d we show the lag-frequency and lag-energy spectra. We define the lag as a function of frequency as $lag(f)=arg[C(f)]/(2\pi f)$, where $C(f)=S^*(f)H(f)$ is the complex valued cross spectrum, and $H$ is the Fourier transform of the hard band light curve and $S$ is the Fourier transform of the soft band light curve (Vaughan \& Nowak 1997; Nowak et al 1999). A positive time lag therefore corresponds to the soft band leading the hard, and negative lags to the soft band lagging the hard. 

At all frequencies below $\lesssim6\times10^{-4}Hz$, the soft band leads the hard, because these are the frequencies that are intrinsically generated in the disc and soft excess, which propagate down to the corona. 

This is shown more clearly in the lag-energy spectrum (fig 3d). For each energy bin we calculate the cross spectrum of the light curve of that energy bin with a hard reference band light curve (minus the energy bin light curve if the energy bin lies within the $1.2-4keV$ reference band). We plot the value of the time lag at low frequencies ($2.3\times10^{-5} - 7.3\times10^{-5}Hz$) in red in fig 3d. A negative lag now represents the energy bin leading the hard reference band. The light curves at energies dominated by disc and soft excess emission ($<1keV$) lead the hard reference band, with a lag that decreases as the energy of the bin increases. This matches the disc contribution decreasing with increasing energy and being replaced by the soft excess, which has a shorter time delay between it and the coronal emission, until $\sim1keV$ by which point the coronal emission begins to dominate the total spectrum and the low frequency lag reduces to $0$. 

In blue in fig 3d we plot the value of the time lags at high frequencies ($2.3\times10^{-4} - 7.3\times10^{-4}Hz$), corresponding to frequencies generated in the soft excess. These show a negative lag at low energies. The energy spectrum of the lag matches the energy spectrum of the soft excess (cf. fig 1a). The lag is a maximum for energy bins corresponding to the peak of the soft excess ($\sim0.3keV$). It decreases towards higher energies as the soft excess emission is gradually replaced by power law emission. It decreases towards lower energies because the soft excess emission is replaced by the disc, which doesn't generate strong fluctuations at high frequencies. However, even at its maximum the measured lag is only $\sim200s$, whilst the actual lag time for propagation from the soft excess to the power law is $600s$. This is a result of dilution. The soft band contains contributions from the disc and power law in addition to the soft excess component. The small contribution from the power law (which has zero lag with respect to the power law dominated hard band) reduces the net lag that is actually measured. Since we can only ever measure the \emph{net lag}, the lag measured between two bands should never be taken as representing a 'true' delay between components since it will always be a combination of the lags from each component contributing to that band, weighted by their contributions to the total flux in the band.

In fig 3c and d we also show the lag values measured by ADV14; the shaded regions show the range of the errors. Clearly the propagating model has no way of producing the observed soft lags (negative grey regions and positive blue regions). These are generally attributed to reverberation but there is an alternative way to produce spectral lags/leads and that is by spectral pivoting. We explore this first below. 

\begin{figure*} 
\centering
\begin{tabular}{l|r}
\leavevmode  
\includegraphics[width=8cm]{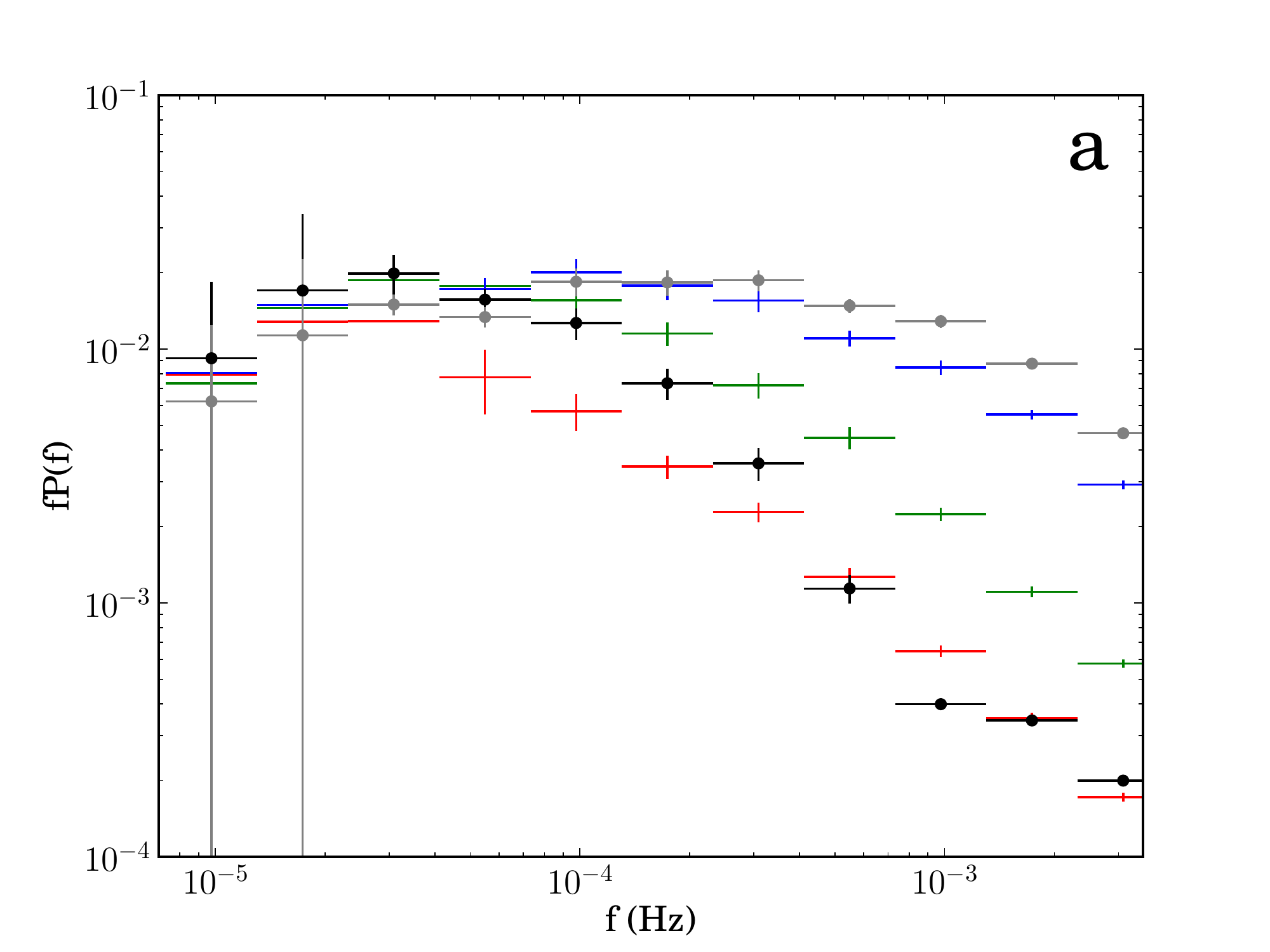} &
\includegraphics[width=8cm]{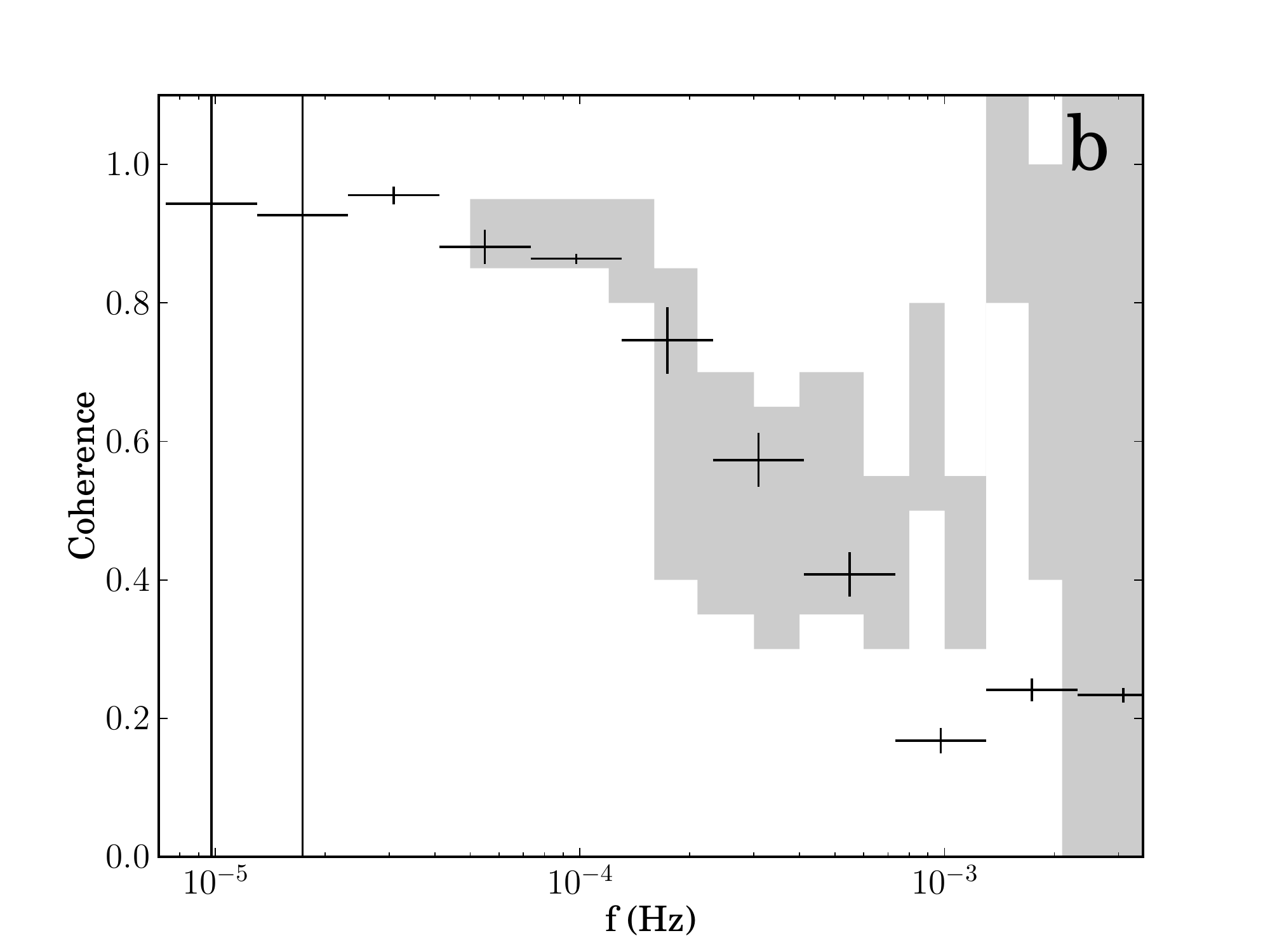} \\
\includegraphics[width=8cm]{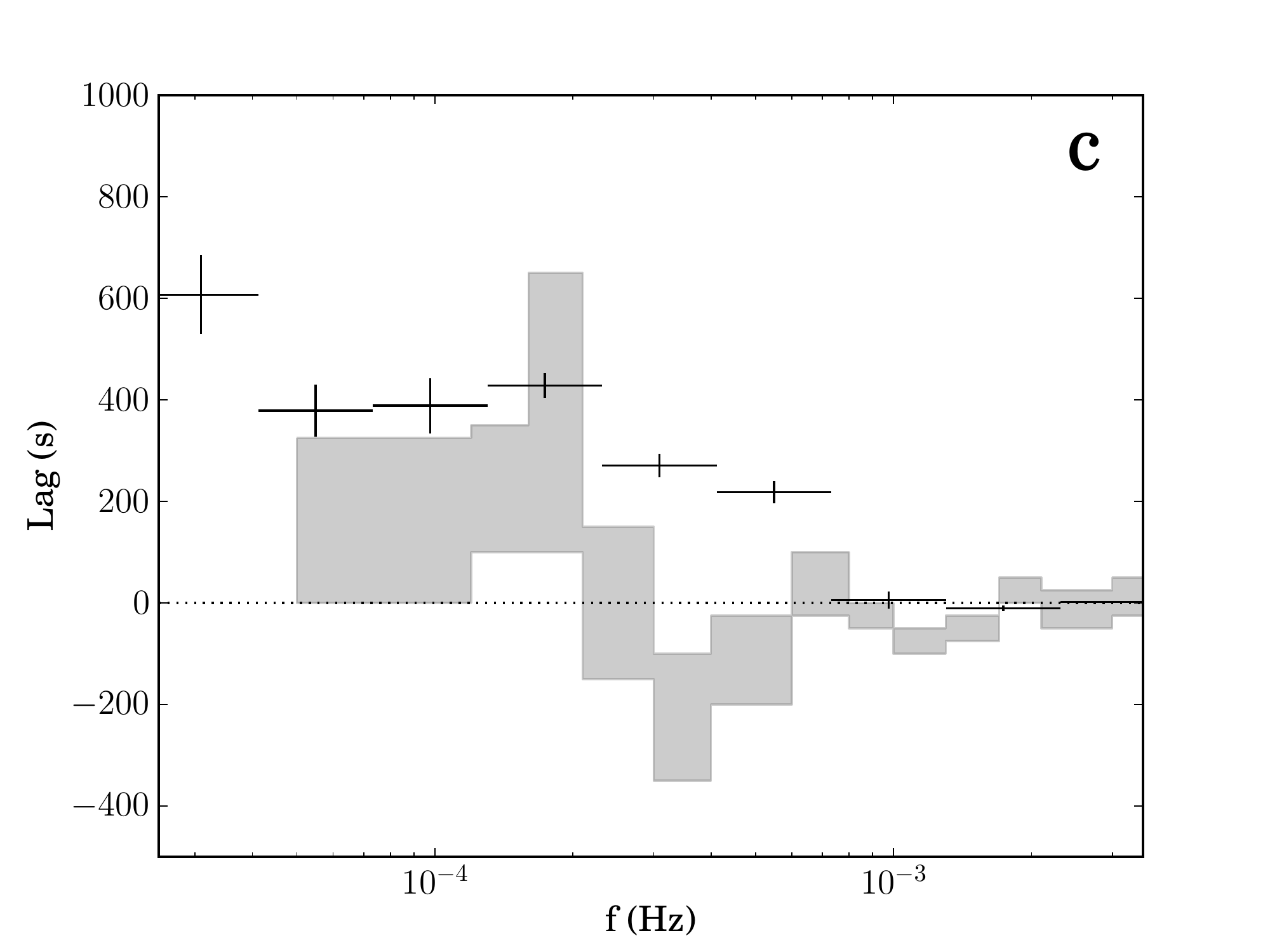} &
\includegraphics[width=8cm]{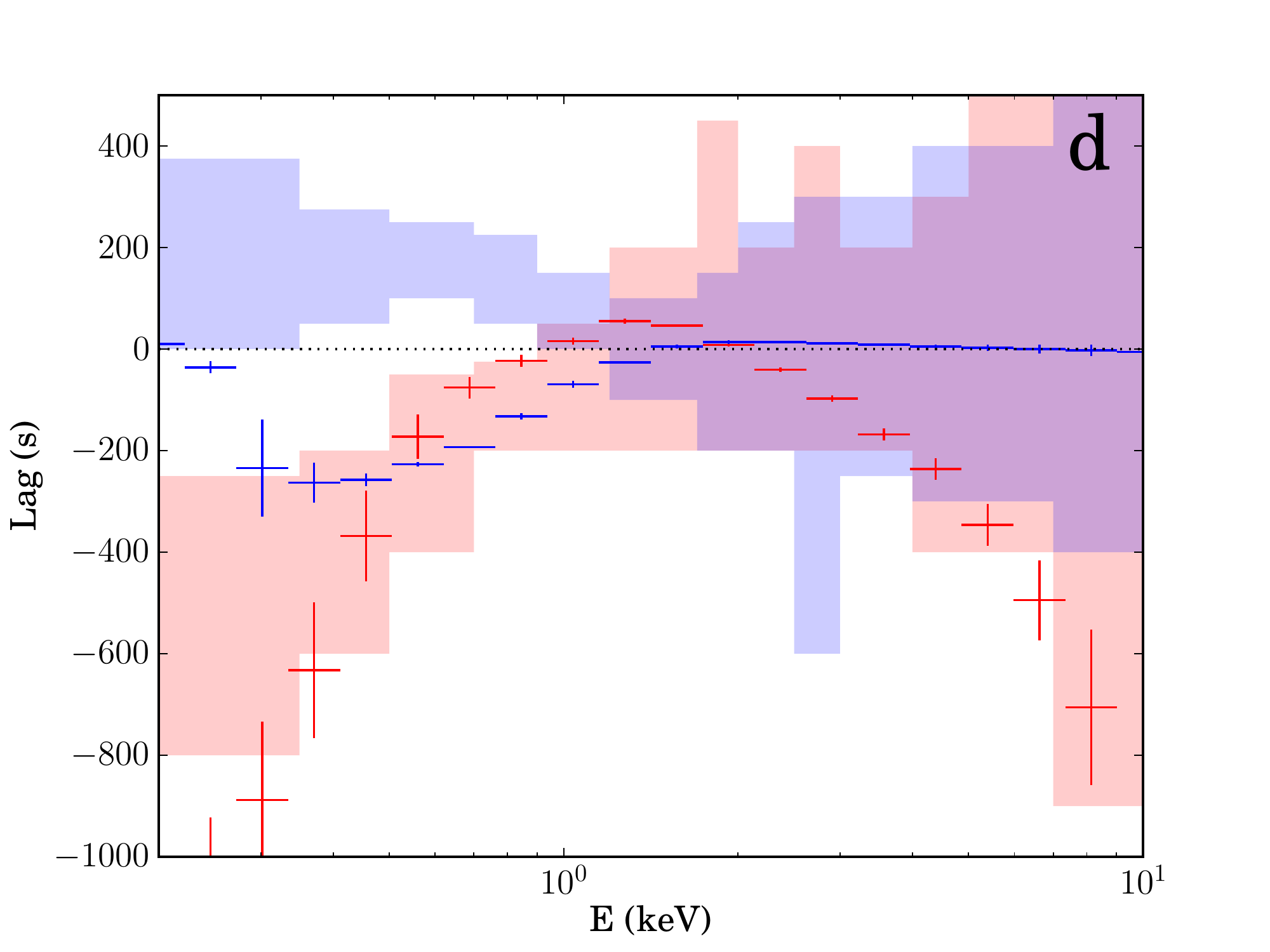} \\
\end{tabular}
\caption{As in fig 3, but now for model with propagation of fluctuations from disc to soft excess ($t_{lag,s}=1000s$) and spectral pivoting of the power law assuming its seed photons are provided by the soft excess ($t_{lag,seed}=t_{lag,p}=600s$).}
\label{fig4}
\end{figure*}

\subsection{The Effect of Power Law Spectral Pivoting}

So far we have taken a simple approach of multiplying the time averaged spectral components by their fluctuations. This is justified for components that retain their spectral shape and only change in normalisation with addition of fluctuations, ie. the disc and soft excess. However the power law has a spectral slope set by the balance of heating and cooling in the corona, so its shape should be affected by fluctuations in the corona and also in its seed photons, which come from the soft excess (J13). In order to allow for this, we use the Comptonisation code {\sc{eqpair}} (Coppi 1999) to replace the simple power law fit from the time averaged spectrum. This calculates the spectrum of the Comptonised emission for a spherical region of size $r$, optical depth $\tau$, seed photon temperature $kT_{seed}$, seed photon power $l_s$ and heating power to the electrons $l_h$. We fix the size of the emission region to $10R_g$ and match the seed photon temperature to that of the soft excess ($0.2keV$). We set the mean values of the optical depth, seed photon power and ratio of heating to cooling power to $\tau_0=1$, $l_{s,0}=1000$, $l_{h,0}/l_{s,0}=1$ in order to reproduce the slope of the time averaged power law spectrum, and then allow them to vary according to the coronal and soft excess fluctuations as: 

\begin{equation}
\tau(t) = \tau_0\dot{M}_{p,int}(t)
\end{equation}
\begin{equation}
l_s(t) = l_{s,0}\dot{M}_s(t-t_{lag,p})
\end{equation}
\begin{equation}
\frac{l_h(t)}{l_s(t)} = \frac{l_{h,0}(t)}{l_{s,0}(t)}\frac{\dot{M}_{p,int}(t)}{\dot{M}_s(t-t_{lag,p})}
\end{equation}

\noindent where $\dot{M}_{p,int}$ are the intrinsic fluctuations generated in the corona as before. $\dot{M}_p$ therefore consists of a combination of the intrinsic fluctuations in coronal power and the fluctuations in seed photon flux from the soft excess. This assumes that $t_{lag,p}$ is not the lag time for fluctuations to physically propagate from the soft excess into the corona, but instead the light travel time for seed photons. 
This has clear motivation from the similarity of timescales between the propagation and 
reprocessing, as determined by ADV14. It is not unlikely that there would be both transmission of fluctuations through varying seed photon flux and also through physical propagation of accretion rate fluctuations, with a slightly longer lag time. However we can only measure one net lag, and since it is closer to a light travel time that suggests seed photon propagation makes a strong contribution. Therefore for simplicity we assume all the transmission of fluctuations is via the seed photon flux in this model.
We normalise the resulting spectrum to have a total power equal to the input electron heating power plus the seed photon power. This approach allows us to account for spectral pivoting of the power law resulting from the relationship between the corona and its seed photons from the soft excess. 

Fig 4 shows the resulting power spectra, coherence, lag-frequency and lag-energy spectra. The power spectra, coherence and lag-frequency spectra are very similar to the non-pivoting case. However the lag-energy spectrum at low frequencies has changed (fig 4d, red points). There is a negative lag at low energies, tending to zero at $\sim1keV$. But instead of remaining at zero, as in the non-pivoting case, a negative lag returns above $\sim1keV$, increasing in strength with increasing energy. This is due to the pivoting of the power law spectrum.

Poutanen \& Fabian (1999) showed how such pivoting from spectral evolution could cause time delays between hard and soft photons. They used a model for short timescale variability in black hole binary
systems where a magnetic reconnection event caused particle acceleration above the disc, heating
the electrons. These are cooled by Compton cooling off the copious seed photons from the disc, resulting in
a soft spectrum. If the reconnection event expands upwards then the flux of seed photons drops and the spectrum hardens. This spectral change from soft to hard gives a hard lag as seen in the data, but its origin
is from the spectral evolution of a single region rather than delays between spectra emitted from 
different regions.  

In our model, the coronal spectrum is initially hard and an increase in seed photons from the soft excess causes the coronal emission to soften. Hence this appears as a soft lag (ie. harder energy bins increasingly 'lead') in the lag-energy spectrum. The hard reference band extends from $1.2-4keV$. Above $4keV$, the 'lead' increases with energy as the effect of spectral pivoting increases. However, because the softening occurs on timescales associated with fluctuations in the soft excess this does not lead to soft lags at high frequencies but at low frequencies. The leads are also at hard energies, whilst the data require soft leads. The high frequency lag-energy spectrum looks the same as in the non-pivoting case (cf. fig 3d). 

The observed low frequency lag-energy spectrum of PG1244+026 does not appear to show this characteristic pattern of soft lags at high energies (ADV14). In fig 4d we also show the range of the low frequency lag measured by ADV14 shaded in pink. This is more consistent with remaining at zero lag, although the errors are large. This suggests the power law component in PG1244+026 does not pivot, but merely changes in normalisation like the other components. To produce a change in normalisation without pivoting in the spectrum requires $l_h$ and $l_s$ to change together. This suggests a situation where the corona is connected to the soft excess
so that a mass accretion rate fluctuation in the soft excess can increases the soft flux, but can also 
propagate into the corona (perhaps through evaporation) to produce a correlated increase in the power in hot electrons ($l_h$).

\subsubsection{Covariance spectrum}

In fig 5 we show the $4-10keV$ covariance spectrum (for comparison we also show the covariance spectrum from the non-pivoting propagation model as dashed lines). This shows the spectrum of the variability that is correlated with variations in the $4-10keV$ band. Red shows the spectrum of the variability that is correlated at low frequencies ($2.3\times10^{-5} - 7.3\times10^{-5}Hz$) and blue shows the correlated variability at high frequencies ($2.3\times10^{-4} - 7.3\times10^{-4}Hz$). The covariance is calculated as in J13 (see also Wilkinson \& Uttley 2009). Briefly, a light curve is generated for each energy bin and also for the $4-10keV$ reference band (subtracting the light curve of the energy bin if that bin lies within the reference band). The light curves are then Fourier transformed and the power set to zero for all frequencies other than the range of interest. An inverse Fourier transform is then applied to transform the filtered periodogram back into a light curve, now containing variability only in a narrow frequency range. The covariance between the filtered energy bin and reference band lightcurves is then calculated as in Wilkinson \& Uttley (2009). 

J13 showed that at low frequencies the spectrum of correlated variability has the same shape as the total spectrum, ie. there is correlated low frequency variability in all components. Our model reproduces this (fig 5, solid red line), by propagation of slower fluctuations from the outer disc down to the inner components. 

In contrast the observed high frequency variability correlated with the $4-10keV$ band drops off below $1keV$, following the shape of the time averaged coronal power law. The high frequency covariance spectrum calculated from our model similarly drops off below $1keV$. However the spectral slope of the correlated variability is different depending on whether the power law is allowed to pivot (solid blue line) or not (dashed blue line), with the slope being slightly harder than the time averaged power law for the model including spectral pivoting. The observed high frequency covariance does seem to suggest a slightly harder slope at the highest energies (J13), which would favour a pivoting model. However the low frequency lag-energy spectrum clearly rules out this scenario. This shows that it is not sufficient for a spectral model to match one aspect of the timing observations. A successful spectral model must correctly reproduce them all and matching one does not necessarily mean it will match the others. 

In all subsequent simulations we choose to use the non-pivoting model since this is what the data require. 

\begin{figure} 
\centering
\begin{tabular}{l}
\leavevmode  
\includegraphics[width=8cm]{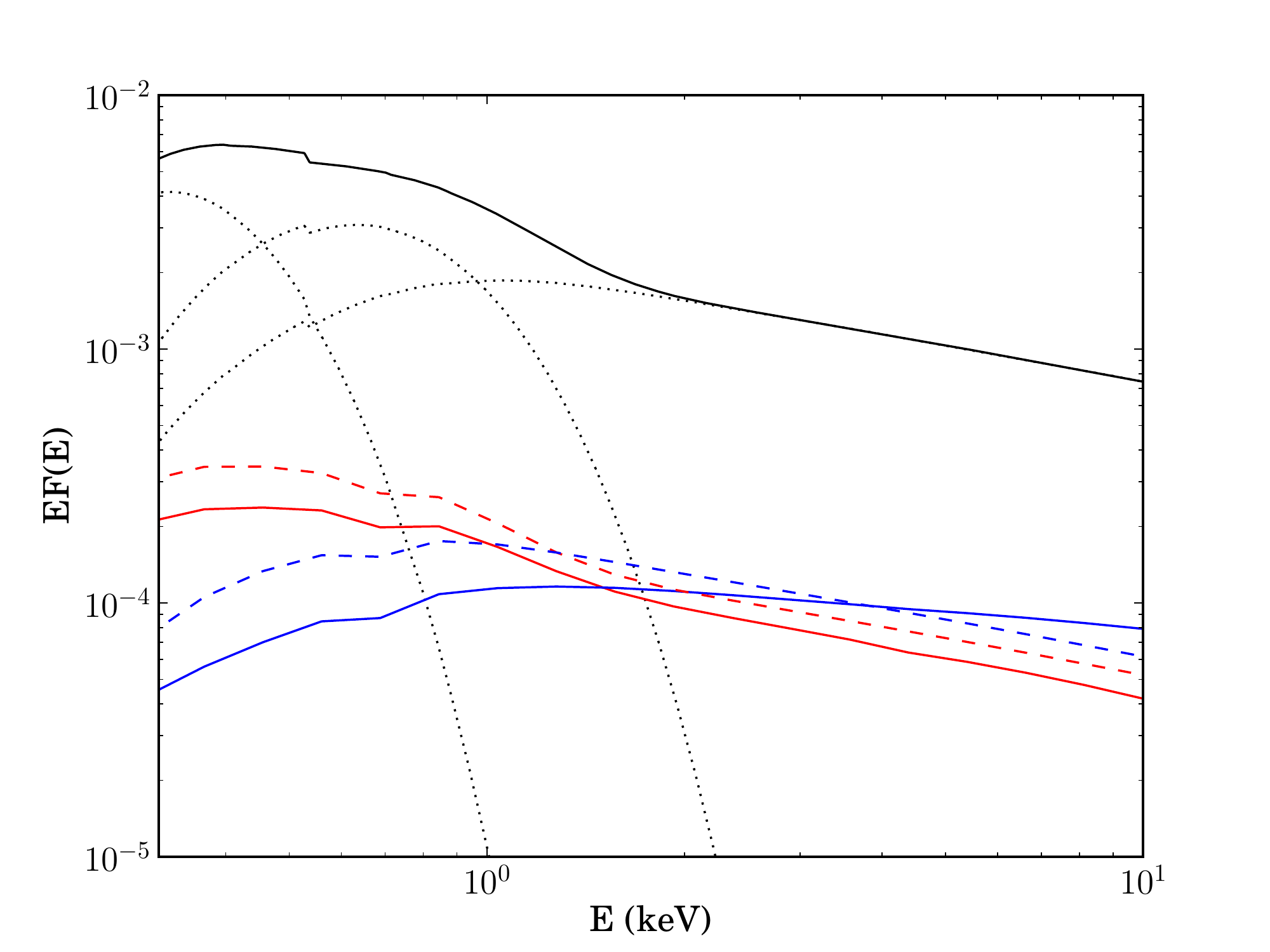} \\
\end{tabular}
\caption{$4-10keV$ covariance spectra for low frequencies ($2.3\times10^{-5} - 7.3\times10^{-5}Hz$, red) and high frequencies ($2.3\times10^{-4} - 7.3\times10^{-4}Hz$, blue) for the propagating model including spectral pivoting of the power law (solid red and blue) and the propagation model without spectral pivoting (dashed red and blue). Black solid line shows total spectrum, dotted lines show model components.}
\label{fig5}
\end{figure}

\begin{figure} 
\centering
\begin{tabular}{l}
\leavevmode  
\includegraphics[width=8cm]{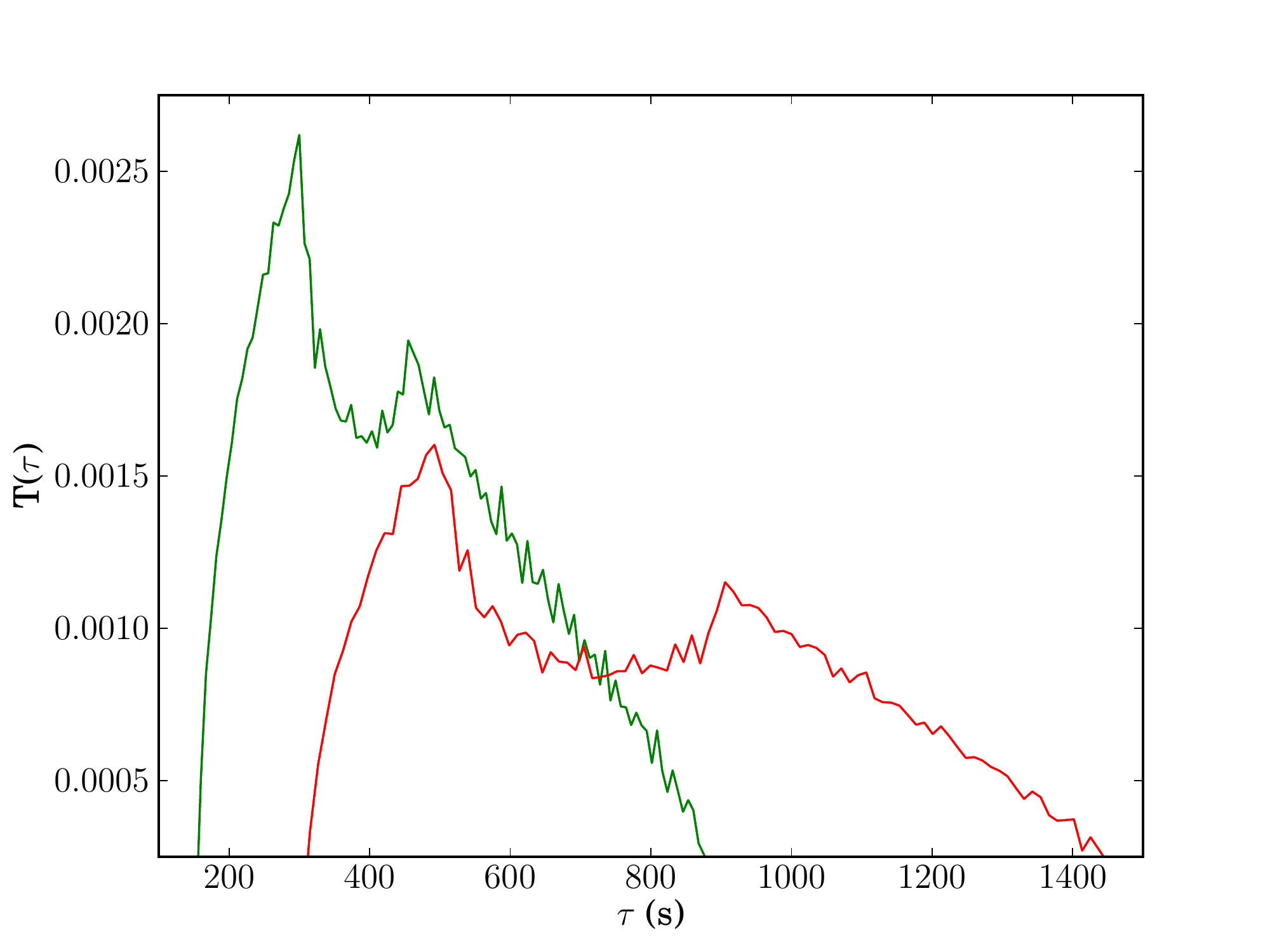} \\
\end{tabular}
\caption{Transfer functions for reflection off the disc ($R=12-20Rg$, red) and soft excess ($R=6-12Rg$, green).}
\label{fig6}
\end{figure}

\begin{figure*} 
\centering
\begin{tabular}{l|r}
\leavevmode  
\includegraphics[width=8cm]{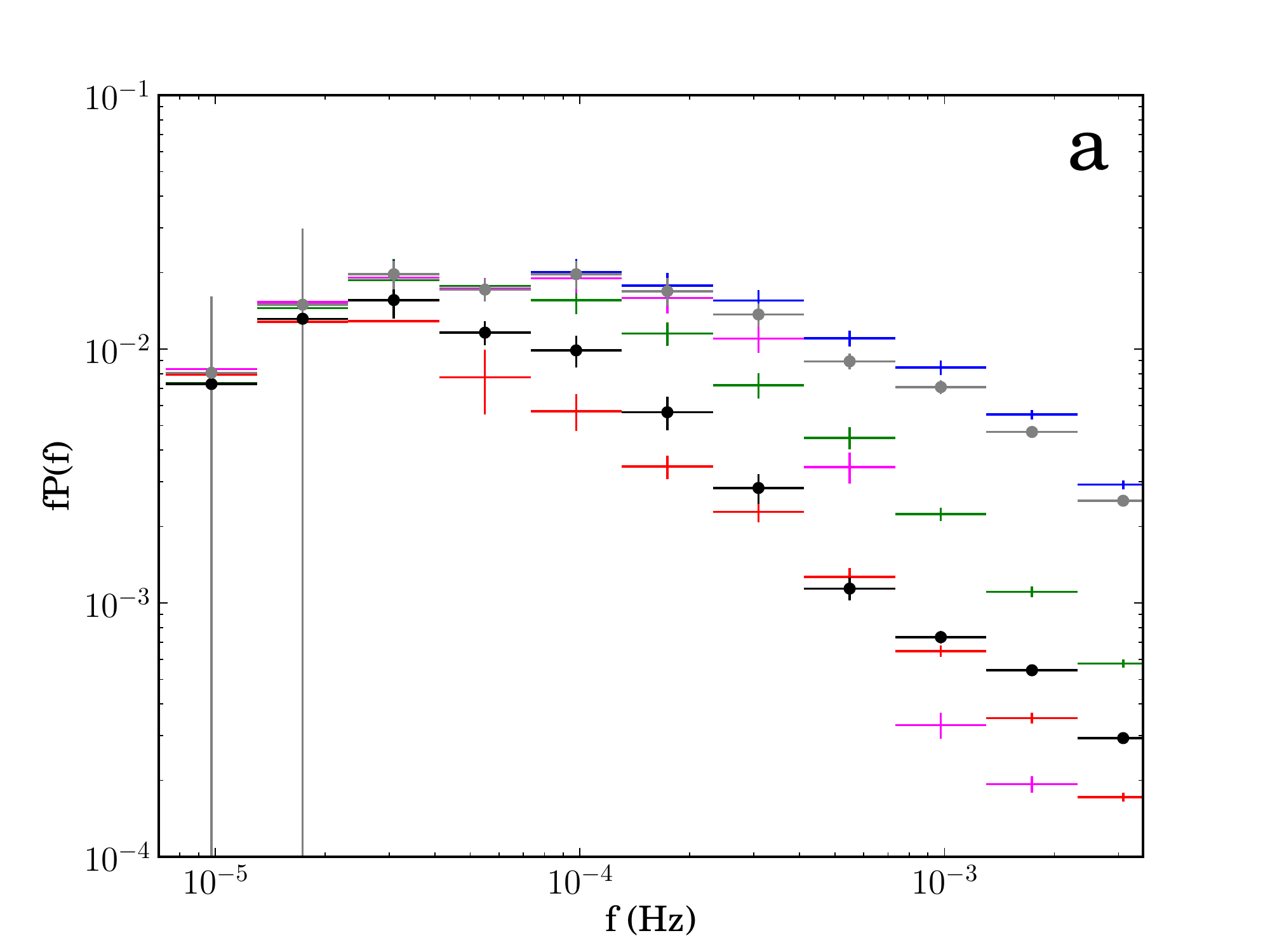} &
\includegraphics[width=8cm]{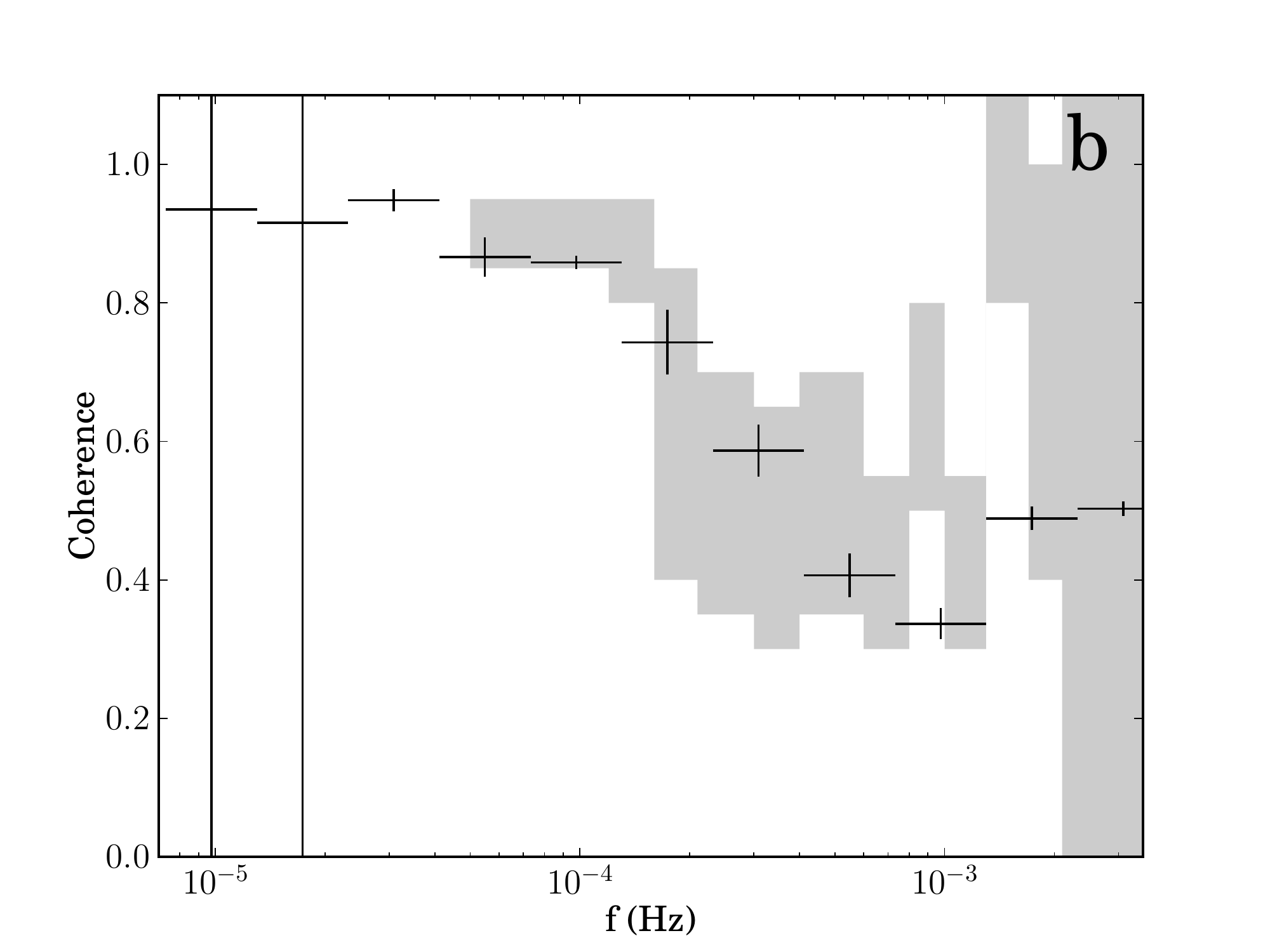} \\
\includegraphics[width=8cm]{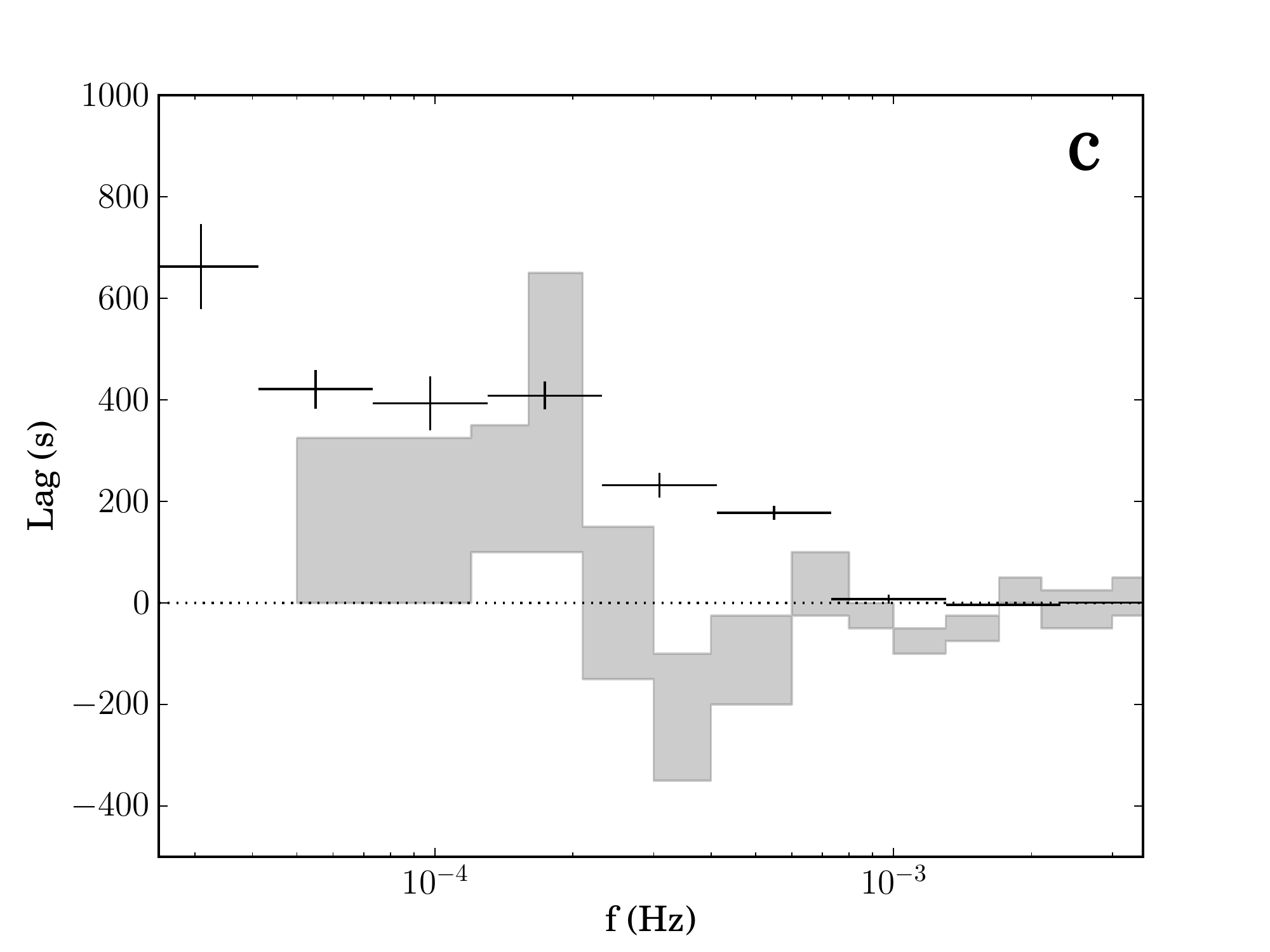} &
\includegraphics[width=8cm]{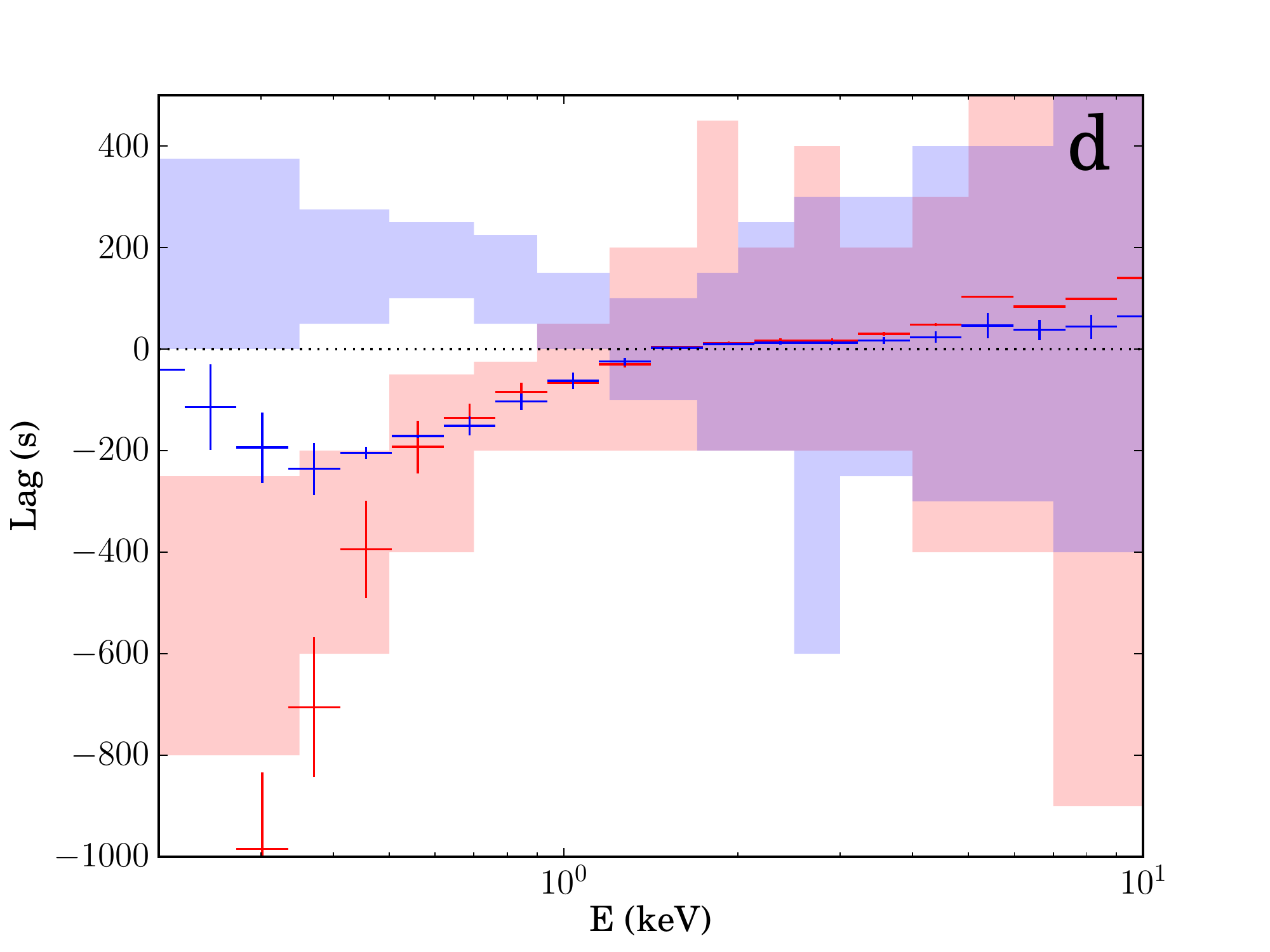} \\
\end{tabular}
\caption{As in fig 3, but now for model with propagation of fluctuations from disc to soft excess ($t_{lag,s}=1000s$) to coronal power law ($t_{lag,p}=600s$) including reflection off the disc ($R=12-20R_g$, magenta).}
\label{fig7}
\end{figure*}

\subsection{Reflection}

\begin{figure*} 
\centering
\begin{tabular}{l|r}
\leavevmode  
\includegraphics[width=8cm]{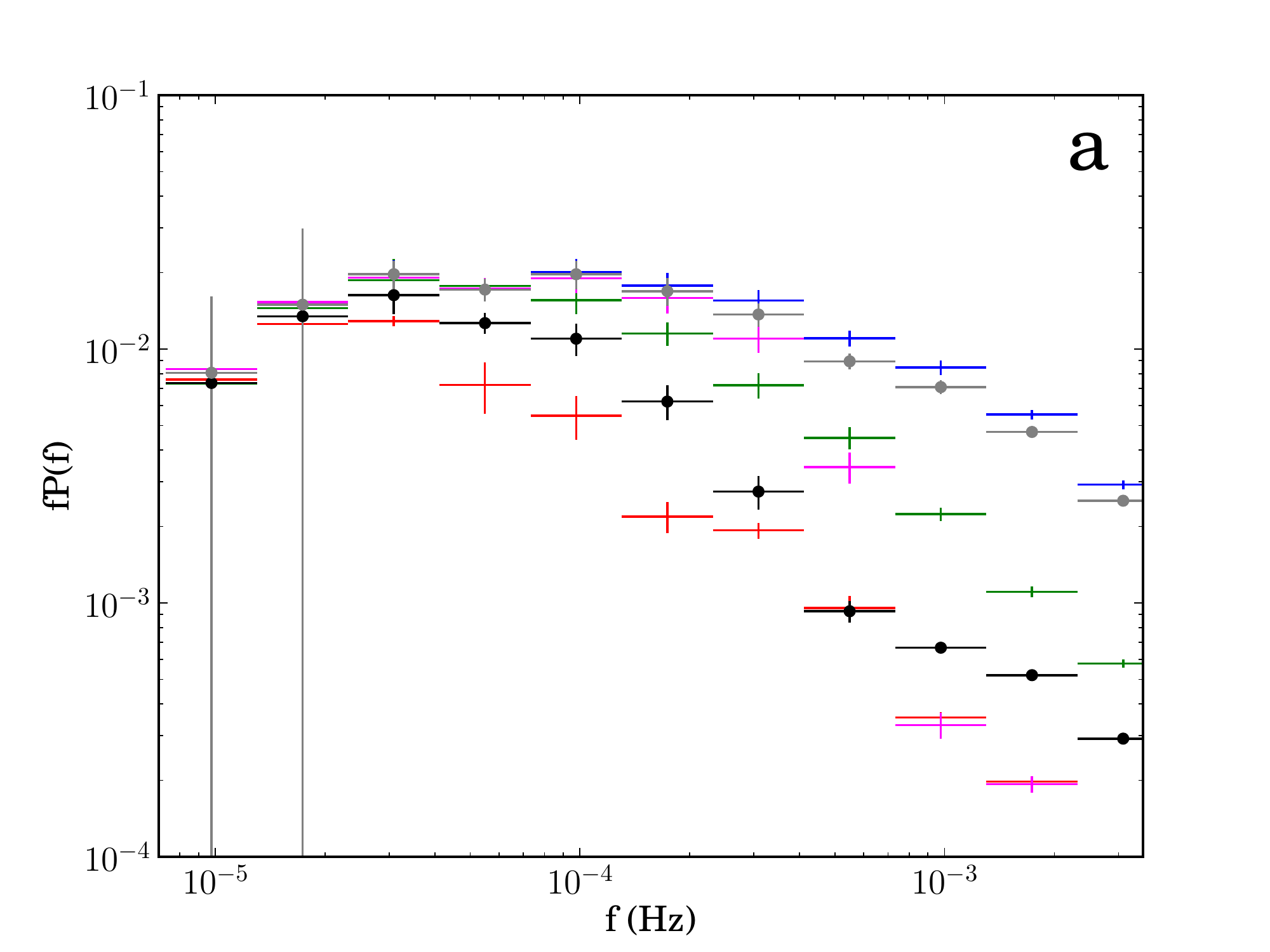} &
\includegraphics[width=8cm]{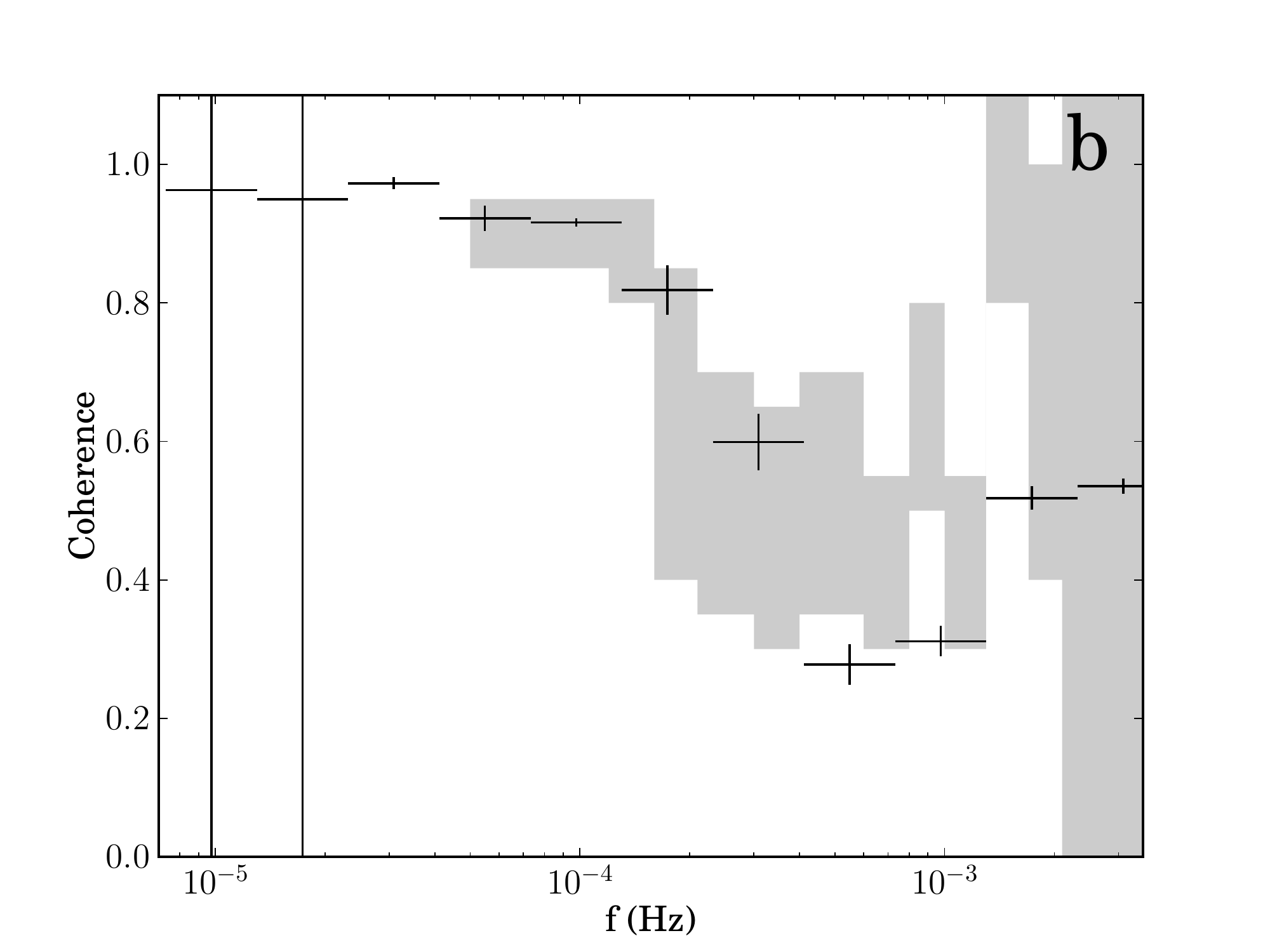} \\
\includegraphics[width=8cm]{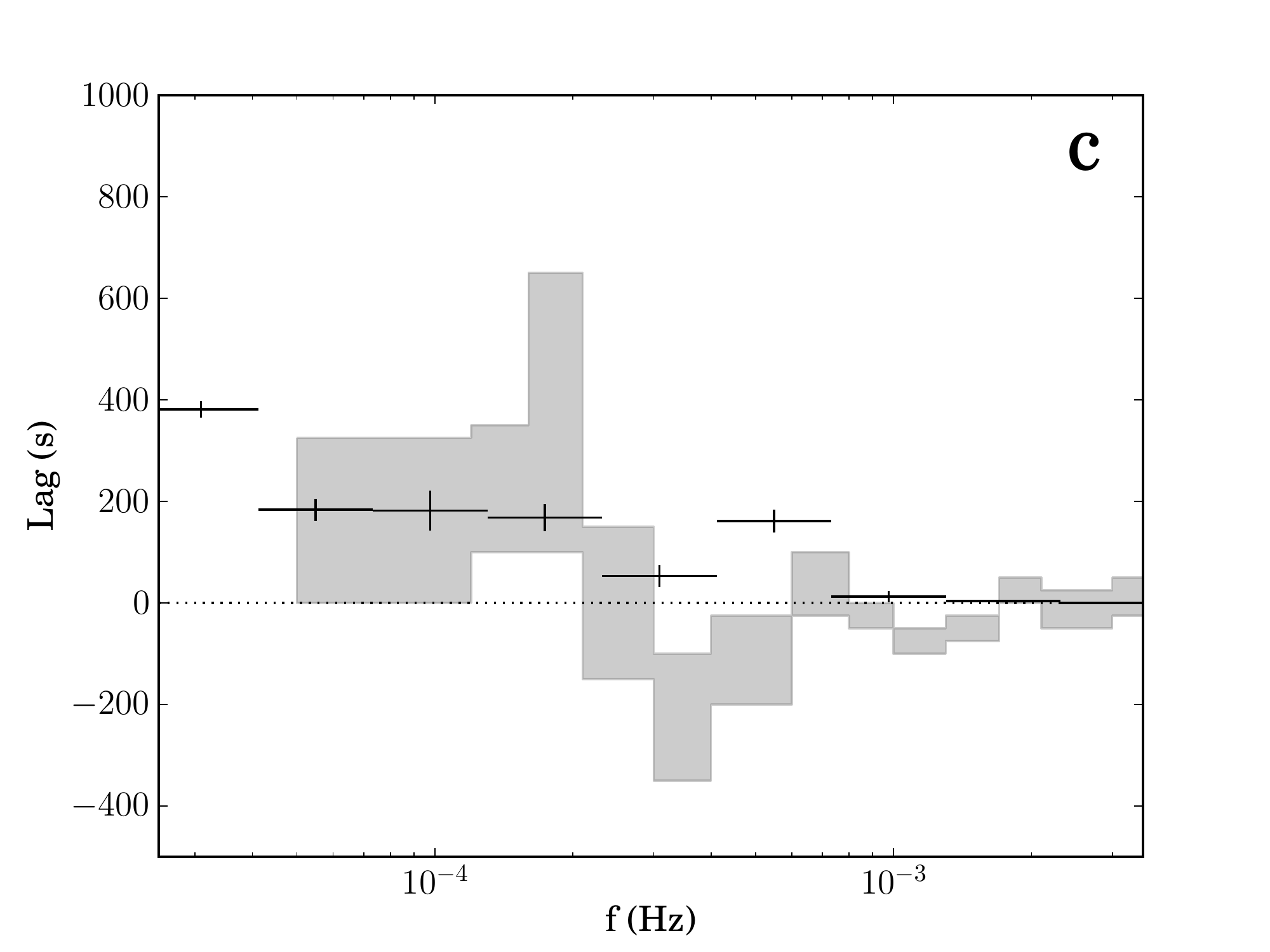} &
\includegraphics[width=8cm]{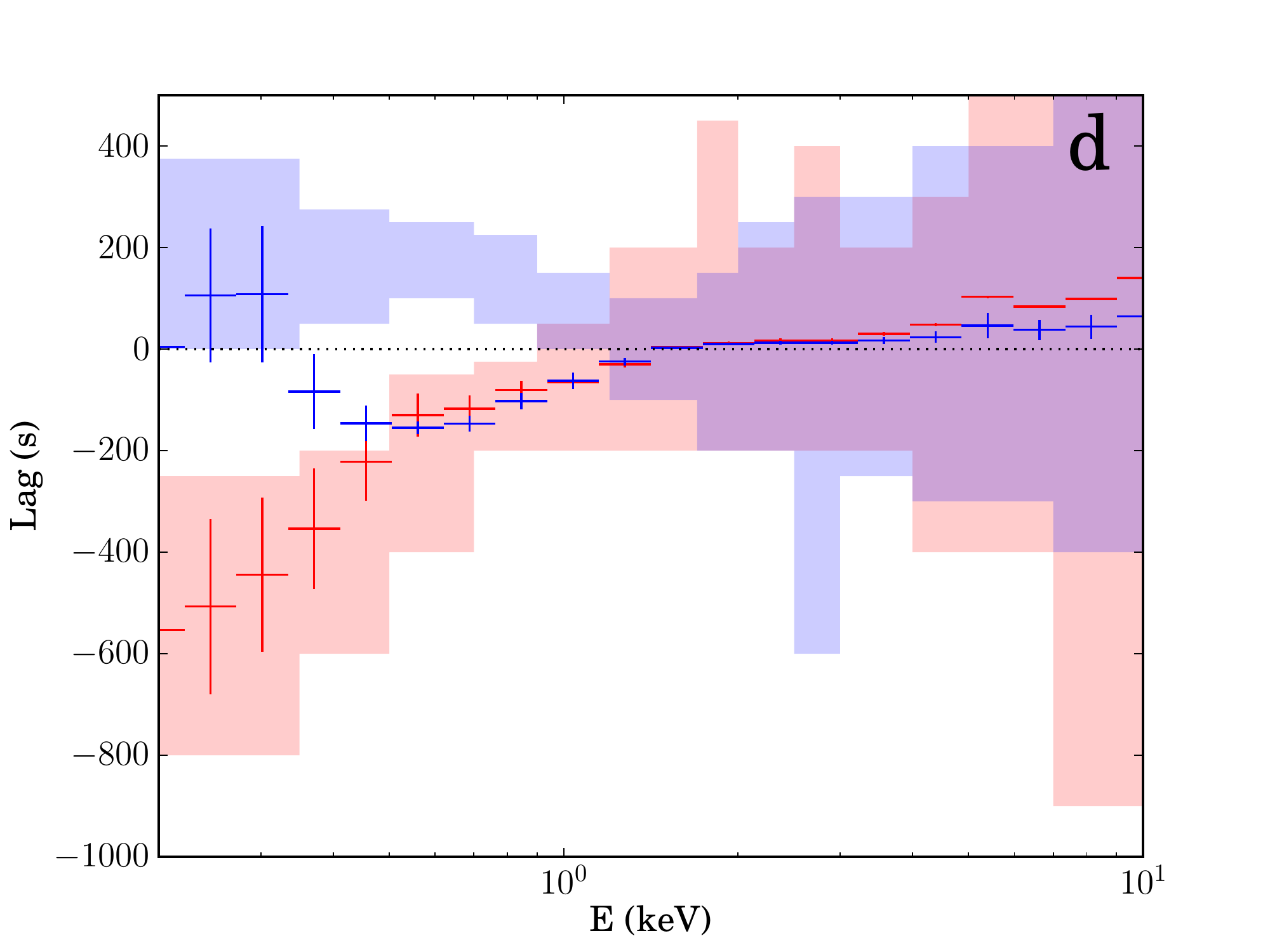} \\
\end{tabular}
\caption{As in fig 3, but now for model with propagation of fluctuations from disc to soft excess ($t_{lag,s}=1000s$) to coronal power law ($t_{lag,p}=600s$) including reflection and reprocessing on the disc ($R=12-20R_g$).}
\label{fig8}
\end{figure*}

In order to fit the time averaged spectrum we require some of the coronal power law to be reflected off the disc (fig 1a, magenta component). The reflected component should therefore respond to variations in the coronal emission, with a light travel time delay. However there will be a range of time delays, as reflection will occur first from inner parts of the disc and take longer to travel to larger radii. If we are viewing the disc at some angle, not face on, the near side of the disc will appear to respond before the far side. The fastest fluctuations in the coronal emission will therefore be smoothed out. In order to account for this we calculate the transfer function of the disc. 

We assume reflection occurs from the innermost parts of the disc, between $12-20R_g$, since below $12R_g$ the disc is replaced by the soft excess, and beyond $20R_g$ the solid angle subtended by the disc is rather 
small. This range of radii matches the range used in our spectral fit. 

The time delay ($\tau$) for light reflected off a point on the disc at radius $r$ from a central source is (Welsh \& Horne 1991): 

\begin{equation}
\tau = \frac{r}{c}(1-\sin i \cos \phi)
\end{equation}

\noindent where $i$ is the inclination of the axis of the disc to the line of sight and $\phi$ is the angle between the point on the disc and the projection of the line of sight onto the disc. Since we have not specified the geometry of the corona (beyond confining it to the central regions $<10R_g$) we do not include any general relativistic corrections on photon light travel times. Given that we measure a low BH spin from our spectral fit this is a reasonable approximation.

We assume the disc of PG1244+026 is inclined at $30^\circ$ with respect to our line of sight, consistent with it being classed as a 'simple' NLS1, assuming that some of the complexity of the 'complex' NLS1s
(Gallo 2006) is from absorption/scattering in a disc wind. We calculate the transfer function of the disc by summing the contribution to the time delay from each azimuth of the disc over all relevant radii, and show the result in fig 6 (red line). 

The fluctuations in the reflected component will therefore be the fluctuations of the coronal emission, convolved with this transfer function, ie: 

\begin{equation}
\dot{M}_{refl}(t) = \int^{\tau_{max}}_0 T(\tau)\dot{M}_p(t-\tau)d\tau
\end{equation}

Fig 7a shows the power spectrum of the reflected component (magenta), along with the other spectral components. There is a clear drop off in power in the reflected component above $\sim3\times10^{-4}Hz$, corresponding to the highest frequency the transfer function can transmit. 

Comparing the lag-energy spectra in fig 3d with fig 7d, shows that including reflection has caused the highest energy bins ($>5keV$) to lag slightly behind the  $1.2-4keV$ reference band at low frequencies (red points). This is the region where reflection makes the largest contribution to the total spectrum (see fig 1a), due to the presence of the iron line at $\sim6.7keV$. 

We find the reflection component alone is not enough to generate the observed high frequency soft lags (fig 7c) as its contribution to the soft band is too small to overcome the soft lead from propagation.

\subsection{Reprocessing on the Disc}

Only a fraction of the coronal flux incident on the disc is reflected. The fraction that is not reflected will thermalise in the disc and be reprocessed. This happens in the same physical location as reflection, so will have the same transfer function, but (unlike reflection) the reprocessed emission is concentrated in the soft X-ray band. We first assume it is reprocessed on the BB disc, so that a fraction of the total flux in the BB disc spectrum will come from reprocessed flux. We assume that half of the coronal emission goes up - the half we observe as power law emission - and the other half goes down, back towards the disc. The total reprocessed flux is then the power law flux minus the flux of the reflected component: $L_{rep} = (\Omega/2\pi) L_{p}-L_{ref}$, where
$\Omega/2\pi=0.65$ is the solid angle of the reflector as measured from the spectral fit (table 1). 
The fraction of total disc flux that is due to reprocessed emission is then $f_{rep}=L_{rep}/L_d\simeq 0.3$ for our chosen spectral decomposition. The thermalisation time is of order the Compton time for an AGN disc ($\sim \sigma_Tnc\sim0.5s$ for $n\sim10^{14}$, Stepney 1983), hence we assume it is negligible compared to the timescale of coronal fluctuations.

The fluctuations in the spectrum from the BB disc therefore consist of a sum of the intrinsic disc fluctuations, which have fractional variability $F_d$ around a BB of luminosity $f_{int}L_d = (1-f_{rep})L_d\simeq 0.7L_d$, and the reprocessed fluctuations, which follow the coronal fluctuations (smoothed out by a transfer function) around a BB of luminosity $f_{rep}L_d$, ie:

\begin{equation}
\begin{split}
\dot{M}_d(t) &= f_{int}\dot{M}_{d,int}(t)+f_{rep}\int^{\tau_{max}}_0 T(\tau)\dot{M}_p(t-\tau)d\tau \\
&= f_{int}\dot{M}_{d,int}(t)+f_{rep}\dot{M}_{rep}(t)
\end{split}
\end{equation}

Fig 8a-d show the power spectra, coherence function, lag-frequency and lag-energy spectra, now including reflection and reprocessing on the disc. Comparison of the lag-frequency spectrum (fig 8c) with fig 7c shows that the low frequency hard lags have been reduced. This is due to dilution of the intrinsic disc fluctuations, which produce the hard propagation lags, by the reprocessed disc component, which produces soft lags. The net lag at low frequencies is simply the intrinsic hard lag minus the reprocessed
soft lag, taking into account the relative proportions of each. So for there to be a net hard lag at low frequencies requires that the propagation lags
dominate over the reprocessing lags. This is easily achieved if the intrinsic flux dominates over the reprocessed flux (as in this case). However it can still be achieved in situations where the reprocessed flux dominates, if the propagation lag is sufficiently longer than the shorter light travel time reprocessing lag to compensate for the smaller intrinsic fraction. But there must \emph{be} a component capable of producing the propagation lag. Thus the existence of hard lags at low frequency rules out models where the soft X-ray excess is produced by reflection and a soft jet power law, as
we will show explicitly below (Section 4).

\begin{figure} 
\centering
\begin{tabular}{l}
\leavevmode  
\includegraphics[width=8cm]{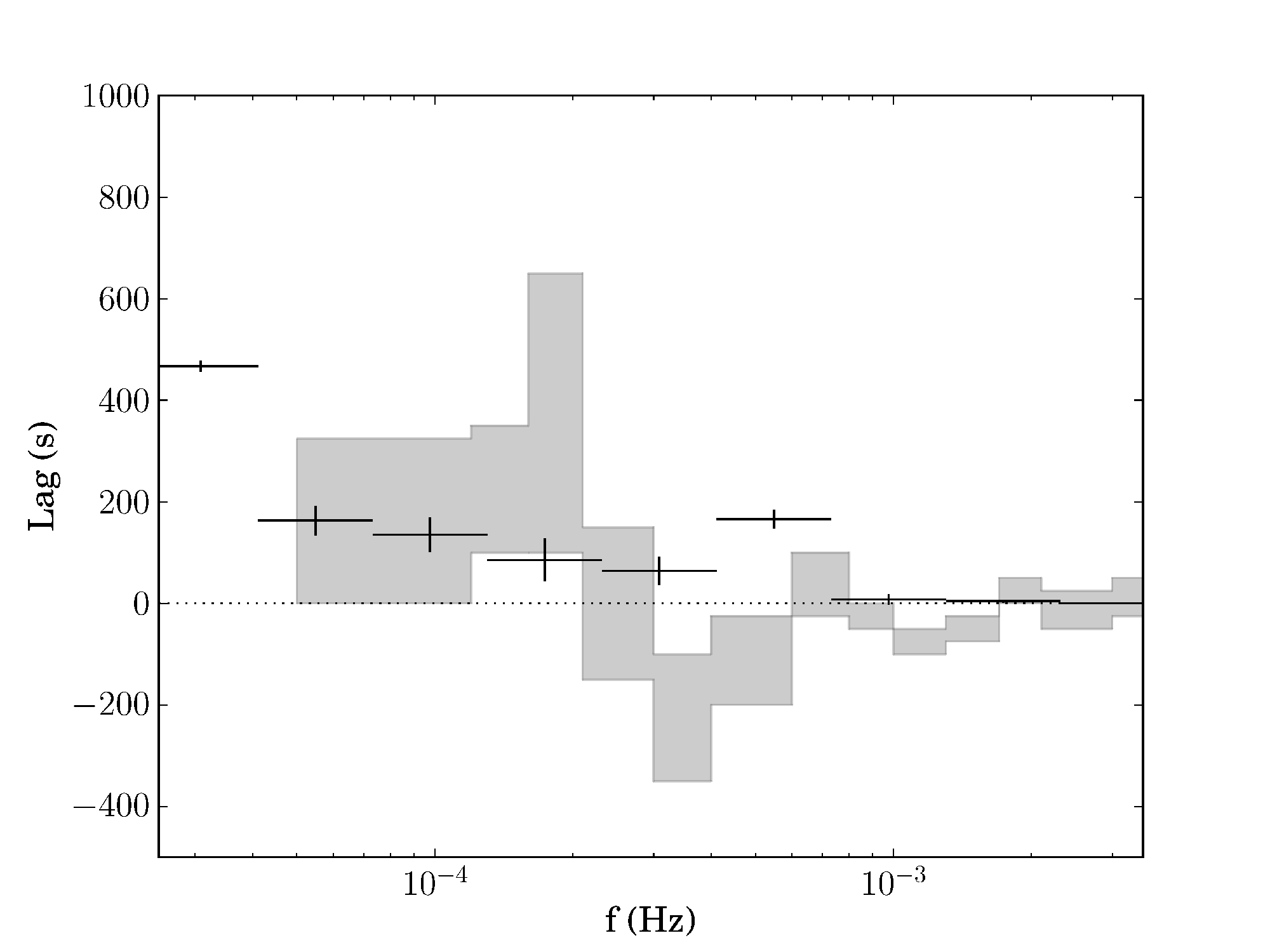} \\
\end{tabular}
\caption{Lag-frequency spectrum, including reflection and reprocessing off the disc and propagation of reprocessed fluctuations. Shaded grey regions show range of error on the lags measured by ADV14 for PG1244+026.}
\label{fig9}
\end{figure}

Fig 8d shows the lag-energy spectrum. The high frequency lags (blue) are almost identical to fig 7d, except for the lowest three energy bins ($<0.4keV$). These show a small lag behind the $1.2-4keV$ hard reference band. Comparison with the spectrum in fig 1a shows that these are the only energies at which the disc emission dominates over the soft excess. Between $0.4-1keV$ the lag-energy spectrum shows the energy bins leading the hard reference band, due to propagation lags generated by the soft excess. In contrast, the \emph{observed} high frequency lag-energy spectrum shows all energy bins from $0.2-1keV$ lagging the hard reference band (fig 8d, shaded blue regions). 

ADV14 suggest this could be achieved by allowing the reprocessed fluctuations to propagate from the disc down into the soft excess. The illuminating coronal flux heats up the disc, which then re-emits the radiation as reprocessed flux. The process of disc heating will alter the viscous frequency, allowing matter to propagate faster. In this way, fluctuations in the illuminating continuum can become accretion rate fluctuations in the disc, which can propagate inwards just like the intrinsic fluctuations. 

To test the effect of this we now allow both the intrinsic disc fluctuations and fluctuations generated by reprocessing to propagate into the soft excess (and on to the corona). Since the coronal emission is now affected by the delayed propagated fluctuations which are in turn affected by the time lagged coronal emission we must iterate to find a self consistent solution. We first calculate the intrinsic disc, soft excess and power law fluctuations at each time. We then allow these fluctuations to propagate inwards and calculate the fluctuations of the reflected/reprocessed component as a function of time ($\dot{M}_{rep}(t)$). The BB component should consist of a sum of these two components: the intrinsic fluctuations (weighted by $f_{int}$) and the reprocessed fluctuations (weighted by $f_{rep}$), ie. $\dot{M}_d(t) = f_{int}\dot{M}_{d,int}(t)+f_{rep}\dot{M}_{rep}(t)$. We repeat the calculation with this new $\dot{M}_d$, now including the reprocessed fluctuations, and again calculate $\dot{M}_{rep}$. This is fed into the next iteration and the process repeated until successive iterations result in a total fractional change in the power law fluctuations of less than $10^{-3}$.

Fig 9 shows the resulting lag-frequency spectrum, which is nearly identical to fig 8c and still lacks a high frequency soft lag. We find any propagated reprocessed fluctuations are not strong enough to overcome the intrinsic soft excess fluctuations, which cannot themselves be reduced without losing the hard propagation lags. In order to replicate the lag-energy spectrum, the reprocessing must occur not on the disc, but the soft excess. 

\begin{figure*} 
\centering
\begin{tabular}{l|r}
\leavevmode  
\includegraphics[width=8cm]{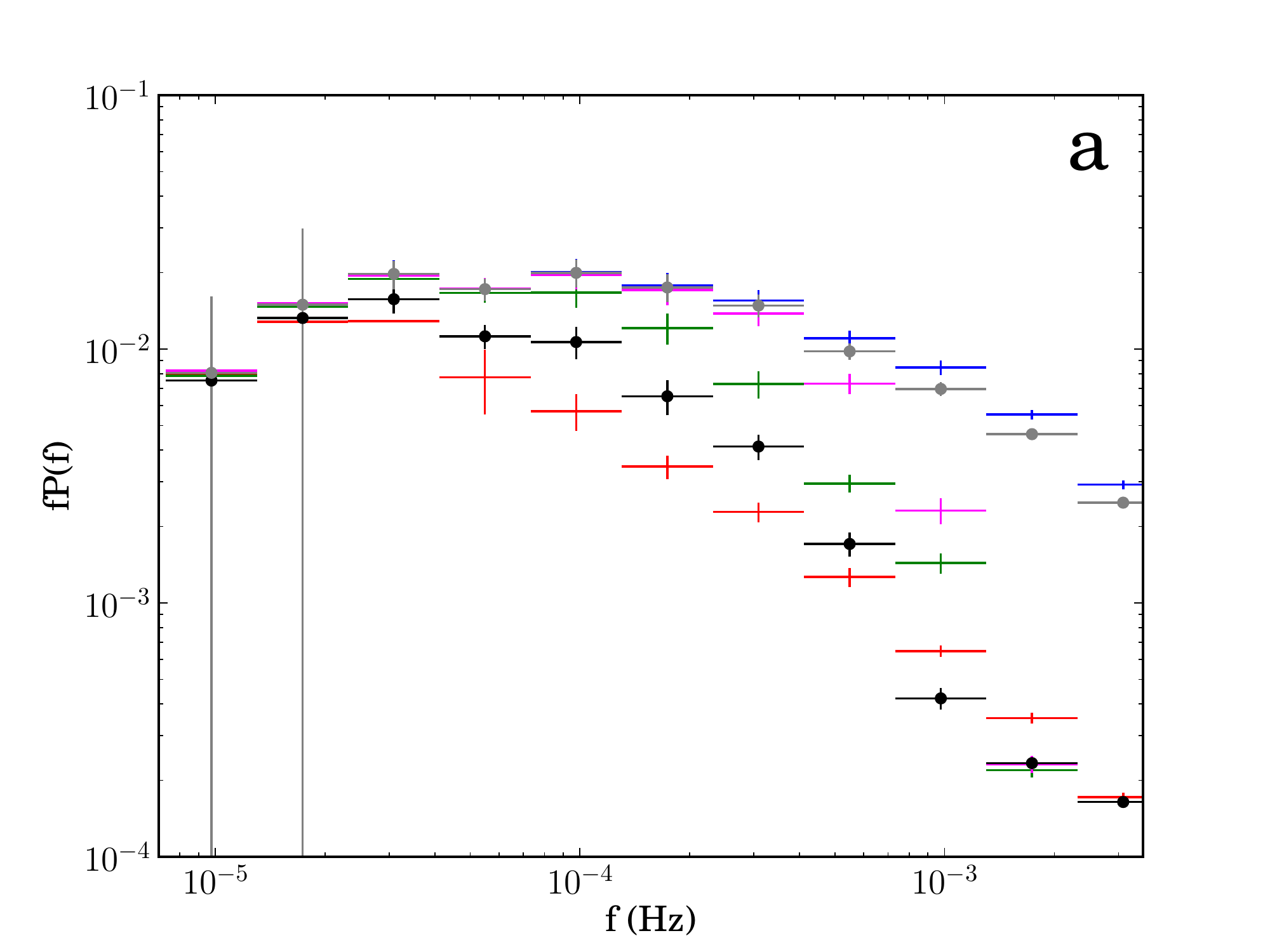} &
\includegraphics[width=8cm]{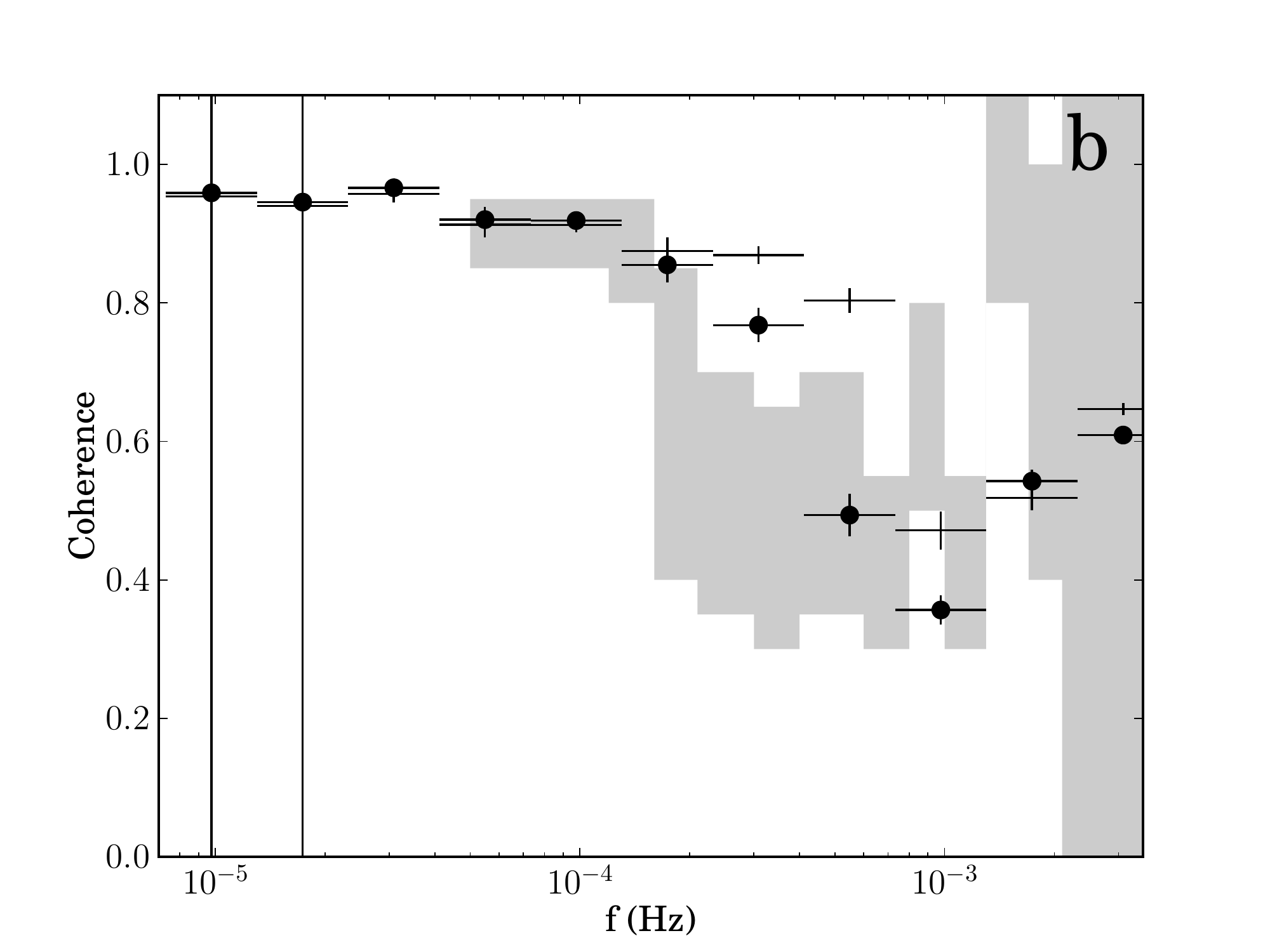} \\
\includegraphics[width=8cm]{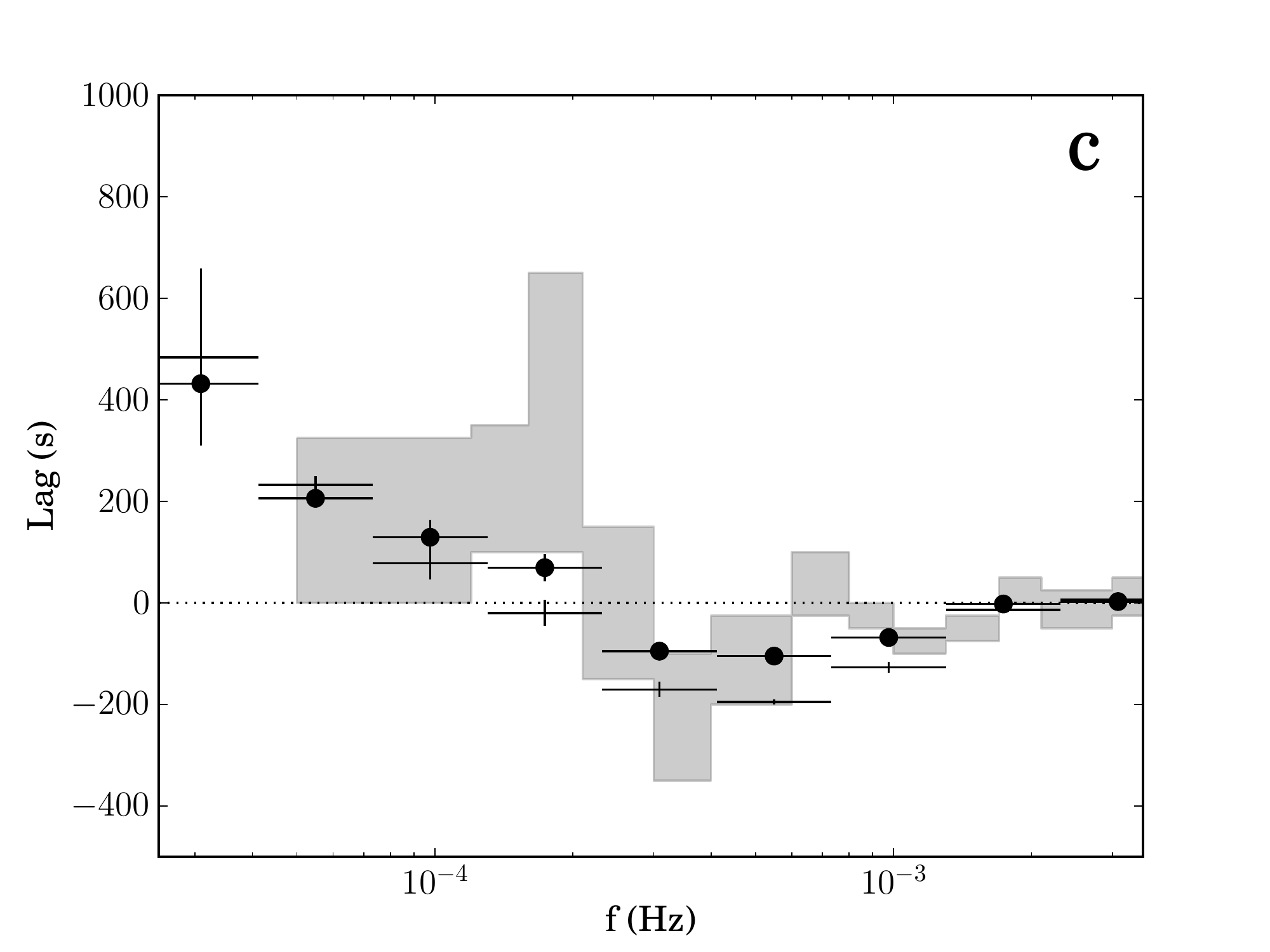} &
\includegraphics[width=8cm]{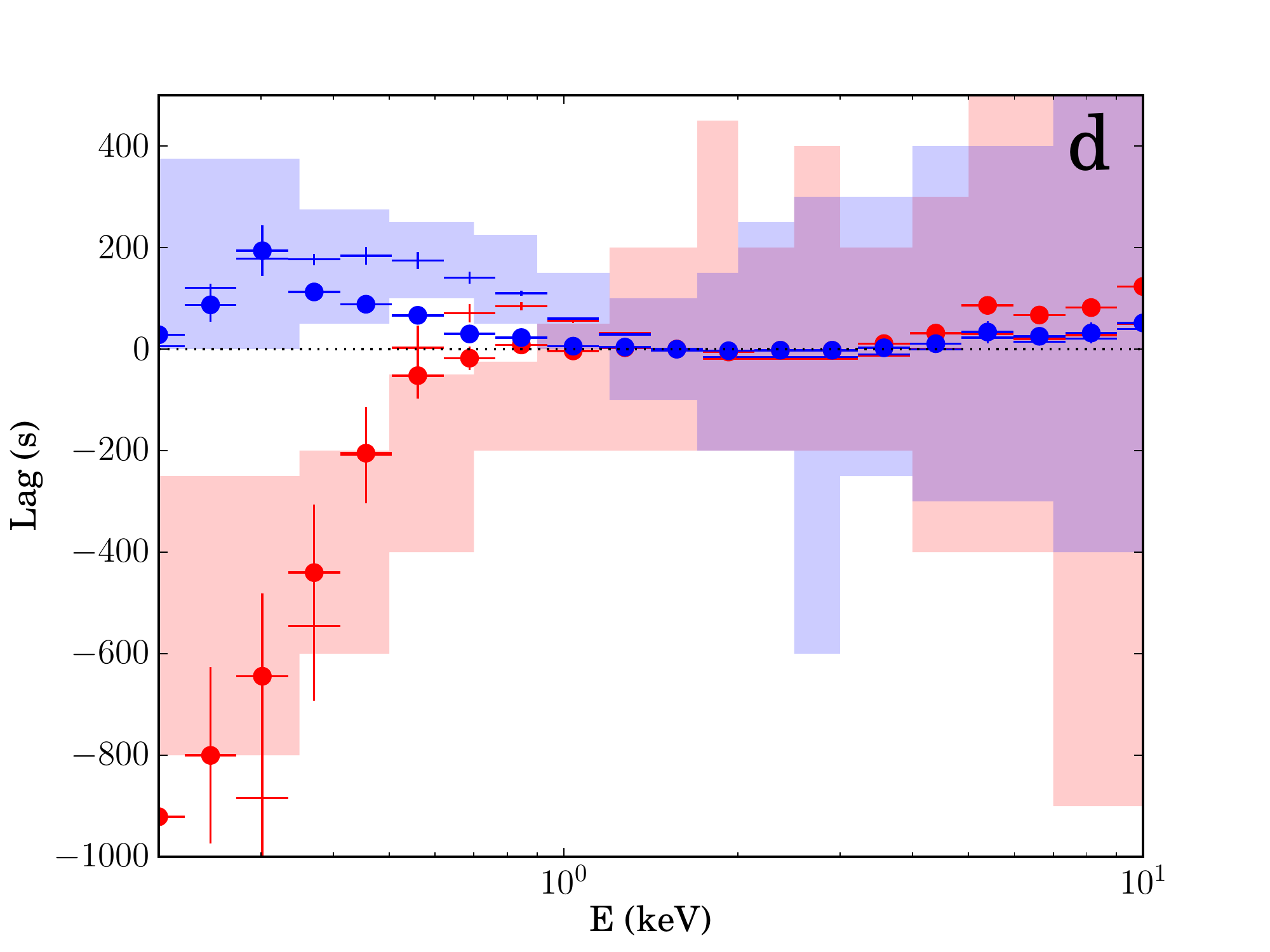} \\
\end{tabular}
\caption{As in fig 3, but now for model with propagation of fluctuations from disc to soft excess ($t_{lag,s}=1000s$) to coronal power law ($t_{lag,p}=600s$) including reflection and reprocessing on the soft excess ($R=6-12R_g$) (crosses). Solid circles show model with half the reprocessing occurring on the disc, half on the soft excess.}
\label{fig10}
\end{figure*}

\begin{figure} 
\centering
\begin{tabular}{l}
\leavevmode  
\includegraphics[width=8cm]{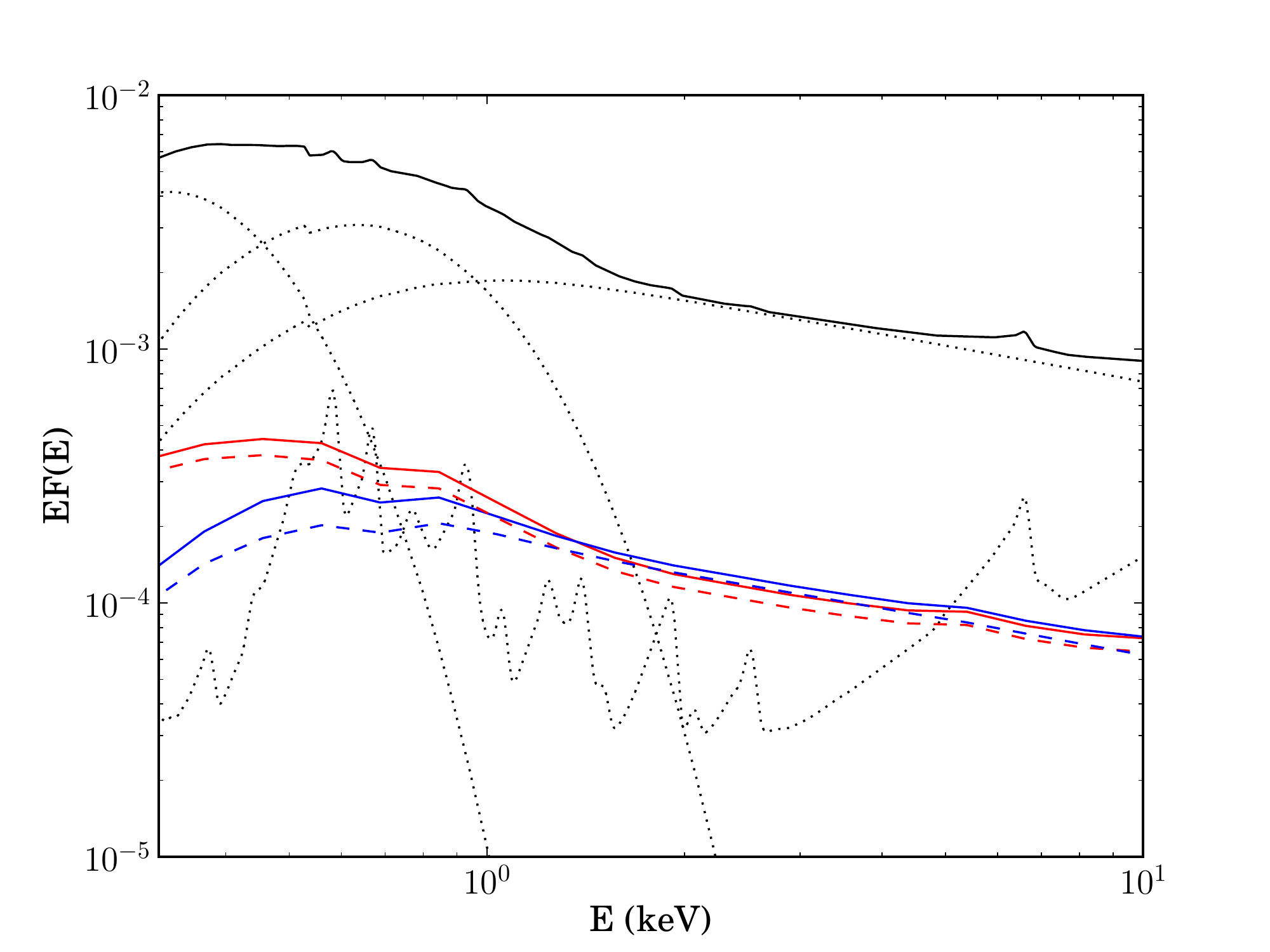} \\
\end{tabular}
\caption{$4-10keV$ covariance spectra for low frequencies ($2.3\times10^{-5} - 7.3\times10^{-5}Hz$, red) and high frequencies ($2.3\times10^{-4} - 7.3\times10^{-4}Hz$, blue) for the model including reprocessing on the soft excess (solid red and blue) and model including reprocessing on both the soft excess and the disc (dashed red and blue). Black solid line shows total spectrum, dotted lines show model components.}
\label{fig11}
\end{figure}

\subsection{Reprocessing on the Soft Excess}

We rerun our model, this time with the reprocessing occurring in the soft excess. We assume this has a reflection spectrum with the same ionisation as used for the disc i.e. 
matching that seen fit to the data (Section 2) since the structure of this region is not known. We change our transfer function to match the size of the soft excess region, $\sim6-12R_g$ (fig 6, green line). This increases the maximum frequency of the reflected/reprocessed power (fig 10a, magenta points). Consequently the coherence remains high up to higher frequencies (compare crosses fig 10b with fig 8b). 

Fig 10c (crosses) shows the lag-frequency spectrum which shows hard propagation lags at low frequencies ($\lesssim10^{-4}Hz$) and a soft reprocessing lag between $10^{-4}\sim10^{-3}Hz$. This is in much better agreement with the observed lags of PG1244+026 (ADV14). 

Fig 10d shows the lag-energy spectrum. The high frequency lag (blue crosses) shows the soft energy bins lagging the hard reference band for nearly all energies below $\sim1keV$, in much better agreement with the observations. Only the very lowest energy bins, where the disc dominates, now lack a strong soft lag. It is likely there is reprocessing on both the disc and soft excess, though clearly reprocessing in the soft excess region dominates in the band of our observations.  

The soft excess is less luminous than the disc, as a consequence the fraction of reprocessed to intrinsic emission is much larger with $f_{int}=0.3$, $f_{rep}=0.7$ while when reprocessing is confined to the disc the fractions are roughly reversed. Allowing for some reprocessing on the disc as well as the soft excess would reduce the reprocessed fraction. Nevertheless, even in this limiting case, the model still produces soft leads in addition to reverberation lags. This is because the propagation time lags between soft excess and corona are longer than the light travel time delay for reverberation, so can be more heavily diluted without losing the net lead at low frequencies. 

However, a very high reprocessing fraction on the soft excess will affect the covariance spectrum. In particular, the spectrum of the correlated variability at high frequencies will be extended to lower energies. We calculate the covariance spectra for the soft excess reprocessing model to check they do not disagree with the data and show them in fig 11 (solid red and blue lines). The low frequency correlated variability still matches the shape of the total spectrum, in agreement with the observations. The high frequency correlated variability now rolls over at $\sim0.6keV$ rather than $1keV$, due to the amount of reprocessing on the soft excess. The dashed red and blue lines in fig 11 show a model where half of the flux available for reprocessing is reprocessed on the soft excess and the other half is reprocessed on the disc. This increases the energy of the roll over. Hence both the high frequency lag-energy spectrum and high frequency covariance spectrum suggest there is some reprocessing on both the disc and the soft excess. In fig 10b, c and d we also show the coherence, lag-frequency and lag-energy spectra for the combined disc and soft excess reprocessing model (circles). This shows that a model with a combination of propagating fluctuations, from the disc through a separate soft excess component to the coronal power law, together with reprocessing on both the disc and the soft excess component, can capture all the main features of the data. 

We note that including reprocessing on the disc as well as the soft excess does not exactly match the shape of the high frequency lag-energy spectrum. This is because the disc reprocessing is likely to be concentrated on the inner edge of the disc, which our simple model doesn't account for. In other words, the data suggest the reprocessed emission should have a spectrum similar to that of the soft excess component, but extending to slightly lower energies to include the innermost radii of the disc.

\section{Comparison with a Reflection Model for the Soft X-ray Excess}

Reflection dominated models have a strong soft X-ray
excess from a combination of line and continuum. However, the
observed soft X-ray excess is smooth, so this requires strong
relativistic smearing. In 'complex' NLS1 such as 1H0707-495, this
smearing is so extreme as to require strongly centrally concentrated
emissivity ($\propto r^{-7}$) onto a high spin black hole (Fabian et al 2009). 
This makes all size scales, and hence reflection/reprocessing lag
times shorter. 

However, while this can reproduce the soft lags, a single power law
which varies only in normalisation and its constant ionisation
reflection cannot match the soft lead at low frequencies.  Chainakun
\& Young (2012) showed that the mismatch was made worse by including
the radial and time variability of the ionisation of the reflector
which should arise from the extremely centrally concentrated and variable
illumination. This means that more complex continuum models are
required to match the spectral-timing data, in particular to match the 
soft lead.

\begin{table*}
\begin{tabular}{lllll}
\hline
Component & Parameter & \S4.1/\S4.3 & \S4.2 & \S4.4 \\
\hline
Galactic absorption & $N_h$ ($10^{22}cm^{-2}$) & 0.019 & 0.019 & 0.019 \\
Intrinsic absorption & $N_h$ ($10^{22}cm^{-2}$) & 0.019 & 0.029 & 0.047 \\
Hard power law & $\Gamma$ & 2.26 & 2.3 & 2.3 \\
 & norm & $6.9\times10^{-4}$ & $7.3\times10^{-4}$ & $4.9\times10^{-4}$ \\
Soft power law & $\Gamma$ & 3.5 & 3.5 & 3.0 \\
 & norm & $1.6\times10^{-3}$ & $1.5\times10^{-3}$ & $9.3\times10^{-4}$ \\
kdblur & index & 3.4 & 3.9 & 4.5 \\
 & $r_{in} (R_g)$ & 3.2 & 3.1 & 3.0 \\
rfxconv & relative refl norm & -2.6 & -3.2 & -2.8 \\
 & $\log(x_i)$ & 2.7  & 2.7 & 3.0 \\
Blackbody & $kT (keV)$ & - & 0.027 & 0.032 \\
 & norm & - & $4.0\times10^{-3}$ & $3.6\times10^{-3}$ \\
\hline
\end{tabular}
\caption{Parameters for the spectral models shown in fig 12-15, with reflection dominating the soft X-ray excess emission. a). Model used in sections 4.1 and 4.3: {\sc{wabs*zwabs(powerlaw + powerlaw +  kdblur*rfxconv*powerlaw)}}. b). Model used in section 4.2: {\sc{wabs*zwabs(bbody + powerlaw + powerlaw + kdblur*rfxconv*powerlaw)}}. c). Model used in section 4.4: {\sc{wabs*zwabs(bbody + powerlaw + powerlaw + kdblur*rfxconv(powerlaw + powerlaw))}}. For all models $Z_{Fe}$ was fixed to 1.0.}
\label{table2}
\end{table*}

\begin{figure*} 
\centering
\begin{tabular}{l|r}
\leavevmode
\includegraphics[width=8cm]{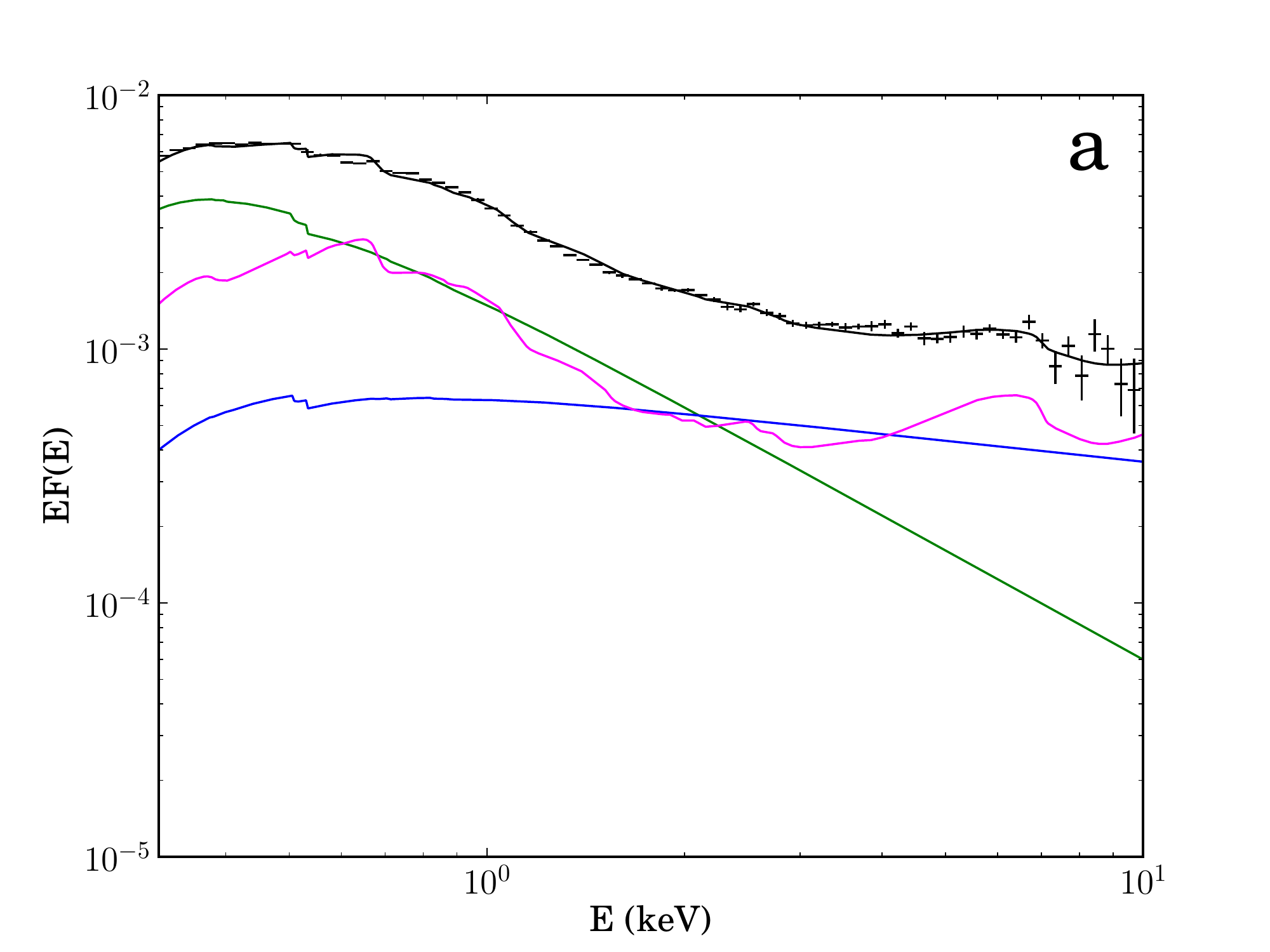} &
\includegraphics[width=8cm]{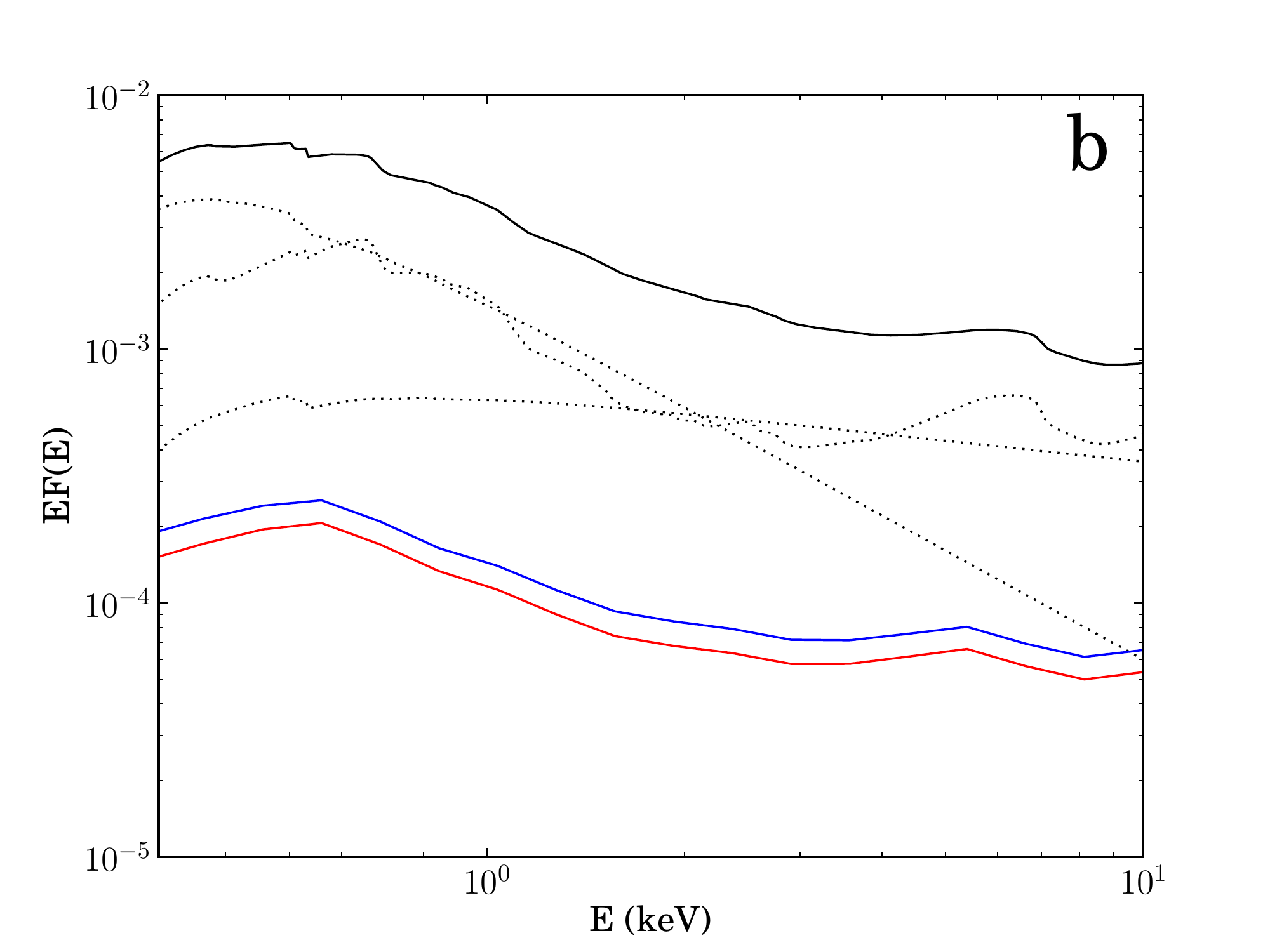} \\
\includegraphics[width=8cm]{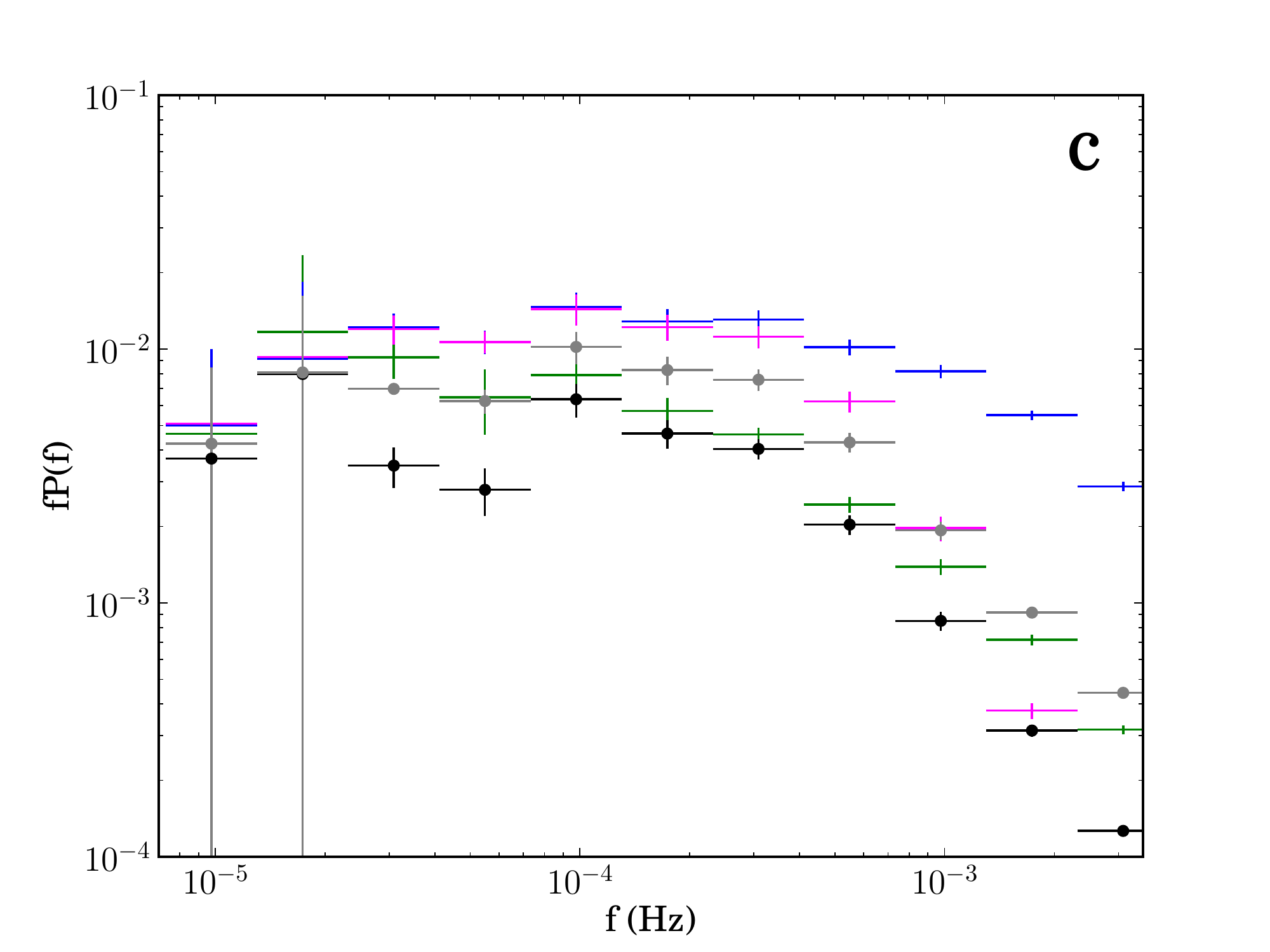} &
\includegraphics[width=8cm]{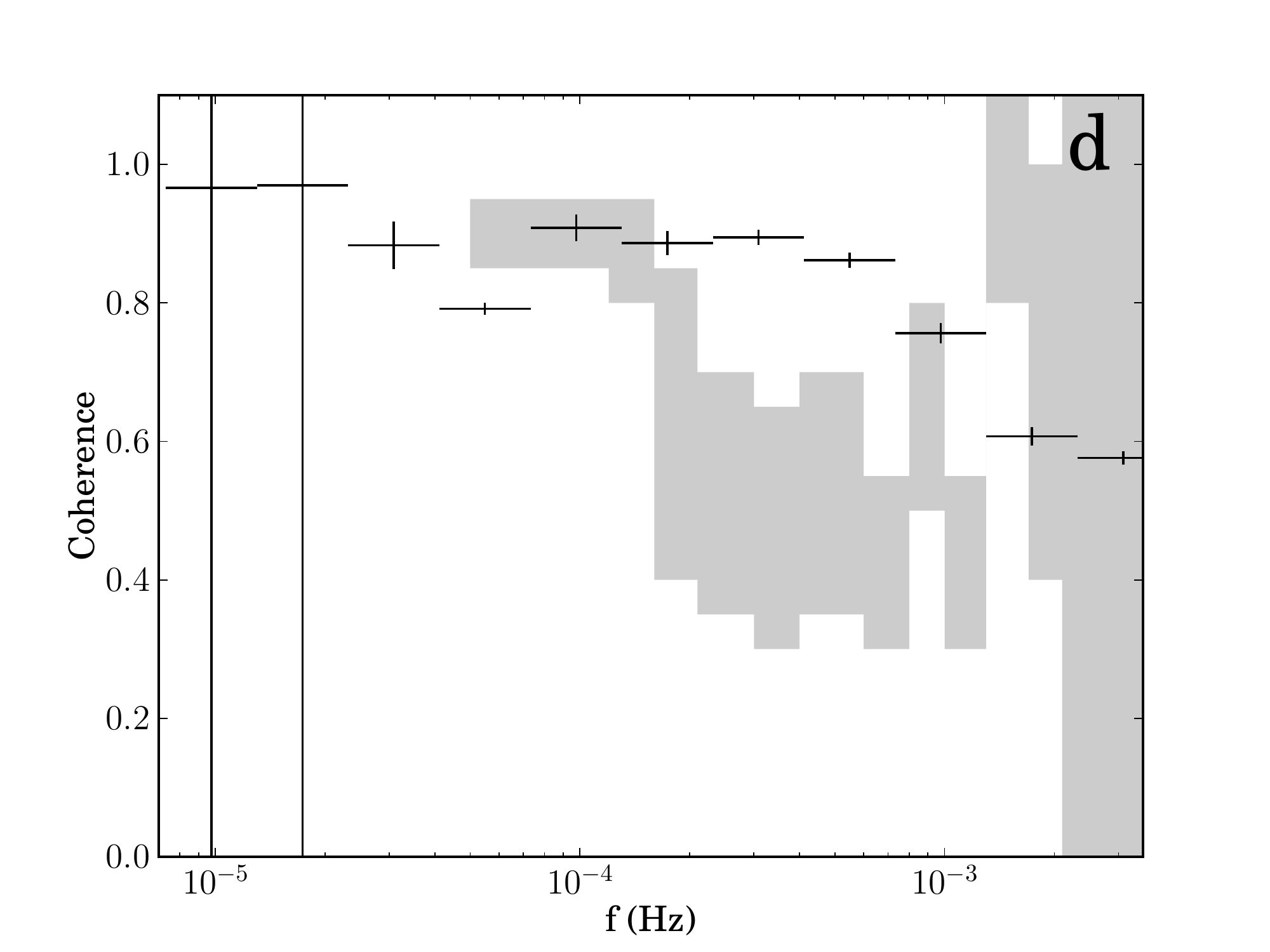} \\
\includegraphics[width=8cm]{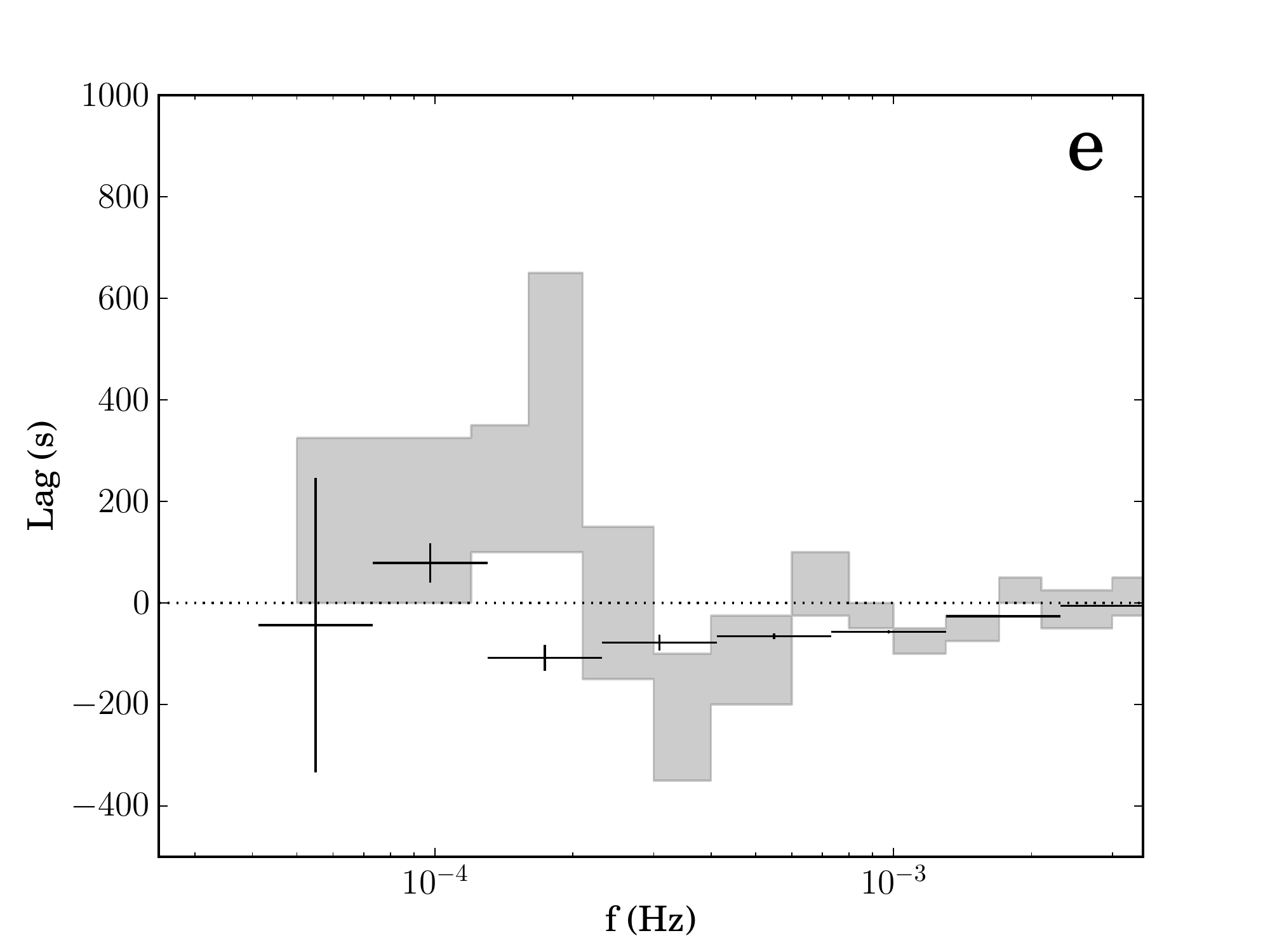} &
\includegraphics[width=8cm]{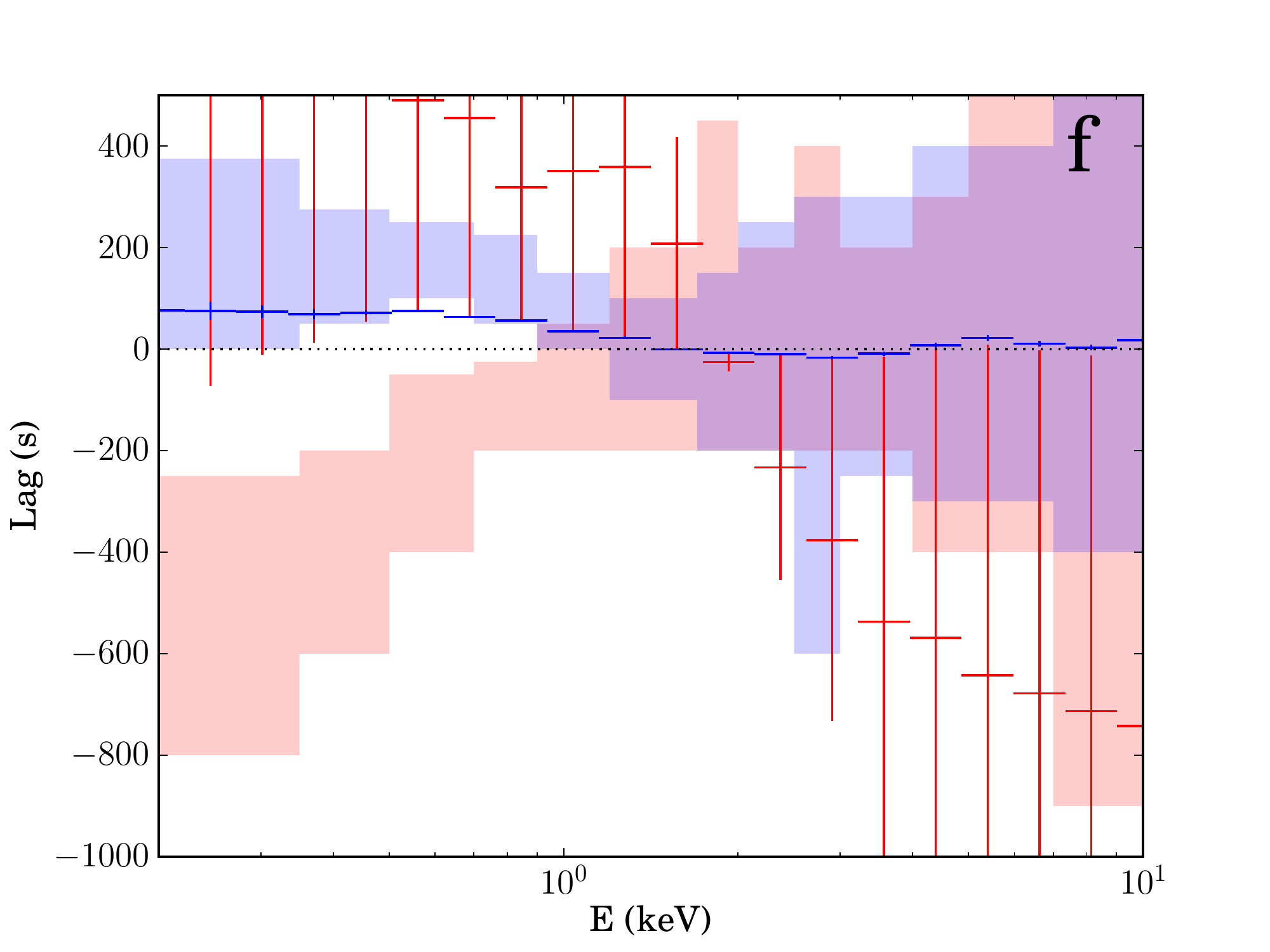} \\
\end{tabular}
\caption{Model with a hard coronal power law and its reflection with additional soft power law from the jet. $t_{lag,jet}=10^{4}s$ for propagation of fluctuations from the corona to the jet. Reflection from $R=1-12R_g$. a). Spectral decomposition: soft jet power law (green), hard coronal power law (blue) and reflection of the hard power law (magenta), with total shown in black. Data points show time averaged spectrum (OBS ID: 0675320101). b). $4-10keV$ covariance spectra for low frequencies ($2.3\times10^{-5} - 7.3\times10^{-5}Hz$, red) and high frequencies ($2.3\times10^{-4} - 7.3\times10^{-4}Hz$, blue). Black solid line shows total spectrum, dotted lines show model components. c). Power spectra: jet (green), coronal power law (blue), reflection (magenta), soft band ($0.3-1.0keV$, black), hard band ($2-10keV$, grey). d). Coherence between hard ($1.2-4keV$) and soft ($0.3-0.7keV$) bands. e). Lag-frequency spectrum between hard ($1.2-4keV$)and soft ($0.3-0.7keV$) bands. f). Lag-energy spectrum calculated using $1.2-4keV$ reference band: low frequency lag ($2.3\times10^{-5} - 7.3\times10^{-5}Hz$), red points; high frequency lag ($2.3\times10^{-4} - 7.3\times10^{-4}Hz$), blue points. Shaded regions show approximate range of error on values measured by ADV14 for PG1244+026.}
\label{fig12}
\end{figure*}

\subsection{Hard coronal power law and its reflection with additional soft power law from the jet: K13}

K13 decompose the spectrum as a soft power law, together with a harder
power law and its reflection (fig 12a, see table 2 for spectral
parameters). The soft power law then gives an additional component
which could give a source for the soft lead. However, they interpret
this as being from jet, but fluctuations propagate through the
accretion flow, and then up the jet, so the soft emission from the jet
should always lag the harder emission from the accretion flow.  In
this model there is no component of the soft emission which leads the
power law as reflection also always lags, so this cannot give an
origin for the soft lead seen in the data at low frequencies. We show
this explicitly below.

We assume the intrinsic power law variations have a broad power
spectrum consisting of four Lorentzians centred at
$f_{visc,p}=3\times10^{-5}$, $1\times10^{-4}$, $3\times10^{-4}$ and
$1\times10^{-3}Hz$, representing the accumulation of fluctuations
generated at different radii. This is necessary to replicate the broad
range of power observed in the hard band power spectrum. Since there
are no disc/soft excess components, all these frequencies must be
generated intrinsically in the power law. The jet power law peaks in
the soft band. Since the observed soft band power spectrum has less
high frequency power than the hard band, we assign the intrinsic jet
fluctuations a power spectral Lorentzian at $1\times10^{-4}Hz$. Since
the data require a drop in coherence above $10^4Hz$, we allow
fluctuations to propagate from the corona up into the jet with a lag
time of $t_{lag,jet}=10^4s$. This means any fluctuations faster than
$\frac{1}{t_{lag,jet}}$ should be smoothed out, causing a drop in
coherence at $10^{-4}Hz$. We calculate the disc transfer function
between $r_{in}=1R_g$ and $r_{out}=12Rg$, to better match the smaller
radii used to calculate the reflection spectrum in K13. We assume the
ionisation state of the reflector remains constant with both radius and time 
(Chainakun \& Young 2012).

Fig 12 shows the resulting power spectra, coherence, covariance,
lag-frequency and lag-energy spectra. The power spectra for the hard
and soft bands are nearly identical (fig 12c). This is in disagreement
with the observations, which show more high frequency power in the
hard band than the soft band (J13). This is because all three
components - soft power law, hard power law and reflection -
contribute strongly to both the hard and soft energy bands. To achieve
a drop in high frequency power in the soft band requires a long lag
time between the hard power law and the soft jet, to smooth out the
high frequency variability. But because the hard band consists of the
same three components this also produces the same effect in the hard
band. Hence the incoherence between the hard and soft power laws,
required to limit the high frequency power in the soft band, also
restricts the amount of high frequency power in the hard
band. Reducing the lag between power law and jet to increase high
frequency power in the hard band only serves to give a worse match to
the soft band power spectrum. Producing different hard and soft band
power spectra requires at least some of the components to be confined
to one band. This spectral decomposition does not meet this criteria.

The similarity of the hard and soft band power spectra means the spectra of the correlated variability at low frequencies and high frequencies are both identical to the total spectrum (fig 12b). This is in disagreement with the data, which require the spectrum of the variability correlated at high frequencies to drop off below $1keV$. 

In fig 12d we show the coherence. For this model the coherence remains high up to very high frequencies ($\sim10^{-3}Hz$). This is despite our attempts to limit the coherence with a long propagation time to the jet, and again is a result of all three components contributing strongly to both hard and soft bands. In particular the small radii used for reflection allow the reflection component to respond at high frequencies, which are coherent between hard and soft bands because reflection dominates the soft X-ray excess and also the iron line region at $\sim6.7keV$. Consequently it is the roll over in the reflected component power spectrum at $\sim4\times10^{-4}Hz$ that determines the drop in coherence, not our jet lag time. This is in disagreement with the observations, which show a drop in coherence at $\sim10^{-4}Hz$ (ADV14). 

The lag-frequency spectrum shows this model cannot reproduce the observed soft leads at low frequencies (fig 12e). This is shown more clearly in the lag-energy spectrum (fig 12f). At high frequencies the lag-energy spectrum picks out the shape of the reflection spectrum (blue points). This roughly matches the observed high frequency hard lags, although the lag times are a little short due to the very small radii required to produce the relativistically smeared reflection. In contrast the low frequency lag spectrum does not match the data at all. There is no slowly varying soft component in this model that leads the hard power law. The only slowly varying component is the soft jet power law which can only lag, since fluctuations go through the accretion flow before travelling up to the jet. Hence at low frequencies there is a soft lag which is inconsistent with the soft lead seen in the data. Above $\sim2keV$, the hard energy bins increasingly lead the hard reference band as energy increases. This is a result of the contribution from the soft jet power law to the total flux decreasing and causing less dilution of the hard power law. 

\begin{figure*} 
\centering
\begin{tabular}{l|r}
\leavevmode
\includegraphics[width=8cm]{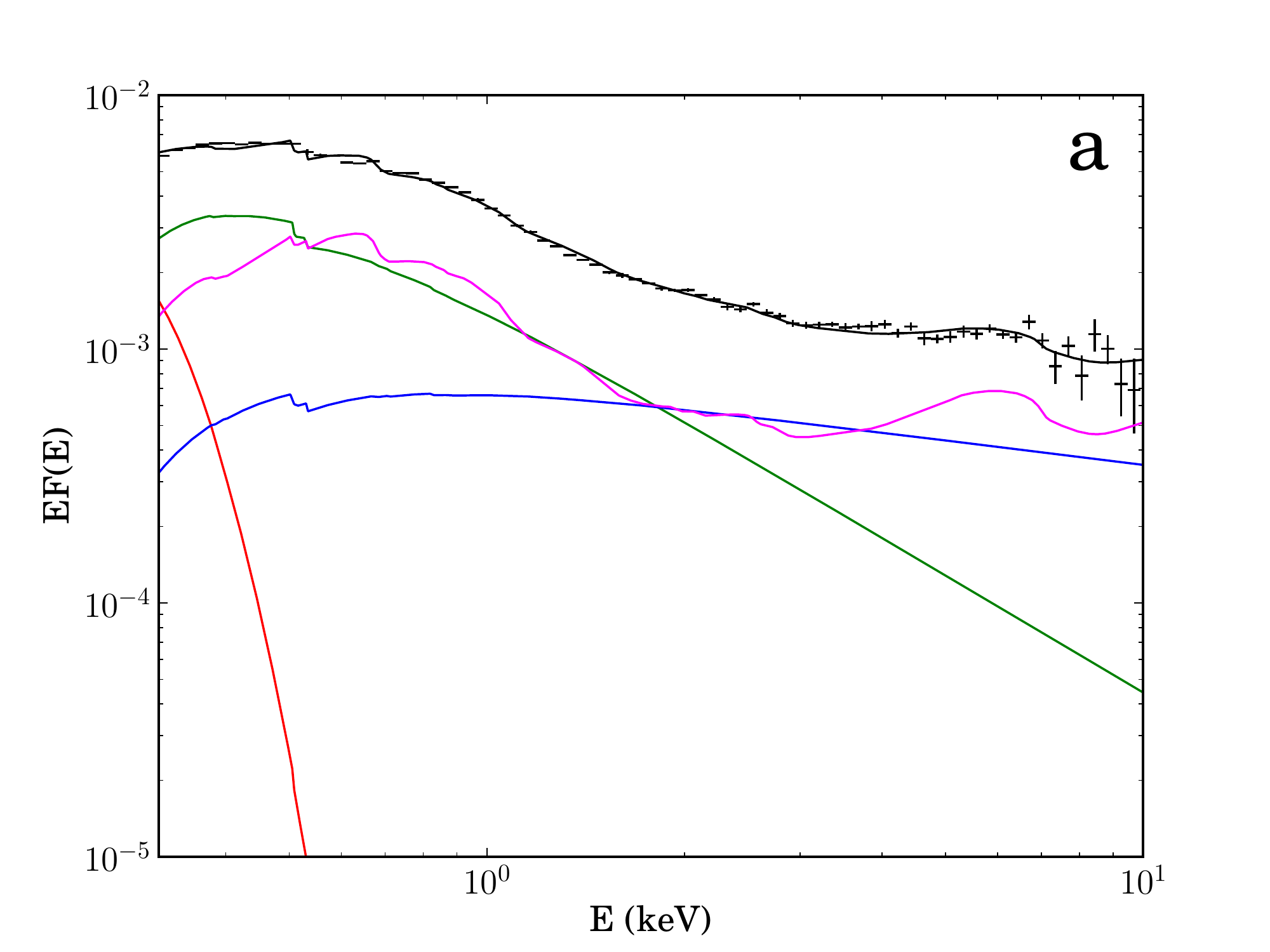} &
\includegraphics[width=8cm]{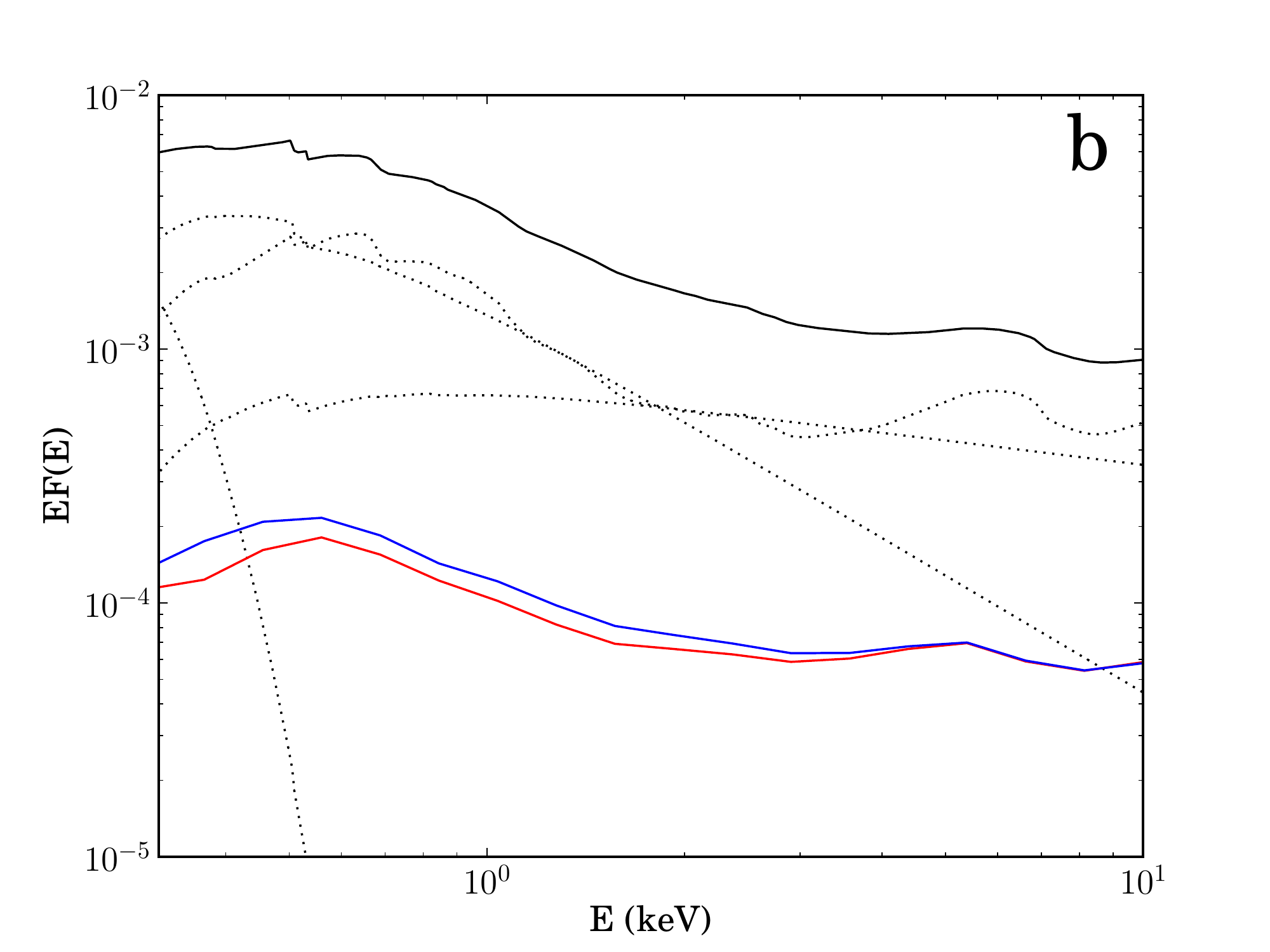} \\
\includegraphics[width=8cm]{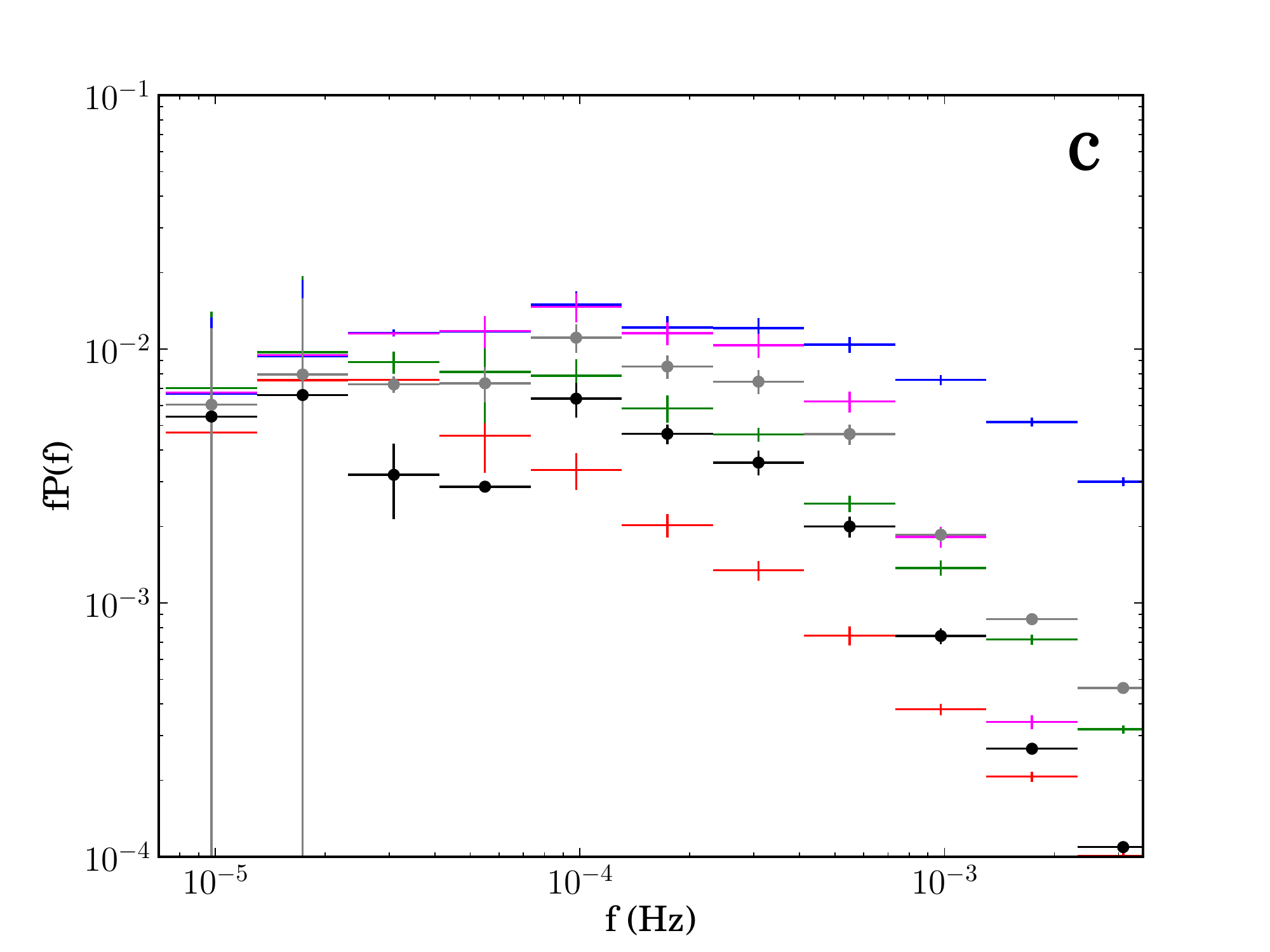} &
\includegraphics[width=8cm]{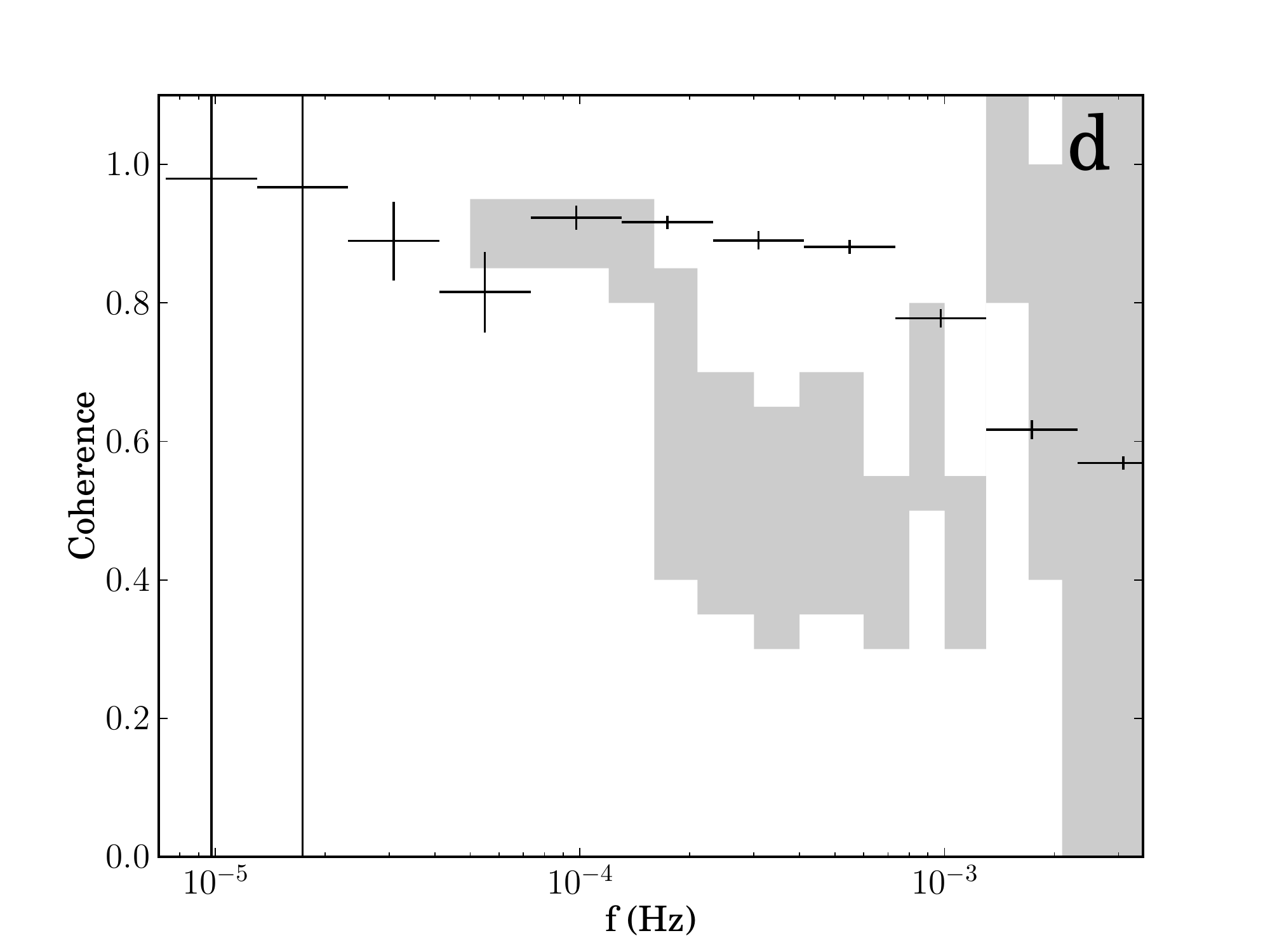} \\
\includegraphics[width=8cm]{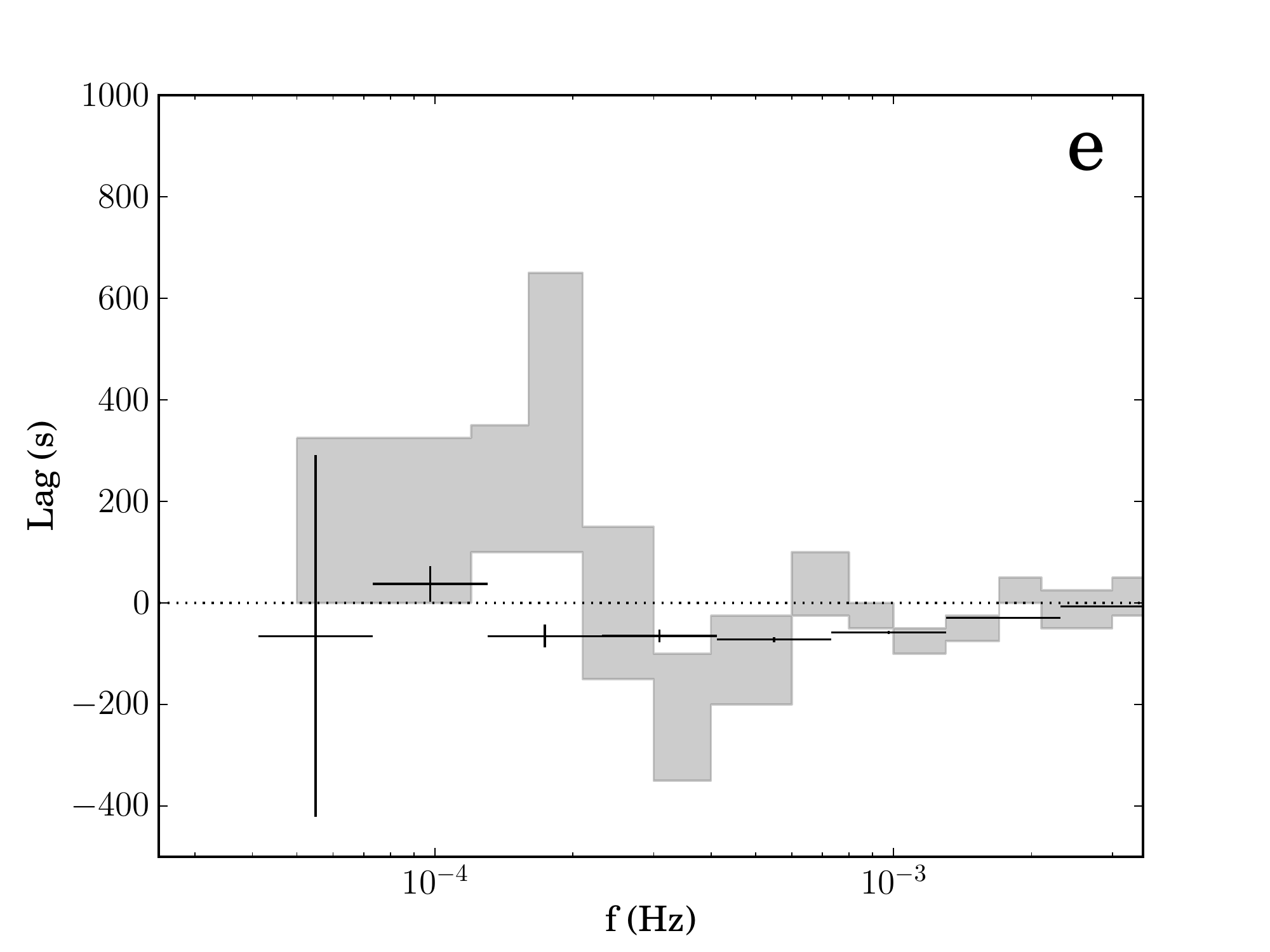} &
\includegraphics[width=8cm]{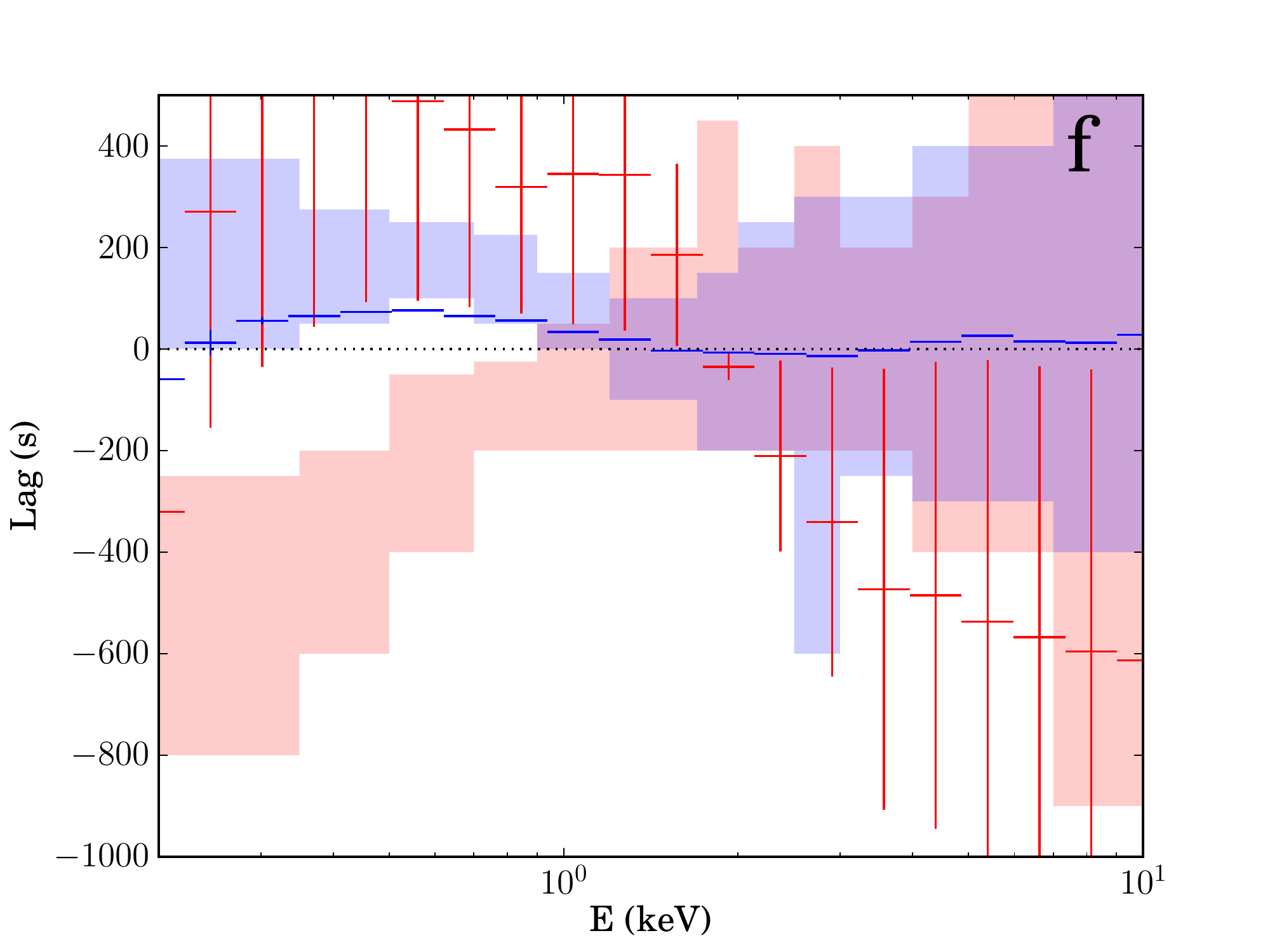} \\
\end{tabular}
\caption{As in fig 12, but for model with a hard coronal power law and its reflection with additional soft power law from the jet and a disc component (red). $t_{lag,p}=600s$ for propagation of fluctuations from the disc to the corona and $t_{lag,jet}=10^4s$ for propagation from the corona to the jet. Reflection from $R=1-12R_g$.}
\label{fig13}
\end{figure*}

\subsection{Hard coronal power law and its reflection with additional soft power law from the jet and a disc component}

We rerun the model of the previous section, this time assigning the lowest frequency fluctuations ($f_{visc}=3\times10^{-5}$Hz) to a disc component (fig 13a, see table 2 for spectral parameters), and allowing them to propagate down to the power law ($t_{lag,p}=600s$) and on to the jet ($t_{lag,jet}=10^4s$). 

\begin{figure*} 
\centering
\begin{tabular}{l|r}
\leavevmode
\includegraphics[width=8cm]{spectrum11.pdf} &
\includegraphics[width=8cm]{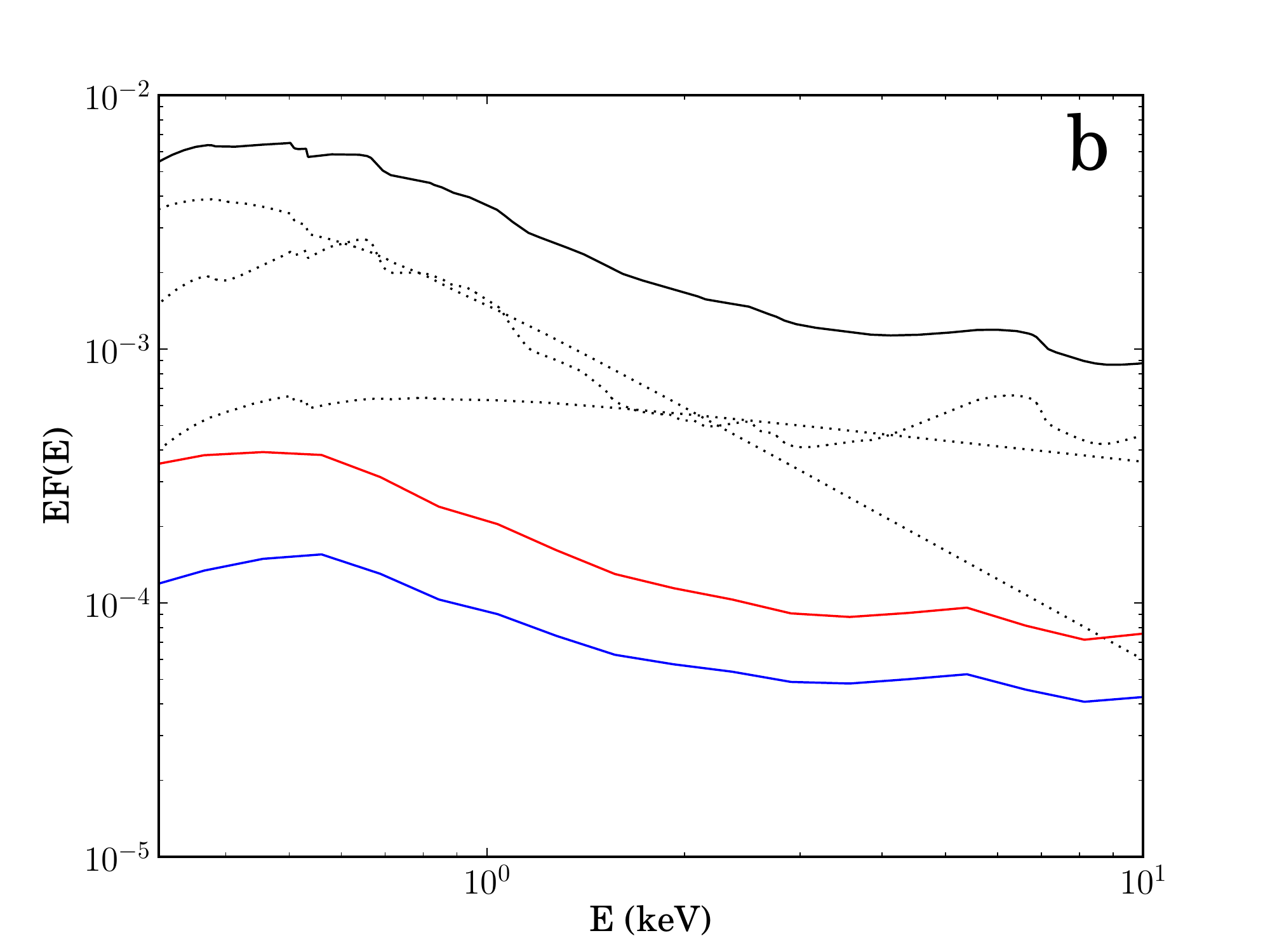} \\
\includegraphics[width=8cm]{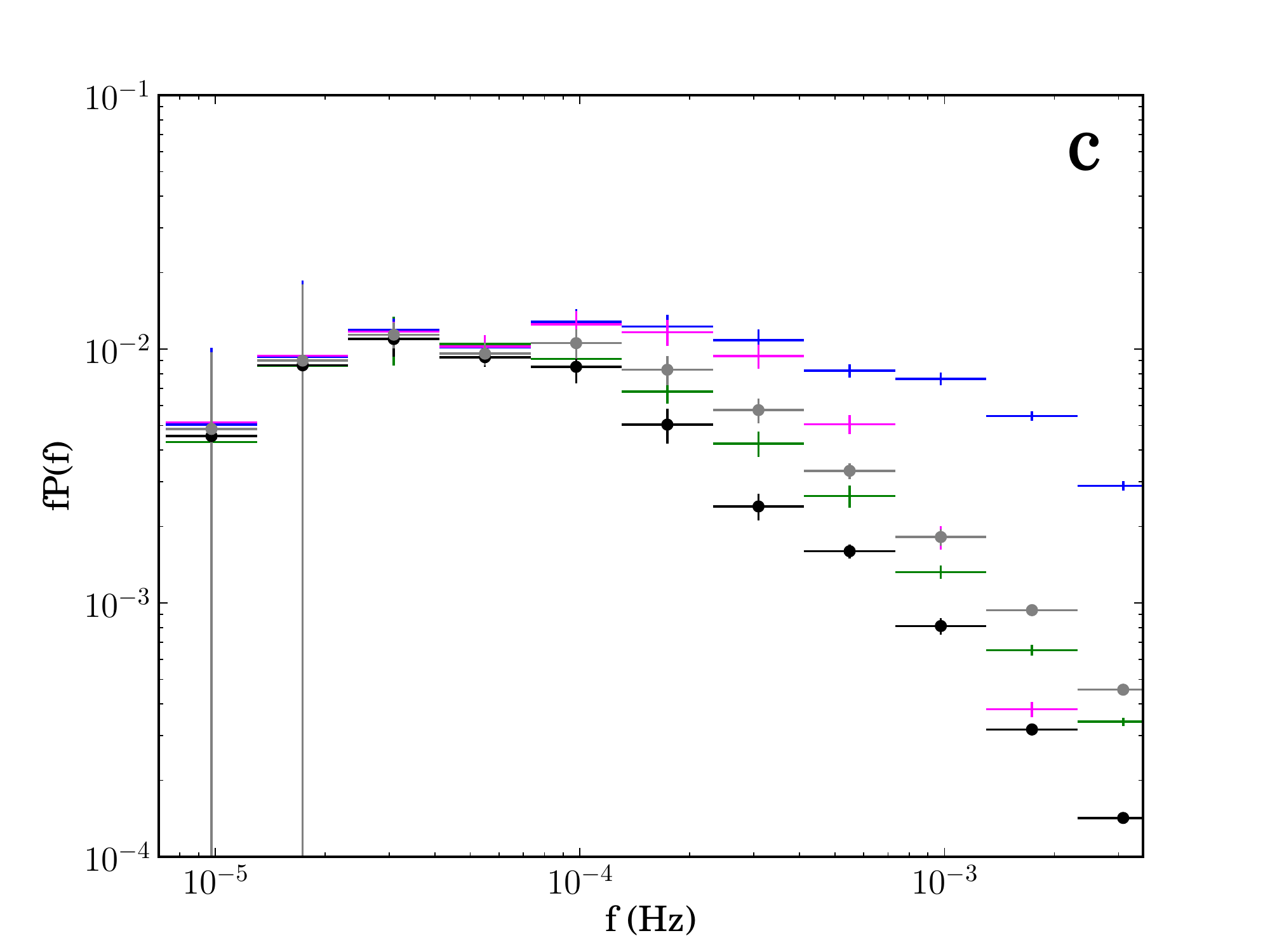} &
\includegraphics[width=8cm]{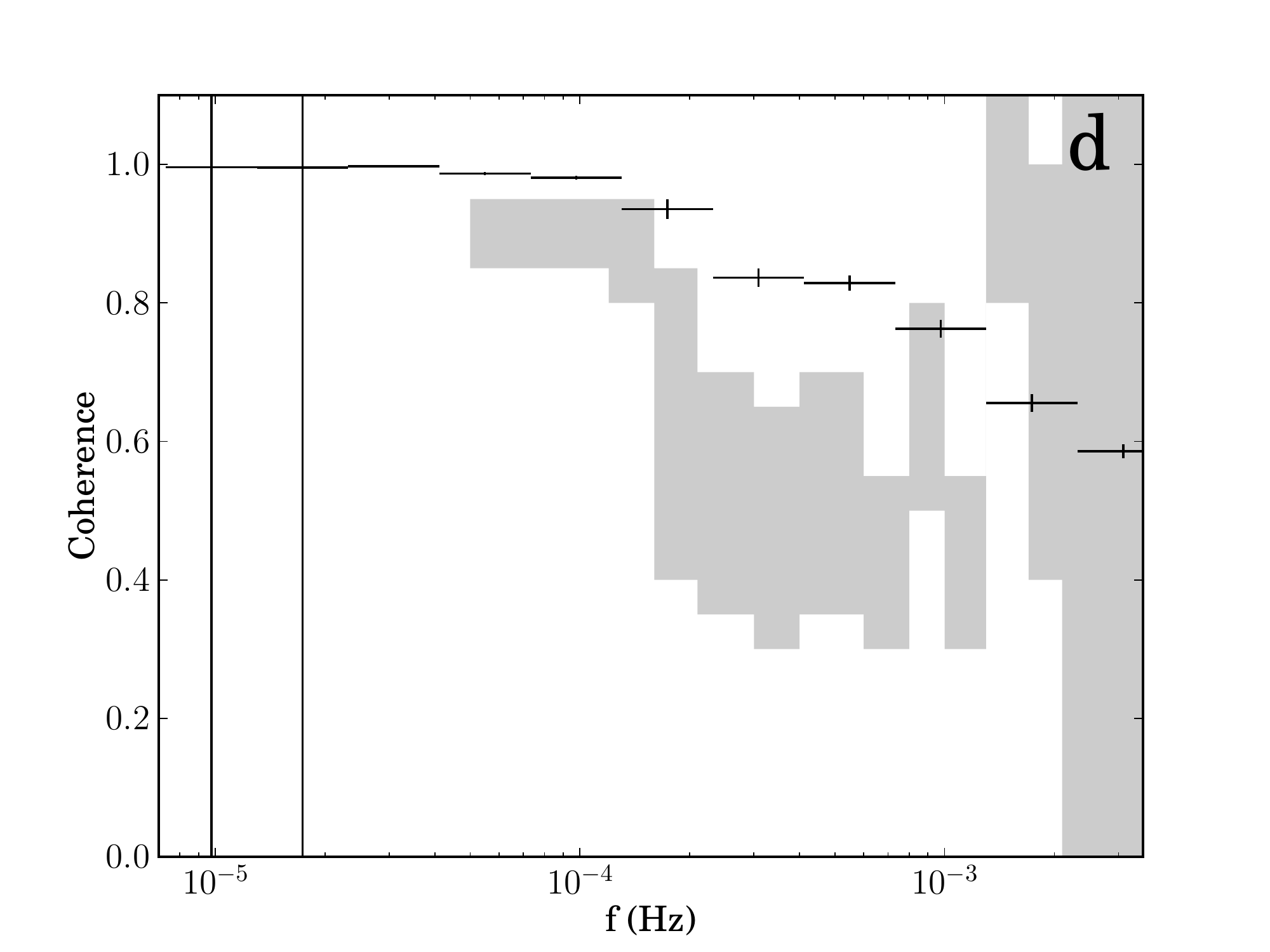} \\
\includegraphics[width=8cm]{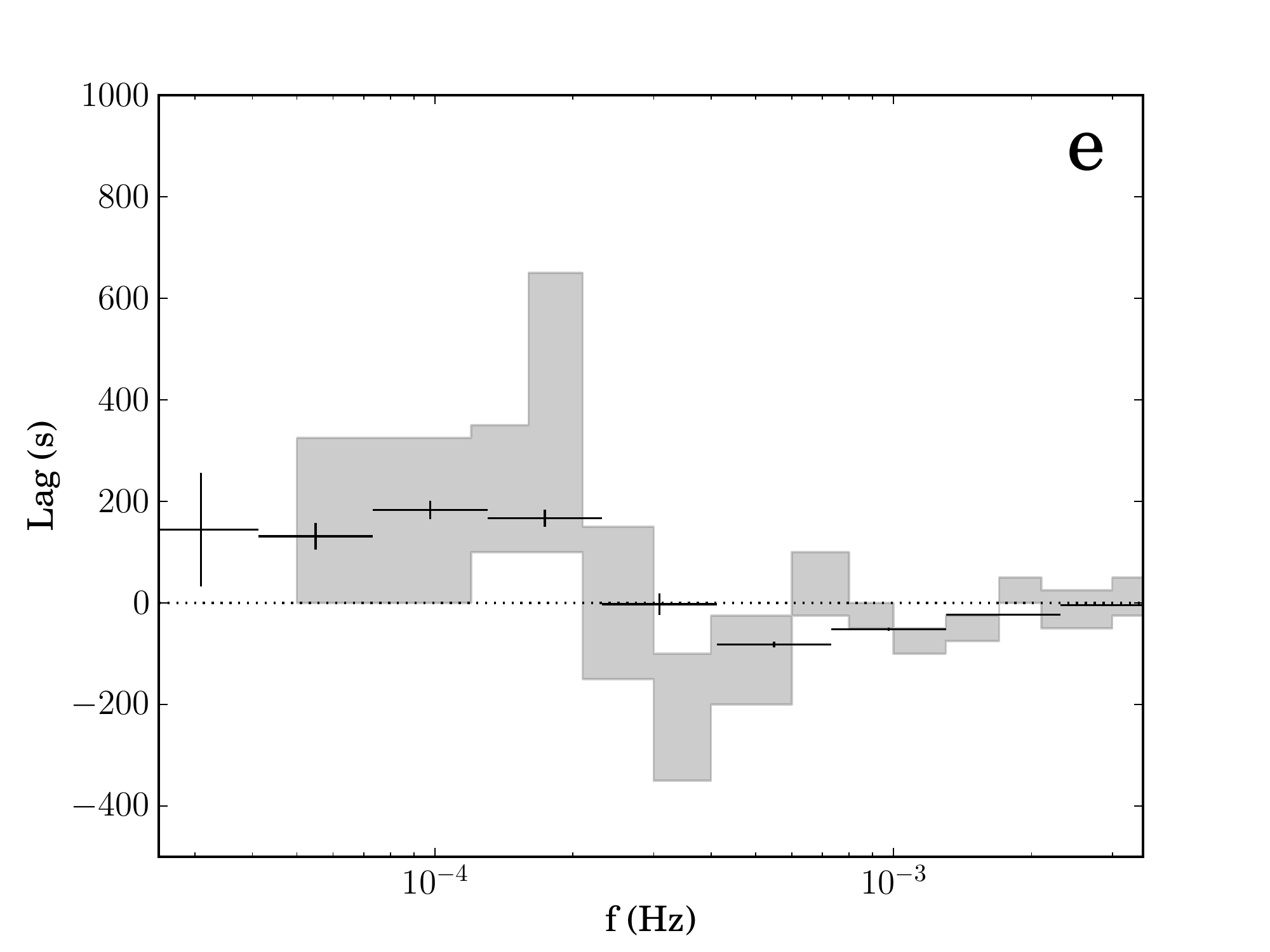} &
\includegraphics[width=8cm]{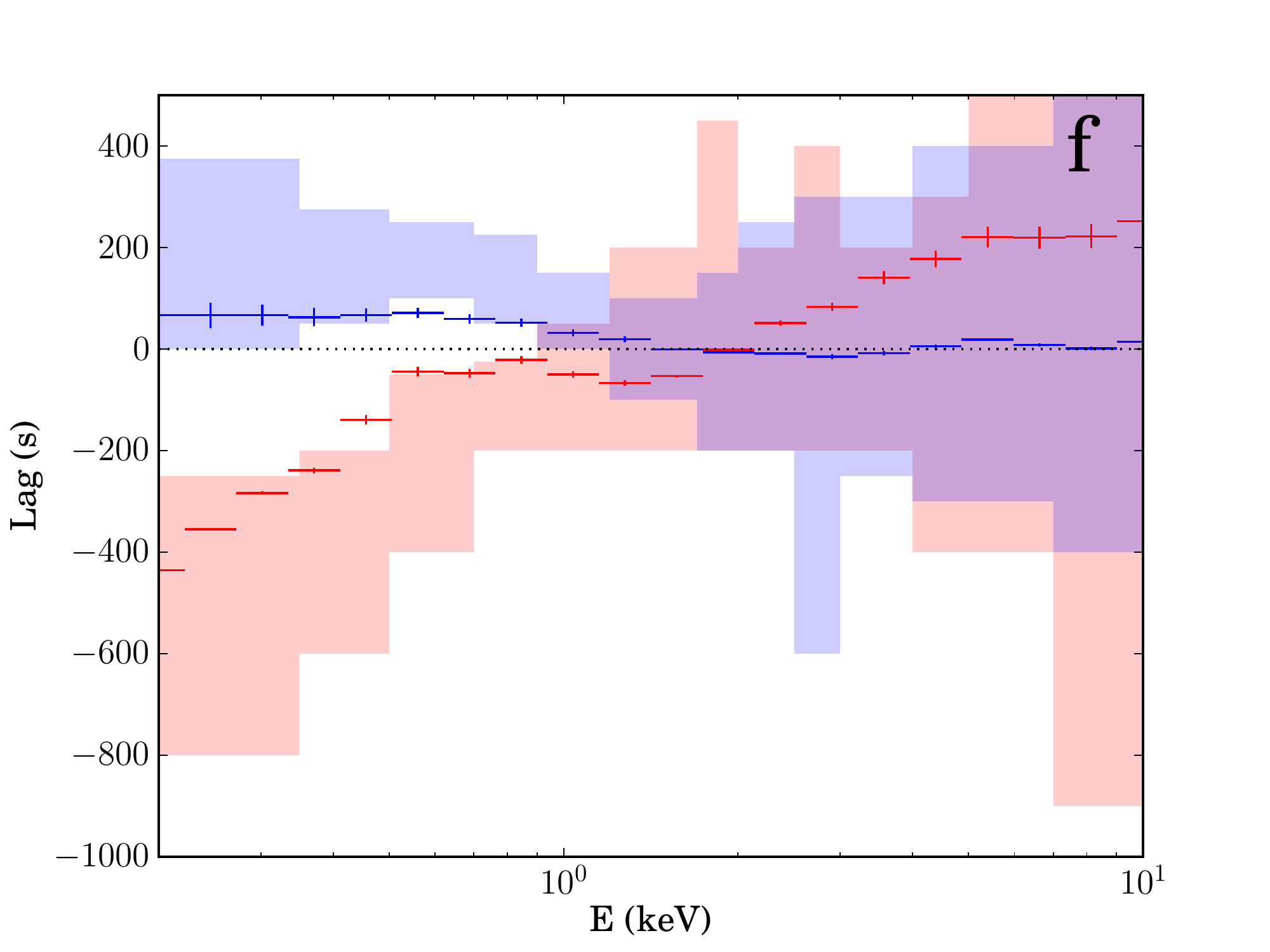} \\
\end{tabular}
\caption{As in fig 12, but for model with a hard coronal power law and its reflection with additional soft power law from the accretion flow. $t_{lag,p}=1200s$ for propagation of fluctuations from the soft power law to the hard. Reflection from $R=1-12R_g$. Low frequency lag-energy spectrum now calculated between $7.3\times10^{-5} - 2.3\times10^{-4}Hz$, high frequency between $4.1\times10^{-4} - 1.3\times10^{-3}Hz$.}
\label{fig14}
\end{figure*}

Fig 13 shows the resulting power spectra, coherence, covariance, lag-frequency and lag-energy spectra. The soft lags of the lowest energy bins $\lesssim 0.4keV$ have been diluted but only the lowest energy bin has been diluted enough to show a soft lead at low frequencies (fig 13f). Even if the propagation lag time to the jet component is reduced (to minimise dilution of the disc lead), the disc makes little contribution to the spectrum above $0.4keV$, so cannot replicate the observed soft leads up to $\sim1keV$. The dominance of the jet spectrum together with reflection, both of which lag the power law, prevent there being any soft leads at these energies. Hence extending the model to include a separate very soft component from the disc does not help, because the disc
makes too small a contribution to the soft band to give a large enough effect, and its soft lead is swamped even
at low frequencies by the soft lag from the reflection component.

\subsection{Hard coronal power law and its reflection with additional soft power law from the accretion flow}

Instead, the lags match more easily to the data if the soft power law is produced between the disc and the hard
power law region, i.e. it comes from the accretion flow rather than the jet (fig 14a).

We assume the soft power law represents the outer parts of the accretion flow, with a power spectrum consisting of two low frequency Lorentzians ($f_{visc,s}=3\times10^{-5}$ and $1\times10^{-4}Hz$). We allow these fluctuations to propagate down into the inner harder power law, which generates intrinsic fluctuations at $f_{visc,p}=3\times10^{-4}$ and $1\times10^{-3}Hz$. The inner harder power law produces the reflection. 

Fig 14 shows the resulting power spectra, coherence, covariance, lag-frequency and lag-energy spectra. The lag-frequency spectrum (fig 14e) now shows a soft lead below $2\times10^{-4}Hz$ and a soft lag above $\sim3\times10^{-4}Hz$. However, producing this soft lead requires a lag time between the soft and hard power laws double that of the previous separate soft excess model, due to the strong reflection component causing much more dilution. The lag-energy spectrum taken at the frequency of the soft lead shows the energy bins lead the hard band at energies below $\lesssim0.6keV$, where the contribution from reflection decreases (fig 14f, red points). The highest energy bins $>2keV$ lag the hard reference band at low frequencies with the lag increasing with increasing energy. This is because at $2keV$ the contributions from the hard and soft power laws are equal and above this energy the fraction of total flux contributed by the leading soft power law decreases while the fraction contributed by the lagging hard power law increases. The data do not show a systematically increasing low frequency lag at high energies, despite the large errors. This suggests whatever soft component is leading the hard power law is confined to the soft band and does not contribute significant flux at energies above $2keV$. In other words, the low temperature optically thick soft excess component of section 3, which is confined to the soft band, provides a better match to the data than an extended soft power law.

At high frequencies, the lag-energy spectrum keeps the shape of the reflection spectrum (fig 14f, blue points). However the small radii required for relativistically smeared reflection mean shorter lag times, which do not match the observed high frequency soft lags as well as the longer lag times produced by the previous soft excess model, where reflection/reprocessing occurred at larger radii (Section 3). The roll over at $0.6keV$ of the reflection spectrum also occurs at too low an energy to reproduce the observed decrease in correlated high frequency variability at $1keV$ in the covariance spectrum (fig 14b). 

Whilst changing the order in which fluctuations propagate through the components has given a better match to the lags compared to the original model in section 4.1, the spectral decomposition itself has not changed. Therefore this model has the same problems replicating the power spectra and coherence as the first model did, ie. the coherence remains higher than the observations at all frequencies (fig 14d) and the hard and soft band power spectra are still too similar, simply because both bands contain strong contributions from all three spectral components. 

\begin{figure*} 
\centering
\begin{tabular}{l|r}
\leavevmode  
\includegraphics[width=8cm]{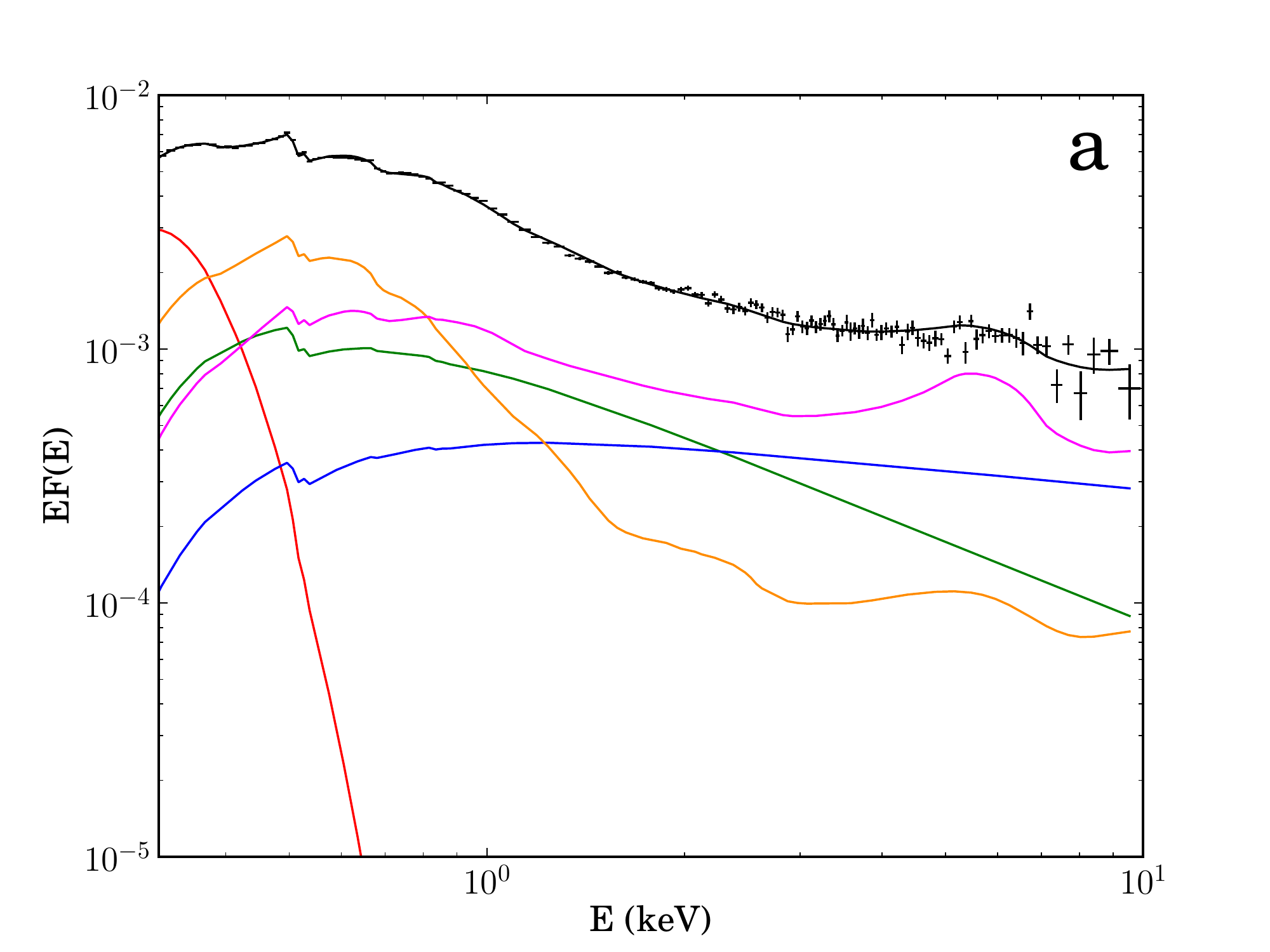} &
\includegraphics[width=8cm]{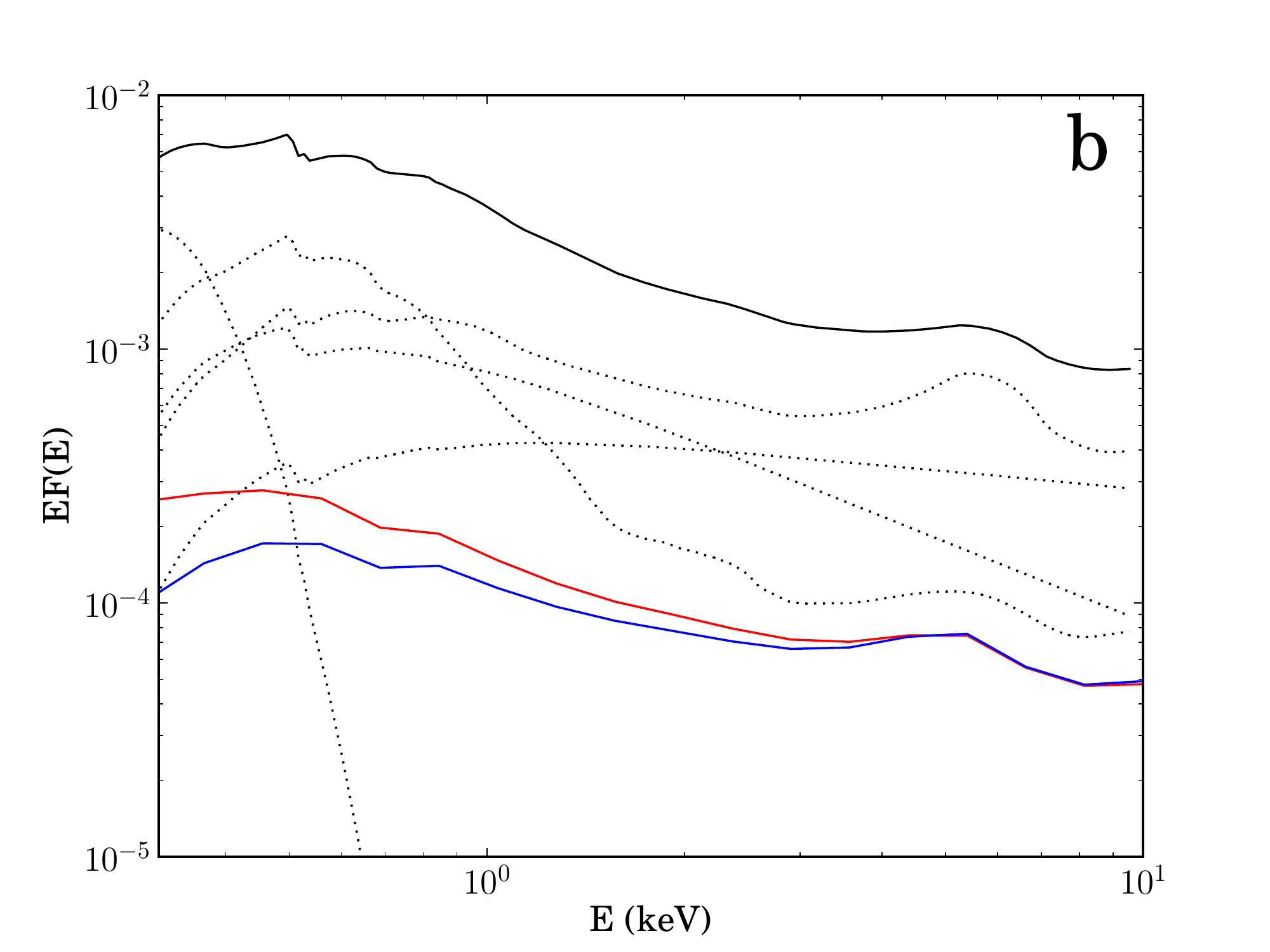} \\
\includegraphics[width=8cm]{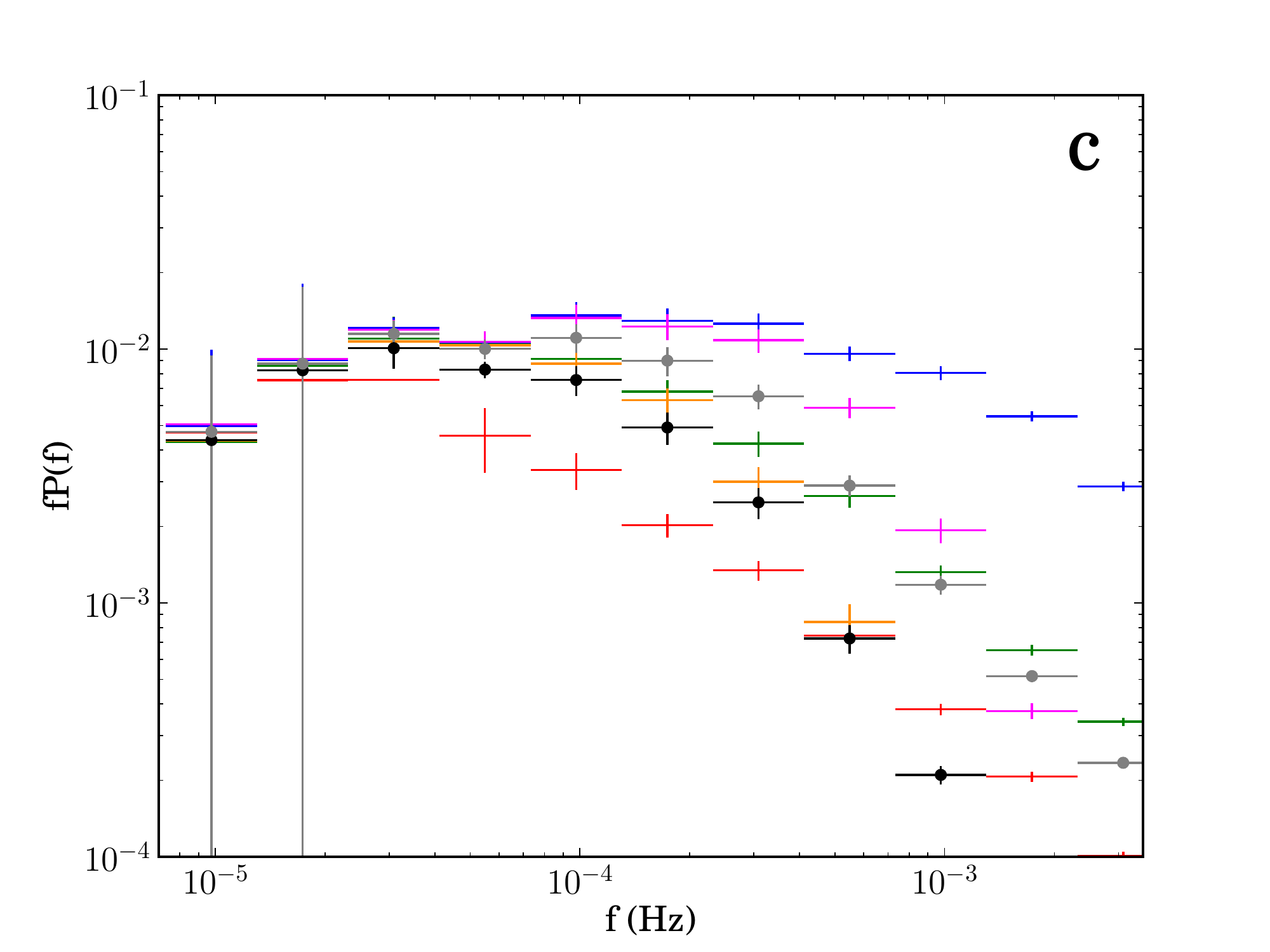} &
\includegraphics[width=8cm]{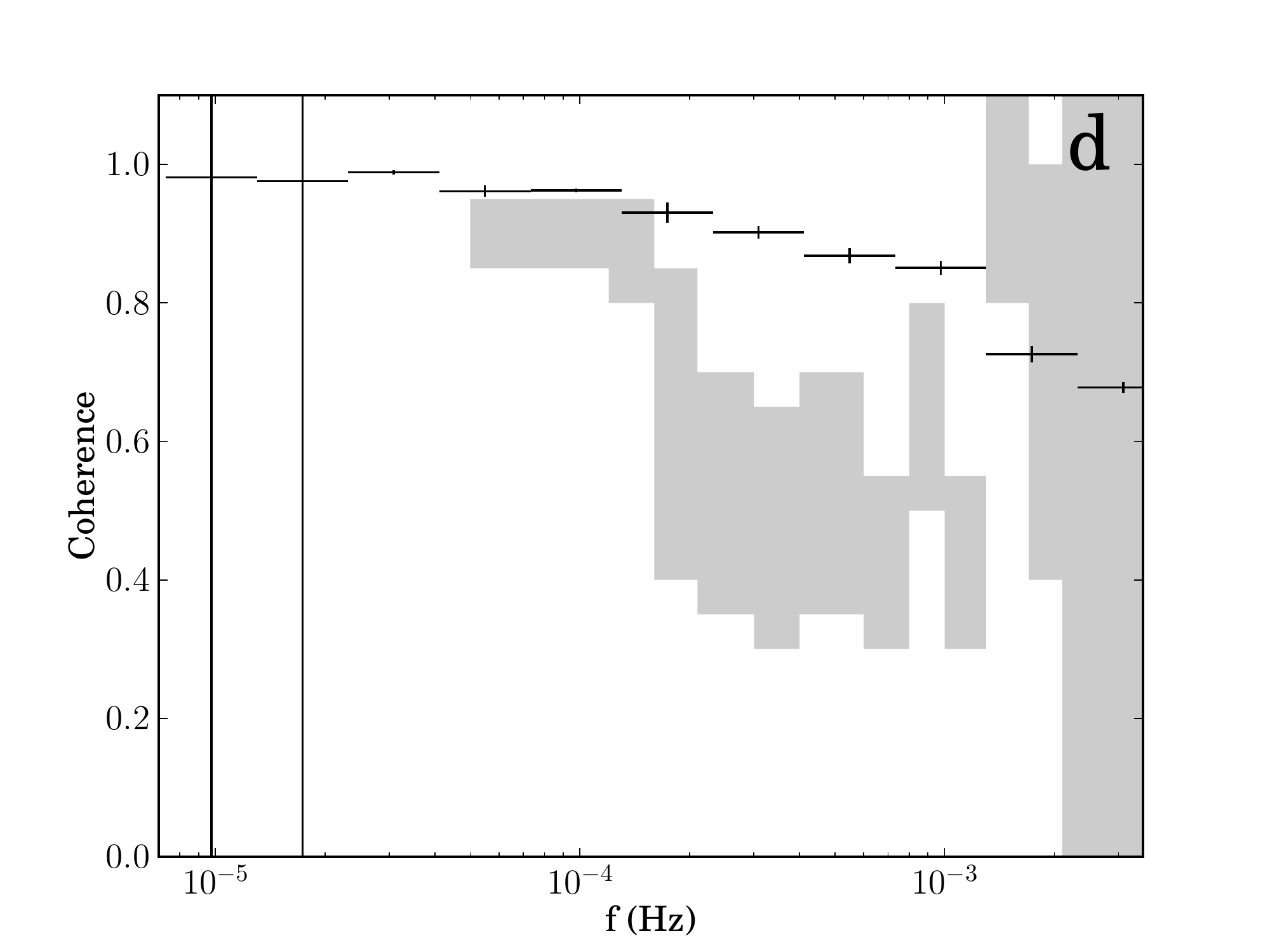} \\
\includegraphics[width=8cm]{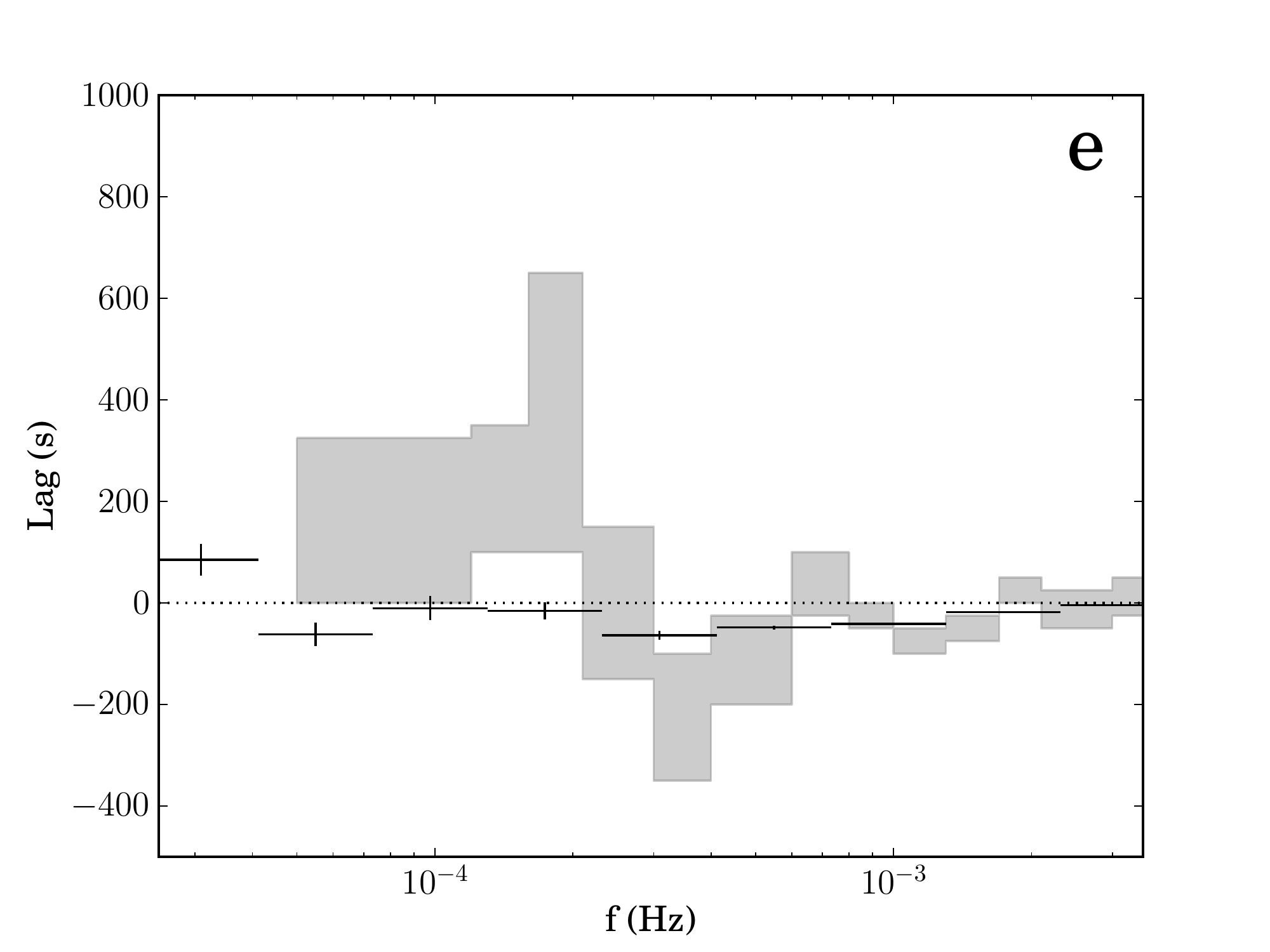} &
\includegraphics[width=8cm]{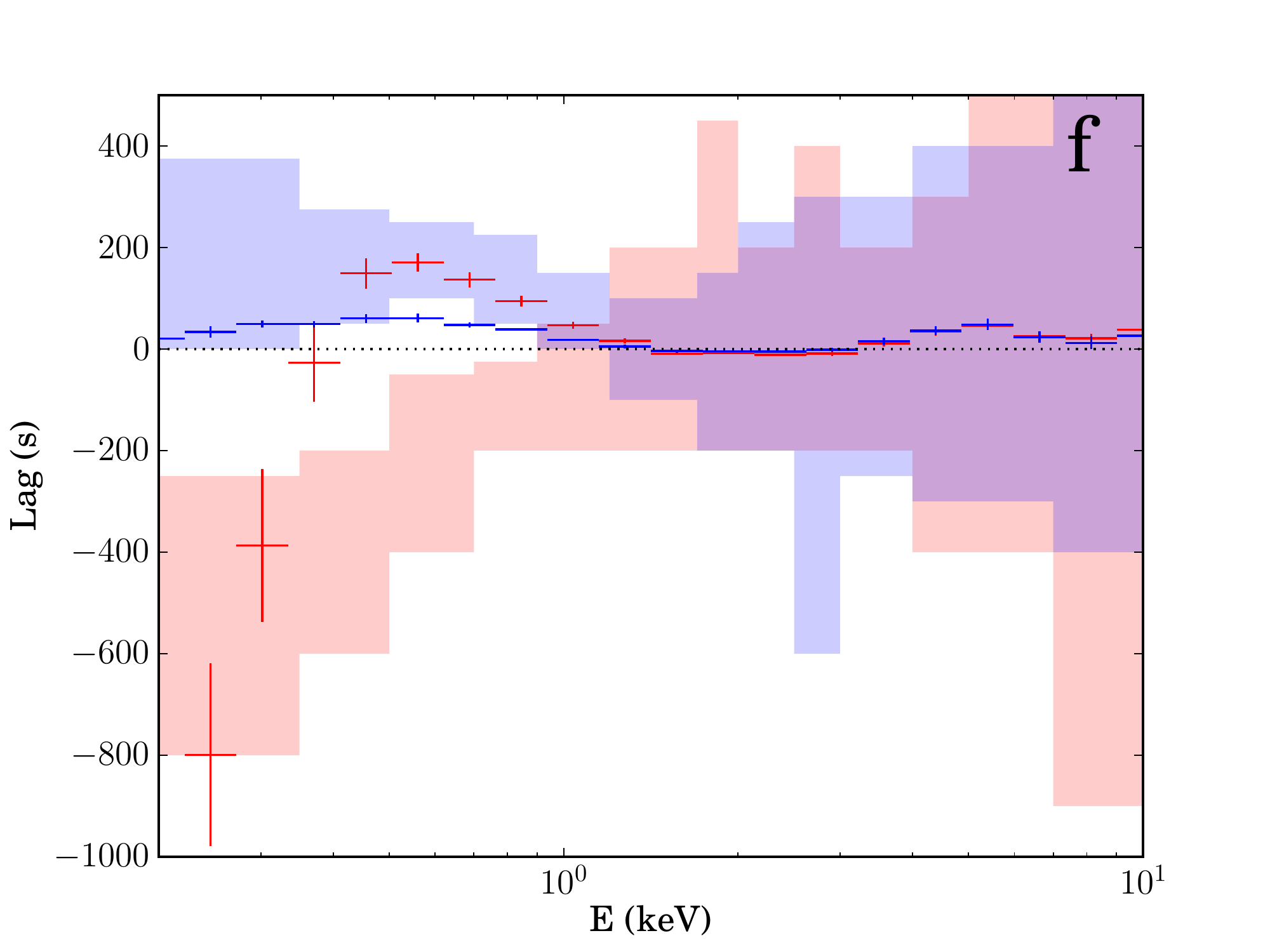} \\
\end{tabular}
\caption{As in fig 12, but for model with soft and hard coronal power laws plus their reflection (orange ($R=12-20R_g$) and magenta ($R=1-12R_g$) respectively) and a disc component (red). $t_{lag,s}=1000s$ for propagation of fluctuations from the disc to the soft power law and $t_{lag,p}=600s$ for propagation from the soft power law and to the hard.}
\label{fig15}
\end{figure*}

\subsection{Soft and hard coronal power laws plus reflection and a disc component}

Finally we test the more complex model used to describe the NLS1 1H0707-495, consisting of two power laws from the accretion flow, both with reflection spectra, and a contribution from a BB disc. 

As before, we assign the lowest frequency intrinsic fluctuations to the disc ($f_{visc,d}=3\times10^{-5}Hz$) and increase the frequency of the fluctuations from the soft power law ($f_{visc,s}=1\times10^{-4}Hz$) to the hard power law ($f_{visc,p}=3\times10^{-4}$ and $1\times10^{-3}Hz$). Since the soft power law represents the outer parts of the accretion flow, we calculate the disc transfer function for its reflection between $12-20Rg$. We assume the inner hard power law is reflected from the inner regions of the disc ($1-12Rg$). We set $t_{lag,s}=1000s$ for propagation of fluctuations from the disc to the soft power law and $t_{lag,p}=600s$ for propagation of fluctuations from the soft power law and to the hard.

Fig 15 shows the resulting power spectra, coherence, covariance, lag-frequency and lag-energy spectra. Including two reflection spectra swamps the propagation leads at low frequencies (fig 15e). They are only visible in the lag-energy spectrum in the lowest energy bins dominated by the disc (fig 15f, red points). The high frequency lag-energy spectrum (fig 15f, blue points) lacks a strong lag at the lowest energies due to the roll over of the reflection spectra and the lack of any reprocessing on the disc. The soft lags are also still too short to match the data, despite including reflection from slightly larger size scales because there is strong dilution by the other four components. 

Moreover, this model suffers from the same problem as the three previous reflection models: the hard and soft bands both contain strong contributions from (nearly) all components, leading to high coherence at all frequencies and very similar power spectra for both bands (fig 15b, c, d). In contrast the data require low frequencies to be generated in the soft band, high frequencies to be generated in the hard band, longer disc transfer functions to limit the amount of high frequency power transmitted back to the soft band and, importantly, for the main reverberation response to be concentrated in the soft band not spread over both bands.

\section{Discussion and Conclusions}

Time variability gives additional information which can break spectral degeneracies. Any successful
model must be able to fit both spectral and timing properties of the data. Here we show quite 
generally that the switch in behaviour from soft leading hard at 
low frequencies to soft lagging hard at high frequencies favours a model where the majority of the soft X-ray excess is 
not produced by reflection, but consists of a combination of emission from the accretion flow and reprocessed emission. We show this explicitly using the 'simple' NLS1 PG1244+026 as
an example. Reflection dominated models including a soft power law from the jet (K13) do not produce soft leads at 
low frequencies, as fluctuations propagate to the jet only after they have gone through the 
accretion flow. 

If the additional soft power law is from the disc rather than the jet, so the 
fluctuations propagate from the soft power law to the hard, these models can produce soft leads but they do not match the shape of the observed lag-energy spectrum, due to the soft power law extending into the hard band, and they cannot match the other observed timing features of the data (see Section 4). The fact that reflection contributes a large fraction of the flux in both the hard and soft energy bands means these models produce high coherence over a wide frequency range, and very similar hard and soft band power spectra. This is made worse by the small size scales over which the reflection occurs (necessary to produce such strongly relativistically smeared emission), which allow the reflection component to respond at very high frequencies. All this is at odds with the observations, which show a drop in coherence between the hard and soft bands at high frequencies, and much less high frequency power in the soft band compared to the hard. 

Since the argument is quite general, it is likely that the
similar model (disc plus soft and hard power laws and their reflection) used for 'complex' NLS1s e.g. 1H0707-495, cannot explain their data either. All the papers using this model concentrate on 
how this fits the soft lags at high frequencies (Zoghbi et al 2011; Kara et al 2013a; 2013b), but the soft leading at low frequencies also
needs to be explained and any spectral decomposition {\emph{must}} be able to replicate the observed power spectra and coherence. 

Instead, the switch in lag behaviour can be explained if the soft X-ray excess is dominated by  
an intrinsically curved component, such as a low temperature, high optical depth Compton 
component. Fluctuations start in the disc, propagate down through this intrinsic soft excess component 
and then into the power law. Reflection alone does not produce enough soft lag in these models
as the reflected emission makes too small a contribution to the soft X-ray bandpass. However, 
the non-reflected flux should be reprocessed, and this reprocessed emission
reverberates in a similar way to reflection, as is seen in the black hole binaries (Uttley et al 2011). Importantly this response is concentrated in the soft band and high frequencies are filtered out due to the larger radii involved. We explore
two models of this reprocessed emission, one where it thermalises to the disc temperature, and one where
it thermalises to produce part of the soft excess. The low temperature of the disc in our spectral models means that it does not contribute enough flux to the soft X-ray band to turn the propagation soft lead into a lag, but thermalising to the soft excess can match the switch in behaviour of the lag as a function of frequency. In reality there is likely to be some reprocessing on both the disc and soft excess. We show that a model including both gives a good match to all the timing observations. 

An important additional finding from our simulations is that the observed lag-energy spectra appear to show no evidence of spectral pivoting of the coronal power law; the power law spectral slope does not vary on short timescales, and instead only changes in normalisation. This lack of spectral pivoting could be explained if there are correlated changes in coronal power and seed photons, perhaps through evaporation of material from the soft excess into the corona. Alternatively this could suggest that the coronal emission mechanism in NLS1s is non-thermal, so that the seed photons only experience one scattering before escaping from the corona. 

A model for the soft X-ray excess in which it consists of a combination of intrinsic 
emission from the accretion flow, together with reprocessed emission, can fit all the current
spectral and timing properties of PG1244+026. Such models do not require inner disc radii smaller than $\sim6R_g$. In contrast, a reflection dominated model for the soft X-ray excess can only replicate the high frequency soft lags and cannot match the observed power spectra, coherence or covariance. This is a direct consequence of the extremely small radii ($\sim1R_g$) required to produce these spectra and the dominance of the smeared reflection component in both the hard and soft energy bands. 

However, we note that neither model can explain
the possible detection of very high frequency soft lags by ADV14. We show that the model
proposed by ADV14, where the irradiated disc fluctuations propagate down to the soft excess, cannot quantitatively explain this feature. If it is confirmed with a 0.5Ms XMM-Newton observation (PI: Alston, 
data scheduled to be taken this year) then this requires some additional reverberation
signal from smaller size scales.

\section{Acknowledgements}

EG acknowledges funding from the UK STFC. CD acknowledges useful conversations with Will Alston and Simon Vaughan on how to calculate and interpret lag spectra, and with Giorgio Matt on the soft excess in Akn 120.

\appendix
\section{}

For two light curves, $s(t)$ and $h(t)$, with Fourier transforms $S(f)$ and $H(f)$, we define the complex valued cross spectrum as (Vaughan \& Nowak 1997; Nowak et al 1999):

\begin{equation}
C(f) = S^{\ast}(f)H(f)
\end{equation}

\noindent Where $S^{\ast}$ denotes the complex conjugate of $S$. The coherence is then:

\begin{equation}
coh(f) = \frac{\left| \langle C(f) \rangle \right|^2}{\langle\left|S(f)\right|^2\rangle \langle\left|H(f)\right|^2\rangle}
\end{equation}

\noindent Where angle brackets denote an average over multiple light curve segments.

The lag is then calculated as: 

\begin{equation}
lag(f) = \frac{\arg[C(f)]}{2\pi f}
\end{equation}

The covariance between the two lightcurves, if $h(t)$ is the reference band, is calculated in the time domain as (Wilkinson \& Uttley 2009): 

\begin{equation}
cov = \frac{\sum(s(t)-\bar{s})(h(t)-\bar{h})}{(N-1)\sqrt{\sigma_h^2}}
\end{equation}

\noindent Where $N$ is the number of time steps in the light curve, $\bar{s}$ and $\bar{h}$ are the mean fluxes of $s(t)$ and $h(t)$ respectively, and the factor of $\sqrt{\sigma_h^2}$ is required for normalisation. $\sigma_h^2$ is the excess variance of the reference band light curve, defined as:

\begin{equation}
\sigma_h^2 = \frac{\sum(h(t)-\bar{h})^2}{(N-1)}
\end{equation}

We normalise all power spectra ($P(f)=\left|H(f)\right|^2$), such that:

\begin{equation}
\int_0^\infty P(f) df = \frac{\sigma^2}{I^2}
\end{equation}

\noindent Where $\sigma^2$ is the excess variance of the light curve and $I$ is its mean flux.

\label{lastpage}

\end{document}